%% file: arxiv.tex
\documentclass[useAMS,usenatbib]{mnras}
\usepackage{natbib}
\usepackage{graphicx}
\usepackage{aecompl}

\include{smcsedfit_definitions}

\title[The dust budget of the SMC]{The evolved-star dust budget of the Small Magellanic Cloud: the critical role of a few key players}
\author[S. Srinivasan, M. L. Boyer, F. Kemper, M. Meixner, D. Riebel and B. A. Sargent]{S. Srinivasan$^{1}$\thanks{E-mail: sundar@asiaa.sinica.edu.tw}, M. L. Boyer$^{2,3}$, F. Kemper$^{1}$, M. Meixner$^{4}$, B. A. Sargent$^{5}$ and D. Riebel$^{6}$\\
$^{1}$ Institute of Astronomy \& Astrophysics, Academia Sinica, 11F, Astronomy-Mathematics Building,\\ No. 1, Roosevelt Rd., Sec 4, Taipei 10617, Taiwan (R. O. C.)\\
$^{2}$ CRESST and Observational Cosmology Lab, Code 665, NASA Goddard Space Flight Center, Greenbelt, MD 20771 USA\\
$^{3}$ Department of Astronomy, University of Maryland, College Park, MD 20742 USA\\
$^{4}$ Space Telescope Science Institute, 3700 San Martin Drive, Baltimore, MD 21218, USA\\
$^{5}$ Center for Imaging Science and Laboratory for Multiwavelength Astrophysics, Rochester Institute of Technology,\\ 54 Lomb Memorial Drive, Rochester, NY 14623, USA\\
$^{6}$ Department of Physics, United States Naval Academy, 572C Holloway Road, Annapolis, MD 21402, USA\\
}
\begin{document}

\date{Accepted . Received ; in original form }

\pagerange{\pageref{firstpage}--\pageref{lastpage}} \pubyear{2015}

\maketitle

\label{firstpage}

\begin{abstract}
The lifecycle of dust in the interstellar medium (ISM) is heavily influenced by outflows from asymptotic giant branch (AGB) and red supergiant (RSG) stars, a large fraction of which is contributed by a few very dusty sources. We compute the dust input to the Small Magellanic Cloud (SMC) by fitting the multi-epoch mid-infrared spectral energy distributions (SEDs) of AGB/RSG candidates with models from the {\em G}rid of {\em R}SG and {\em A}GB {\em M}odel{\em S} (GRAMS) grid, allowing us to estimate the luminosities and dust-production rates (DPRs) of the entire population. By removing contaminants, we guarantee a high-quality dataset with reliable DPRs and a complete inventory of the dustiest sources. We find a global AGB/RSG dust-injection rate of  $(1.3\pm 0.1)\times 10^{-6}$ \msunperyr, in agreement with estimates derived from mid-infrared colours and excess fluxes. As in the LMC, a majority (66\%) of the dust arises from the extreme AGB stars, which comprise only $\approx$7\% of our sample. A handful of far-infrared sources, whose 24 \mic\ fluxes exceed their 8 \mic\ fluxes, dominate the dust input. Their inclusion boosts the global DPR by $\approx$1.5$\times$, making it necessary to determine whether they are AGB stars. Model assumptions, rather than missing data, are the major sources of uncertainty; depending on the choice of dust shell expansion speed and dust optical constants, the global DPR can be up to $\approx$10 times higher. Our results suggest a non-stellar origin for the SMC dust, barring as yet undiscovered evolved stars with very high DPRs.
\end{abstract}


\section{Introduction}
The origin of dust in the interstellar medium (ISM) is a matter of debate.  The major contributors of dust to the Galactic Disc are
supernovae (SNe) and asymptotic giant branch (AGB) stars, with less than 5\% of the ISM dust arising from other sources such as red
supergiant (RSG) and Wolf-Rayet stars, novae, planetary nebulae, and other massive stars \citep*{Gehrz1989,Tielensetal2005,Draine2009}.
Presolar dust grains found in meteorites point to a supernova origin for carbonaceous grains (graphite) and an AGB origin for silicates \citep[e.g.,][]{Gailetal2009,Nguyenetal2010}.  However, when supernova dust destruction is taken into account, it appears that these dust sources cannot account for the current dust mass observed in the ISM \citep*[e.g.,][]{Dwek1998,Zhukovskaetal2008}.\\

The Magellanic Clouds are ideal for studying the life cycle of dust on stellar to galaxy scales owing to their proximity, viewing angle and the low line-of-sight extinction. With the launches of the {\it Spitzer Space Telescope} \citep[{\it Spitzer};][]{Werneretal2004,Gehrzetal2007} and the {\it Herschel Space Observatory} \citep[{\it Herschel};][]{Pilbrattetal2010}, mid- to far-infrared (IR) observations of entire nearby galaxies including the Clouds became possible, allowing direct and unbiased comparisons between the dust produced by individual stars and the current ISM dust. The {\it Spitzer} Legacy Program SAGE (Surveying the Agents of Galaxy Evolution; PI: M. Meixner) imaged the Large Magellanic Cloud \citep[LMC;][]{Meixneretal2006} and the Small Magellanic Cloud \citep[SMC;][]{Gordonetal2011}, spawning follow-up studies that obtained mid-IR spectroscopic data in the LMC \citep[SAGE-Spec;][]{Kemperetal2010} and SMC \citep[SMC-Spec;][]{Ruffleetal2015} as well as {\it Herschel} observations \citep[HERITAGE;][]{Meixneretal2010,Meixneretal2013}.\\

An accurate estimate of the dust input requires radiative transfer calculations to obtain the dust-production rate (DPR) and dust chemistry of each star in the sample. An alternative to this computationally intensive method is to use observational proxies for the DPR. The LMC and SMC dust budget have been estimated from the SAGE data using mid-IR colours \citep*{Matsuuraetal2009,B2011,Matsuuraetal2013} and mid-IR excesses \citep{Srinivasanetal2009,B2012}. \citet{Gullieusziketal2012} modelled the spectral energy distributions (SEDs) and derived chemical classifications, DPRs, and luminosities for LMC AGB stars; however, their study was restricted to a small area.\\

We have developed the Grid of AGB and RSG ModelS (GRAMS) to analyse the SEDs of large populations of dust-producing stars \citep*{Sargentetal2011,Srinivasanetal2011}. The grid consists of silicate and carbonaceous (amorphous carbon with 10\% by mass of SiC) dust shells around photosphere models for O--rich \citep{Kucinskasetal2005,Kucinskasetal2006} and C--rich \citep{Aringeretal2009} AGB stars. The GRAMS models are able to reproduce the entire range of observed colours for LMC AGB and RSG candidates. By fitting the SEDs of LMC AGB/RSG candidates, \citet{R2012} estimate a total dust budget of \cumulativeDPRallLMC\ \msunperyr. About three-quarters of this dust arises from a small number (5\% of the total) of highly evolved `extreme' AGB stars, of which $\approx$97\% are classified as C--rich. These findings emphasise the need for a complete inventory of the dustiest evolved-star sources.\\

The SMC is a metal-poor star-forming galaxy. Its low metallicity \citep[$Z\approx 0.25 Z_\odot$][]{Garnettetal1995,Peimbertetal2000} makes it a local analog of high-redshift galaxies. The star formation history (SFH) derived from optical photometry by \citet{HarrisZaritsky2004} suggests a recent epoch (2--3 Gyr ago) of enhanced star formation, including `bursts' at 2.5 and 0.4 Gyr, corresponding to perigalactic passages with the Milky Way. An enhancement at $\sim$0.7 Gyr, possibly due to interaction between the Clouds, is also seen in the SFH derived from long-period variable star counts \citep{Rezaeikhetal2014}. Based on dust yield calculations, \citet{Schneideretal2014} showed that the integrated injection from SMC stellar sources (mainly AGB and supernovae) is comparable to the ISM dust mass estimated using {\it Herschel} data \citep[\ismdustmass\ M$_\odot$;][]{Gordonetal2014}.\\

The rate of dust ejection by AGB/RSG stars in the SMC has been estimated by \citet{B2012} and \citet{Matsuuraetal2013} using the 8 \mic\ excess and the $[3.6]-[8.0]$ colour, respectively, as proxies for the DPR. \citet{B2012} calculate a total (AGB+RSG) DPR of \excessbaseddustbudget\ \msunperyr, with carbon stars contributing $\sim$90\% of the dust from cool evolved stars. They find that up to 50\% of this dust is contributed by the so-called far-infrared (FIR) objects, whose status as evolved stars is unclear. \citet{Matsuuraetal2013} find a somewhat higher global DPR of \colorbaseddustbudget\ \msunperyr, $\sim$60\% of which comes from carbon stars. They extend a colour-DPR relation derived from a small number of well-studied sources to the entire population.\\

In this paper, we apply the techniques used by \citet{R2012} to the SMC evolved stars. We produce a reliable dataset of the dustiest evolved stars in the SMC using photometry and spectroscopic information, and constrain their SEDs using multi-epoch data. We then fit GRAMS models to these SEDs and obtain the most precise luminosity and DPR estimates to date. Our fits also determine the chemical type, which we find to be in agreement with colour-based classifications and -- more importantly -- with spectroscopic identifications. Our refined source list, fitting technique and the resulting rates and chemical classifications result not only in an improved estimate of the dust budget, but also a reliable measure of the relative contributions from silicate and carbonaceous dust in the ejecta. We improve upon the procedure of \citet{R2012} by carefully inspecting the SEDs of the reddest sources to ascertain their AGB nature as well as chemical type. We also compile all available information on the FIR objects in order to eliminate young stellar objects (YSOs) and post-AGB stars. Finally, we discuss the implications of our new dust budget estimate for a stellar origin of the dust in the SMC ISM. We present our candidate selection in Section \ref{sec:data} and describe the fitting technique in Section \ref{sec:fitting}. We present our results and discussion in Sections \ref{sec:results} and \ref{sec:discussion}, and summarise our work in Section \ref{sec:conclusion}.

\section{The evolved star sample}
\label{sec:data}
In this section, we describe the steps used to generate our list of evolved stars. We first collect information over the entire wavelength range from the optical to the mid-IR, and use multiple epochs to compute the mean flux at each wavelength. We then classify these candidates based on their near- and mid-IR colours and filter out contaminants using results from previous studies of SMC stars. We reconstruct the candidate list rather than using the conservative \citet[][hereafter: B2011]{B2011} sample to ensure the inclusion of the dustiest AGB/RSG candidates, a small number of which dominate the dust budget.\\

\subsection{The IRAC and MIPS 24 \mic\ point-source lists}
\label{subsec:iracmipssourcelist}
SAGE-SMC consists of two epochs of observations in the IRAC and MIPS bands \citep{Gordonetal2011} separated by three months. The SMC bar region was also observed as part of the {\it Spitzer} Survey of the Small Magellanic Cloud program \citep[S$^3$MC;][]{Bolattoetal2007}. The SAGE-SMC survey covers the wing and tail regions in addition to the bar \citep[for a comparison of the spatial coverage of the SAGE-SMC and S$^3$MC surveys, see fig. 1 in][]{B2011}. The single-epoch SAGE-SMC images were co-added, along with observations from the S$^3$MC program as a third epoch (Epoch 0) in the region of overlapping coverage, to generate mosaic images\footnote{SAGE-SMC data can be downloaded from the {\it Spitzer} Science Center at http://irsa.ipac.caltech.edu/data/SPITZER/docs/spitzermission/\\observingprograms/legacy/sagesmc/}.\\

The SAGE-SMC IRAC pipeline produced two kinds of source lists: a highly reliable catalog (hereafter, ``IRAC Catalog") and a more complete archive (``IRAC Archive"). The IRAC Single Frame + Mosaic Photometry (SMP) Catalog and Archive combined the mosaic photometry and the single-frame photometry Epoch 0+1+2 Catalog and Archive source lists, respectively. During the bandmerging, the point-source lists were also matched\footnote{Note: Unless otherwise specified, when searching for counterparts to our IRAC point sources we use a 1\arcsec\ radius for optical and near-infrared source lists, 2\arcsec\ for mid-infrared wavelengths shorter than 24 \mic, and 3\arcsec\ for 24 \mic.} to data from the Magellanic Clouds Photometric Survey \citep[MCPS;][]{Zaritskyetal2002} and the 2-Micron All Sky Survey \citep[2MASS;][]{Skrutskieetal2006}. The MIPS pipeline also generated mosaic images using the three epochs of MIPS data. At 24 \mic, point-source catalogs were created from the individual AOR mosaics, and the individual AOR point-source catalogs were merged to create the ``Full" list. A high-reliability catalog was extracted from the Full list by placing strict restrictions on the source quality. This MIPS24 Catalog rejected about two-thirds of single-epoch list sources. We refer the reader to Section 2 of \citet{Gordonetal2011} and to the SAGE-SMC Data Delivery Document\footnote{http://irsa.ipac.caltech.edu/data/SPITZER/SAGE-SMC/docs/sage-smc\_delivery\_nov09.pdf} for more details on the IRAC/MIPS pipeline, image processing, and point-source extraction.\\

We revisit the source selection to ensure an accurate accounting of the most prolific dust producers. The AGB/RSG population in the SAGE-SMC data was first discussed in \citetalias{B2011}. In that paper, we selected evolved-star candidates based on their photometry in the SAGE-SMC IRAC Mosaic Archive and the MIPS 24 \mic\ Full List.

This combination allowed for fewer spurious detections and a reduced effect of source variability in the IRAC bands, while including fainter MIPS24 sources. However, this choice suffers from a few issues that lead us to reevaluate our source selection in the current study.\\

The bandmerging process employed by the SAGE IRAC team operates on a purely positional basis, without any constraints on the multi-band flux; this is to avoid any colour biases in the crossband matching. On occasion, this procedure results in a source in one epoch/band being erroneously matched to a neighbouring source in another epoch/band. For example, as a result of the inter-epoch positional variation, the 2MASS association is sometimes matched to neighbouring IRAC sources that may or may not have been extracted as multiple sources in the individual epoch images (Brian Babler \& Marta Sewi{\l}o, private communication). While only $\approx$8000 sources ($<$ 2000 brighter than the tip of the red giant branch in the \ks--band) of the 12 million in the Epoch 1 archive are affected by this problem, it prevents \citetalias{B2011} from identifying two spectroscopically confirmed C-rich AGB stars. Similarly, the 3.6 \mic\ position of a nearby faint object is sometimes matched to a source that is bright in the other IRAC bands. This is much rarer (only $\approx$100 sources with single-epoch magnitudes brighter than $[3.6]$ = 14 mag are affected) but it excludes a spectroscopically identified carbon star from the \citetalias{B2011} Mosaic list. In this paper, we modify the selection scheme by first selecting sources from the IRAC Epoch 1 and Epoch 2 Archive lists, then combining the information from both epochs for each source. This circumvents the above complications with the Mosaic list, and also allows us to incorporate variability effects into the candidate SEDs in a quantitative manner.\\

\citetalias{B2011} identified a small number of sources in their list with 24 \mic\ fluxes greater than their 8 \mic\ fluxes, dubbing them far-infrared (FIR) objects. In a procedure similar to that of \citetalias{B2011}, we inspect the SEDs and the 24 \mic\ image to eliminate most of the 303 FIR objects in the \citetalias{B2011} sample because they are faint, cool photospheres matched to different 24 \mic\ sources that lie along the line of sight (see Section \ref{subsubsec:FIRobjects}). In order to obtain a reliable list of dusty sources, we use the 24 \mic\ photometry from the MIPS Catalog list in this paper instead of the ``Full" list, which includes many spurious sources. Indeed, we find that only five red giant branch (RGB) candidates qualify for FIR status if we use the Catalog photometry.

\subsection{Candidate selection}
\label{subsec:candidateselection}
We begin by extracting IRAC sources from the Epoch 1 and 2 archives, and collect the matching 2MASS and MIPS24 photometry for each source. For each IRAC Epoch 1 or 2 source, we also pick up matching IRAC and MIPS24 photometry from the S$^3$MC sample\footnote{The S$^3$MC data used here was reduced using the SAGE-SMC pipeline for consistency. S$^3$MC is a somewhat deeper survey based on a different observing strategy; it is therefore possible that there are S$^3$MC sources without SAGE-SMC counterparts. However, it is unlikely that a bright, red mass-losing point source would be present in S$^3$MC but not in the SAGE-SMC data. We identified all $\sim$ 27\,000 S$^3$MC-only IRAC point sources and verified that they are all fainter than the tip of the red giant branch in the \ks\ \emph{and} 3.6 \mic\ bands.}. In addition to 2MASS, the SAGE-SMC point-source list is matched to data from the InfraRed Survey Facility \citep[IRSF;][]{Katoetal2007}, which we treat as a second near-infrared epoch where available (see Fig 1 in \citetalias{B2011}). We compute the mean magnitudes in the near- and mid-infrared bands using the multi-epoch information to obtain an {\em epoch-averaged} SED for each source\footnote{For sources detected in only one epoch, we include the single-epoch SED in our list.}. As in \citet{B2012} (hereafter, B2012), we set the SMC distance at 60 kpc \citep{Cionietal2000,KellerWood2006} and we correct for interstellar reddening using the \citet{Glass1999} extinction law with $A_V$ = 0.12 mag and $E(B-V)$ = 0.04 mag \citep*{Schlegeletal1998,HarrisZaritsky2004}.\footnote{The $A_{\lambda}$ values chosen for the {\it U, B, V, I, J, H,} and \ks\ bands are \alambda\ mag respectively. Extinction in the mid-infrared bands is negligible.}\\

The colour-selection criteria we use to populate our list of evolved-star candidates is similar to that of \citetalias{B2011}. We refer the reader to Section 3 of that paper for details. We describe the steps in brief here.

\begin{enumerate}
\item {\em Far-infrared objects:} Defined as sources having a 24 \mic\ flux greater than their 8 \mic\ flux, the FIR objects are even redder than the extreme AGB (x--AGB) stars, and may potentially include AGB stars, although they are dominated by interlopers (young stellar objects and galaxies). Such extremely dusty sources can be faint in the near-IR. For this reason, we first flag all sources with \ks\ $<$ 16 mag and $F_{24} > F_{8}$ as FIR candidates. We remove any remaining sources that are simultaneously fainter than the tip of the RGB (TRGB) in {\em both} the \ks\ and 3.6 \mic\ bands \citepalias[12.7 mag and 12.6 mag respectively;][]{B2011} before selecting the other classes of candidates.
\item {\em O--AGB, C--AGB, and RSG candidates:} The O--AGB population consists of AGB stars with oxygen-rich dust, and includes both low- and high-mass O--rich AGB stars. This definition is based on their location in the \ks\ {\it vs.} {\it J}--\ks\ colour-magnitude diagram \citepalias[see Sections 3.1.1 and 3.1.3 in][]{B2011}, and may therefore not include the dustiest oxygen-rich AGB stars (OH/IR stars). The demarcation between the RSG and O--AGB populations is empirical. The RSGs consist mostly of luminous optically thin sources with small-amplitude variability. C--AGB stars are AGB stars with carbonaceous dust, including the intermediate mass range where the third dredge-up is suited for carbon stars. We also use the near-IR CMD to extract C--AGB candidates, and these may not include the dustiest, most evolved carbon stars.\\
We use the \ks\ {\it vs.} {\it J}--\ks\ CMD to select optically thin AGB as well as RSG stars. As shown in fig. 5 of \citetalias{B2011}, the O--rich AGB candidates are selected from the region bounded by the lines K1 and K2, and the \ks--band TRGB. Sources brighter than the line K0 and redder than the line K2 are classified as C--rich. The choice of K0 instead of the TRGB means that we select a small number of stars fainter than the \ks--band TRGB; to reduce contamination from RGBs, we require in this case that the star be brighter at 3.6 \mic\ than the TRGB at that wavelength. The only departure in our selection procedure from that of \citetalias{B2011} affects the RSG candidates, which were separated from the O--rich AGB stars by a gap of 0.5 mag in the \ks\ {\it vs.} {\it J}--\ks\ CMD (fig. 5 in that paper) to avoid contamination. In the present study, we include the sources in this gap as RSG candidates.
\item {\em Anomalous AGB:} This group consists of dusty low-mass stars that can be O--rich or C--rich. Among the sources classified according to their colours as O--rich, \citetalias{B2011} discovered a subpopulation with redder {\it J}--[8.0] colours, which they labelled as anomalous O--rich AGB (aO--AGB in that paper) stars. \citet{Boyeretal2015} found this sample to consist of evolved, low-mass dusty O--rich and C--rich AGB stars near the third dredge-up mass limit that defines the transition from O--rich to C--rich chemistry. As many of them were found to be carbon-rich, they suggest the term `anomalous AGB' (a--AGB) instead. We adhere to this terminology in the current paper. We identify the a--AGB population using the [8.0] {\it vs.} {\it J}--[8.0] CMD, as in Section 3.1.5 of \citetalias{B2011}.
\item {\em Extreme AGB:} This category consists of the dustiest, most evolved AGB stars, classified empirically by their near- and mid-IR colours. While their dust chemistry is not discernible from their colours alone, they are primarily carbon-rich AGB stars \citep[e.g.,][]{R2012}. However, this population also contains a small number of dusty O--rich (O--AGB as well as RSG) stars. We select x--AGB candidates using the [8.0] {\it vs.} {\it J}--[8.0] and  (for sources without {\it J}--band detections) [8.0] {\it vs.} [3.6]--[8.0] CMDs using Equations 1 and 2 from Section 3.1.2 of \citetalias{B2011} (Equation 2 terminates at [3.6]--[8.0] = 3 mag, and extends horizontally to redder colours).
\end{enumerate}

Using this procedure, we classify \Nrawev\ AGB and RSG candidates. Table \ref{tab:sourcecounts} lists the source count in our initial list by colour class, and compares these numbers to the \citetalias{B2011} dataset. The major changes are a 75\% reduction in the number of FIR objects as a result of using the higher quality MIPS 24 \mic\ Catalog List (only 5 RGB candidates remain; the number of AGB/RSG FIR objects increases from 57 to 85), and a $\approx$20\% increase in RSG candidates due to the inclusion of the 0.5 mag strip as described above. Since we extract our sample from an updated version of the SAGE-SMC Point Source Catalog, our numbers for other source types are slightly higher than those in \citetalias{B2011}. Our candidate list includes all of the 81 {\it Spitzer} IRS staring mode targets \citep{Ruffleetal2015} spectroscopically classified as either RSG or AGB (including the S star BFM 1).\\

We now refine the candidate SEDs by adding information from other programs. In particular, we use the information from various optical through mid-infrared studies to incorporate the effects of variability into the flux uncertainties.

\subsection{Accounting for variability with multiple epochs of data}
\label{subsec:variability}
Ideally, multi-wavelength broadband photometry of long-period variables must be obtained concurrently to accurately reproduce their SEDs. In the absence of such data, we can incorporate the effect of variability into the flux uncertainties to obtain reliable luminosity and dust-production rate estimates. Inflating the uncertainties in this manner is necessary to obtain meaningful fits using the GRAMS models, which do not account for time variability. Our method is described below for each wavelength regime.\\

\subsubsection{Optical and near-IR data}
We collect optical variability information for our sample from the OGLE (Optical Gravitational Lensing Experiment) survey. In particular, we extract {\it V}-- and {\it I}--band mean magnitudes along with {\it I}--band amplitudes from the third release of the OGLE Catalog of Variable Stars \citep[OGLE-III;][]{Udalskietal2008}. We find that \NOGLEmatches\ of our sources have OGLE counterparts. For each source with an OGLE match, we replace the MCPS {\it V} and {\it I}--band fluxes with the mean fluxes from OGLE. Following a procedure similar to \citet[][hereafter, R2012]{R2012}, we add half of the peak-to-peak OGLE amplitude in quadrature to the photometric uncertainty of each of the four optical bands. Similarly, using the photometry of the IRSF counterparts to \NIRSFmatches\ of our sources as a second near-infrared epoch, we estimate the mean flux and use the flux difference between the two epochs to inflate the uncertainty in the {\it J},{\it H}, and \ks\ bands.\\

\subsubsection{IRAC and MIPS~24~\mic\ data}
Our source selection procedure already extracts all available epochs (Epoch 1, Epoch 2, and S$^3$MC or `Epoch 0') of {\it Spitzer} data for each source. We have all three epochs of data for \NSthreeMCmatches\ stars in our sample, and two SAGE-SMC epochs for the rest. We use this information to compute the mean flux from the two or three epochs, and use the minimum and maximum fluxes to estimate the flux amplitude. This amplitude is then folded into the flux uncertainty as described above.\\

\subsubsection{AKARI and WISE data}
We also explored the possibility of incorporating mid-infrared (mid-IR) data from the {\it AKARI} \citep[][]{Itaetal2010} survey of the SMC and the Wide Infrared Survey Explorer \citep[{\it WISE};][]{Wrightetal2010} All-Sky Data Release.\\

\citet{Itaetal2010} present the {\it AKARI} Bright Point Source Catalog for the SMC, which contains about 1800 sources. The {\it AKARI} data includes photometry taken with the Infrared Camera (IRC) at 3.2 (the {\it N3} band), 4.1 ({\it N4}), 7.0 ({\it S7}), 11.0 ({\it S11}), 15.0 ({\it L15}), and 24~\mic\ ({\it L24}). We found {\it AKARI} catalog matches to the positions of our {\it Spitzer} sources using the DAOphot routine DAOmatch \citep{Stetson1987} which iteratively solves for the transformation coefficients and assigns matches. Starting with a 10 pixel (5\arcsec) radius, the routine updates the nearest-neighbour distances down to a 1 pixel (2\arcsec) separation. The {\it AKARI} spatial coverage does not include the entire SMC, but instead comprises twelve $10\arcmin \times 10\arcmin$ regions. Two regions (IDs $=$ 4 and 11) include data at all available {\it AKARI} filters, 5 regions include only {\it {\it N3}}, {\it N4}, {\it S7}, and {\it S11} imaging, and 5 regions include only {\it L15} and {\it L24} imaging. Only \NAKARImatches\ sources in our list have matches in the {\it AKARI} catalog.\\

We use the US Virtual Astronomical Observatory (VAO) to cross-match {\it WISE} photometry to our {\it Spitzer} data with a maximum matching distance of 1\arcsec\ to avoid spurious matches. The VAO uses the {\it WISE} All-Sky Data Release Source Catalog which includes 3.4 ({\it W1}), 4.6 ({\it W2}), 12 ({\it W3}), and 22~\micron\ ({\it W4}) photometry covering the entire SAGE footprint on the SMC. We find {\it WISE} counterparts to \NWISEmatches\ of our evolved-star candidates.\\

The {\it AKARI} {\it N3}, {\it N4} and {\it L24} filters cover a range of wavelengths similar to the IRAC 3.6 \mic, 4.5 \mic\ and MIPS~24~\mic\ filters respectively. This is also true of the {\it WISE} {\it W1}, {\it W2} and {\it W4} bands. These observations could be added as an extra epoch to the corresponding {\it Spitzer} filters. However, we find that the {\it AKARI}/{\it WISE} measurements in these bands are not always consistent with the {\it Spitzer} photometry, even when variability is taken into consideration. In fact, the distribution of normalised flux differences $(F_{\rm AKARI}-F_{Spitzer})/dF_{Spitzer}$ is skewed to positive values for all three pairs of filters mentioned above, with median values $>$3. The same is true for the {\it WISE} photometry, the most extreme case being that of the {\it WISE} {\it W4} band, which has a median normalised flux difference of 12 with respect to the MIPS24 flux. These systematic discrepancies cannot be explained by variability alone, and may be caused by differences in beam size. Indeed, the angular resolution in the two shortest-wavelength {\it AKARI}/{\it WISE} bands is about 2--3 times that of the nearest IRAC band ($\approx$2\arcsec). However, the {\it S11}, {\it L15}, and {\it W3} bands are comparable in resolution ($\approx$5--6\arcsec) to MIPS24. In addition, these bands provide useful constraints on the optical depth and DPRs of our best-fit models, as the 10--22 \mic\ wavelength range contains interesting dust signatures such as the 9.7 and 18 \mic\ silicate and 11.3 \mic\ SiC features.

\subsection{Eliminating non-AGB/RSG stars}
\label{sec:vetting}
We generate a reliable list of {\em dusty} evolved stars through the following filtering procedure. While this list may exclude faint and/or optically thin sources when their nature is unclear, such sources  do not sizeably contribute to the dust budget.\\

\begin{enumerate}
\item Of the sources with OGLE counterparts, only two are not classified as long-period variables (LPVs) -- one is a Cepheid, the other an RV~Tau star -- we eliminate these two objects from our list. 
\item As in Section 2 of \citetalias{B2011}, we reduce the contamination in our sample by excluding all the point sources within 8\arcmin\ and 5\arcmin\ of the Galactic clusters 47 Tuc and NGC 362 during our source selection (when combined, this is $<$0.4\% of the SMC-SAGE survey area; we do not expect it to change our results). 
\item It is possible that some sources are misclassified as AGB/RSG candidates because of the automated nature of the classification. By performing a SIMBAD search for the entire list of IRAC mosaic designations, we exclude \Nmisclassified\ sources -- including one FIR object -- that are either foreground sources or SMC objects with confirmed non-AGB/RSG classifications.

\item We compared our list to the list of point sources targeted with {\it Spitzer} IRS \citep{Ruffleetal2015}, and used the spectroscopic classifications to remove a total of \NSMCSpecnotES\ sources from our sample, including two post-AGB stars, thirteen YSOs, three RCrB stars, three planetary nebulae, a symbiotic star, and a B[e] star. Of the \NSMCSpecnotES\ objects thus removed, 17 are flagged as FIR objects. All of the \NSMCSpecES\ spectroscopically identified by \citet{Ruffleetal2015} as SMC AGB/RSG stars (19 O--rich AGB and 39 carbon stars, 22 RSGs and one S--type star) are in our list, and they are correctly classified by our colour-selection criteria. In Section \ref{sec:results}, we compare our SED--based chemical type with the SMC-Spec classifications.
\item
Following Section 4.4 of \citetalias{B2011}, we eliminate four extreme AGB candidates based on their bright MIPS 70 \mic\ associations. Two of these sources are associated with H\,{\sevensize II} regions, one with a post-AGB star, and one with an emission-line star.
\end{enumerate}

\subsubsection{FIR objects}
\label{subsubsec:FIRobjects}

In this section, we pare down the list of FIR AGB/RSG candidates using the following steps:
\begin{enumerate}
\item {\it Exclusion based on spectroscopic classifications}: Using the \citet{Ruffleetal2015} spectroscopic classifications, we remove 17 non-AGB/RSG stars. We exclude two more sources -- a blue supergiant and a planetary nebula -- based on identifications from previous studies.\\

\begin{figure}
 \includegraphics[width=84mm]{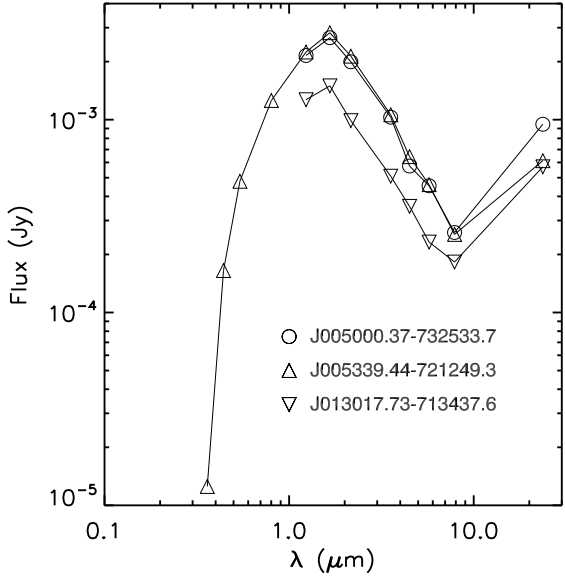}
 \caption{SEDs of three misidentified 24 \mic\ objects from the FIR subsample. At $\lambda \la$ 8 \mic, the SED is consistent with that of a cool photosphere peaking in the near-infrared.\label{fig:debrisdisk}}
\end{figure}

\item {\it Misidentified 24 \mic\ sources:}
Many of the remaining FIR candidates have SED shapes consistent with cool photospheres up to 5.8 or 8.0 \mic, along with an excess at 8.0 or 24 \mic. Such SEDs are consistent with those of debris discs \citep[see, e.g., the SEDs in fig. 3 of][]{Adamsetal2013}. Fig. \ref{fig:debrisdisk} shows three examples of such SEDs. For SMC stars, however, this effect is most likely due to the larger beam size at MIPS 24 \mic\ (6\arcsec, corresponding to $>$ 1 pc). Confusion due to surrounding diffuse emission, blended point sources, or an IR source and a different optical source within the same beam would result in misclassification as a debris disc \citep[e.g.,][]{Sheetsetal2013}. In either case, these sources are not evolved stars and are not relevant to our study. We refer to objects with such SEDs as `misidentified 24 \mic\ sources'.

\begin{figure}
 \includegraphics[width=84mm]{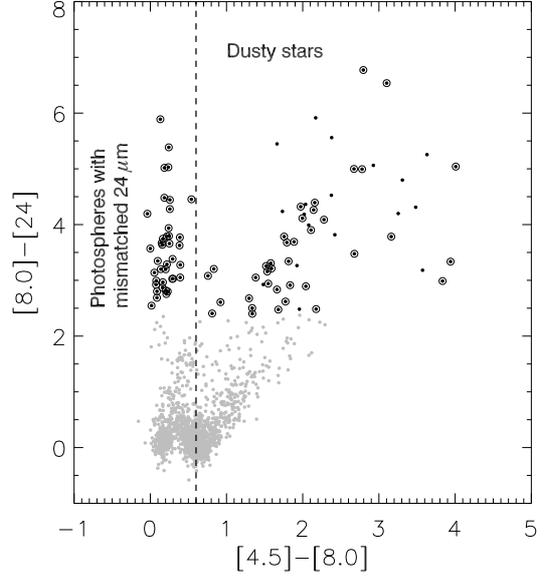}
\caption{A {\it Spitzer} colour-colour diagram to identify genuine dusty evolved stars in the FIR object sample. Sources in our evolved-star sample (black and grey symbols) are classified as FIR objects (filled black circles) if $[8.0]-[24] > 2.39$ mag. Starting with \NrawFIR\ such sources, we eliminate 17 based on spectroscopic identifications (see text). Of the remaining objects (open circles), those bluer than [4.5]--[8.0] = 0.6 mag (dashed line) have SEDs resembling cool star photospheres up to 8 \mic, while a majority of the redder population consists of genuinely rising SEDs. \label{fig:plotccd_FIR}}
\end{figure}

The FIR candidates split into two groups on the [8.0]--[24] {\it vs}. [4.5]--[8.0]  colour-colour diagram (Fig. \ref{fig:plotccd_FIR}), with the bluer population spanning a relative narrow range in [4.5]--[8.0] colour. We verified that all the SEDs in the blue population resemble those of debris discs -- they are intrinsically faint sources that satisfied the FIR criterion due to their 24 \mic\ flux. We remove 37 sources bluer than [4.5]--[8.0] = 0.6 mag from our list.\\

\begin{figure}
 \includegraphics[width=84mm]{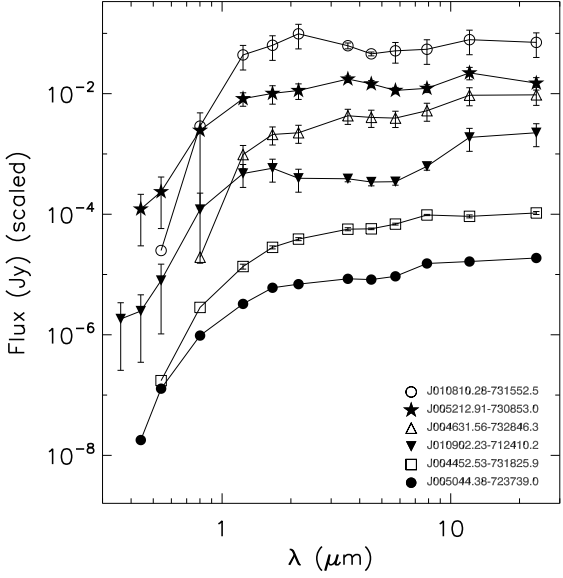}
\caption{FIR objects in Group 1. SEDs are progressively scaled down by a factor of five for clarity. All the SEDs in this group have a very flat shape from the near-IR to 24 \mic, and are all identified as AGB/RSGs based on their mid-IR spectra.\label{fig:FIR_group1}}
\end{figure}

\begin{figure}
 \includegraphics[width=84mm]{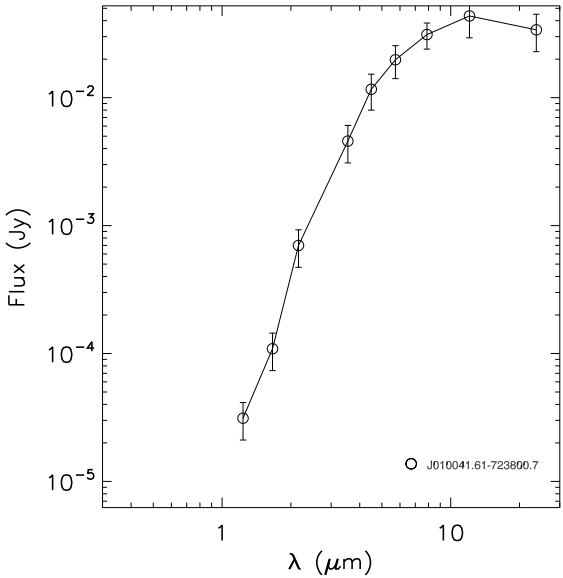}
\caption{Group 2 contains only one object, the carbon star candidate SSTISAGEMA~J010041.6--723800.7.\label{fig:FIR_group2}}
\end{figure}

\begin{figure}
 \includegraphics[width=84mm]{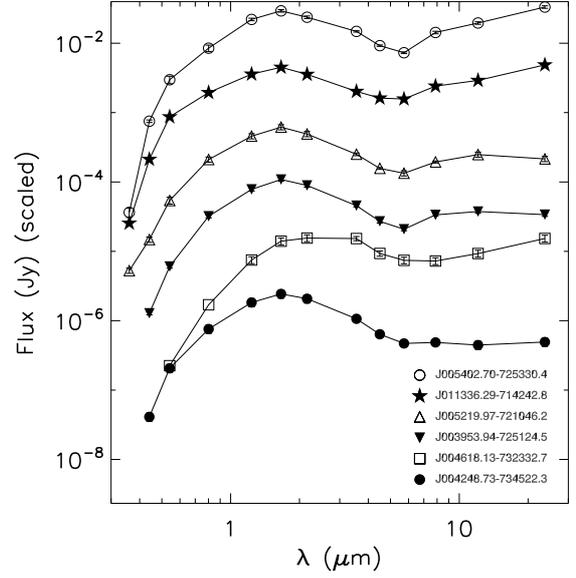}
\caption{FIR object Group 3. SEDs are progressively scaled down by a factor of five for clarity. All Group 3 SEDs have a 24 \mic\ flux comparable to the near-IR peak flux. The group contains two objects identified as AGB stars based on optical spectra.\label{fig:FIR_group3}}
\end{figure}

\begin{figure}
 \includegraphics[width=84mm]{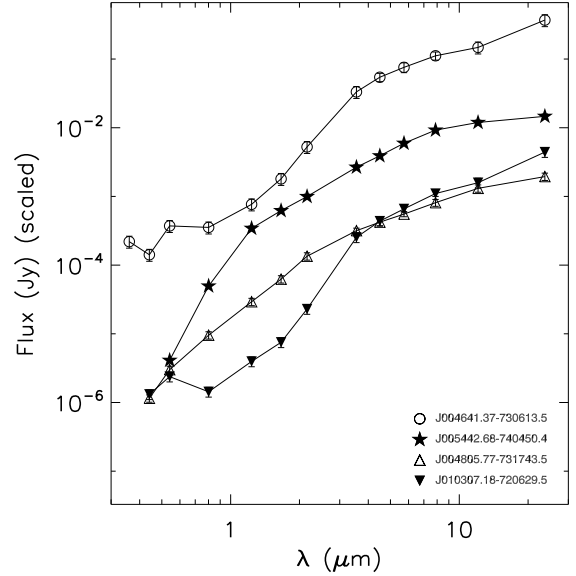}
\caption{FIR object Group 4 contains SEDs that suffer from image blending/diffuse emission at mid-IR wavelengths. Three sources in this group are mid-IR variables, and one (J005442, Object 16 in Table \ref{tab:FIRcategories}) is identified as an AGB star based on its optical spectrum. The SEDs are progressively scaled down by a factor of five for clarity.\label{fig:FIR_group4}}
\end{figure}

\begin{figure}
 \includegraphics[width=84mm]{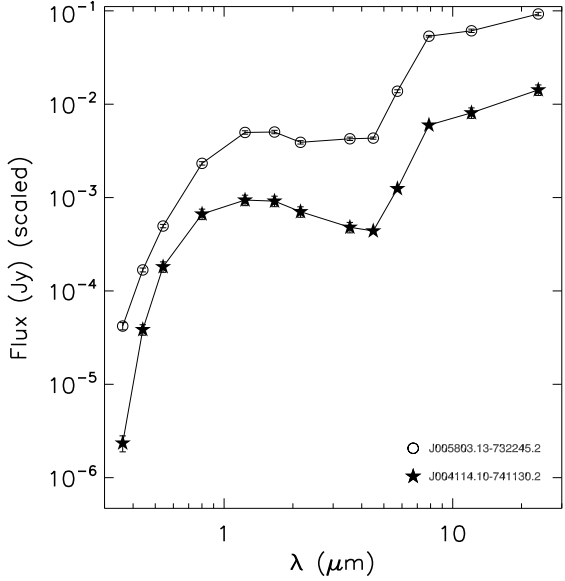}
\caption{There are two SEDs in Group 5, both showing a brighter, second peak beyond $\sim$ 5.8 \mic. The second SED is scaled down by a factor of five for clarity.\label{fig:FIR_group5}}
\end{figure}

\begin{figure}
\includegraphics[width=84mm]{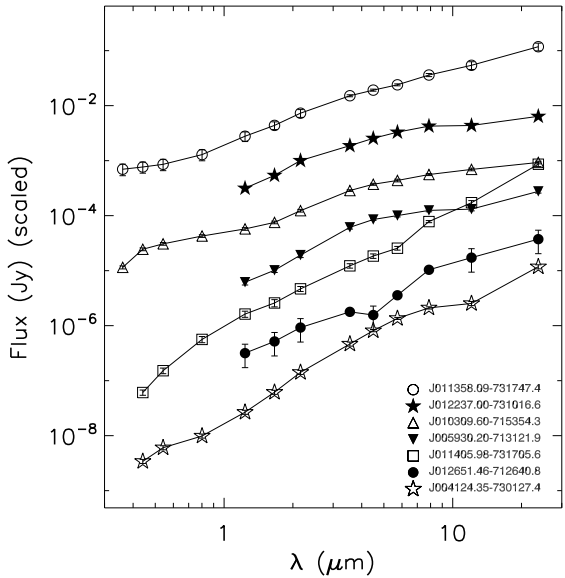}
\caption{All the SEDs in Group 6 show a steady rise from the optical to 24 \mic. The SEDs are progressively scaled down by a factor of five for clarity.\label{fig:FIR_group6}}
\end{figure}

\begin{figure}
\includegraphics[width=84mm]{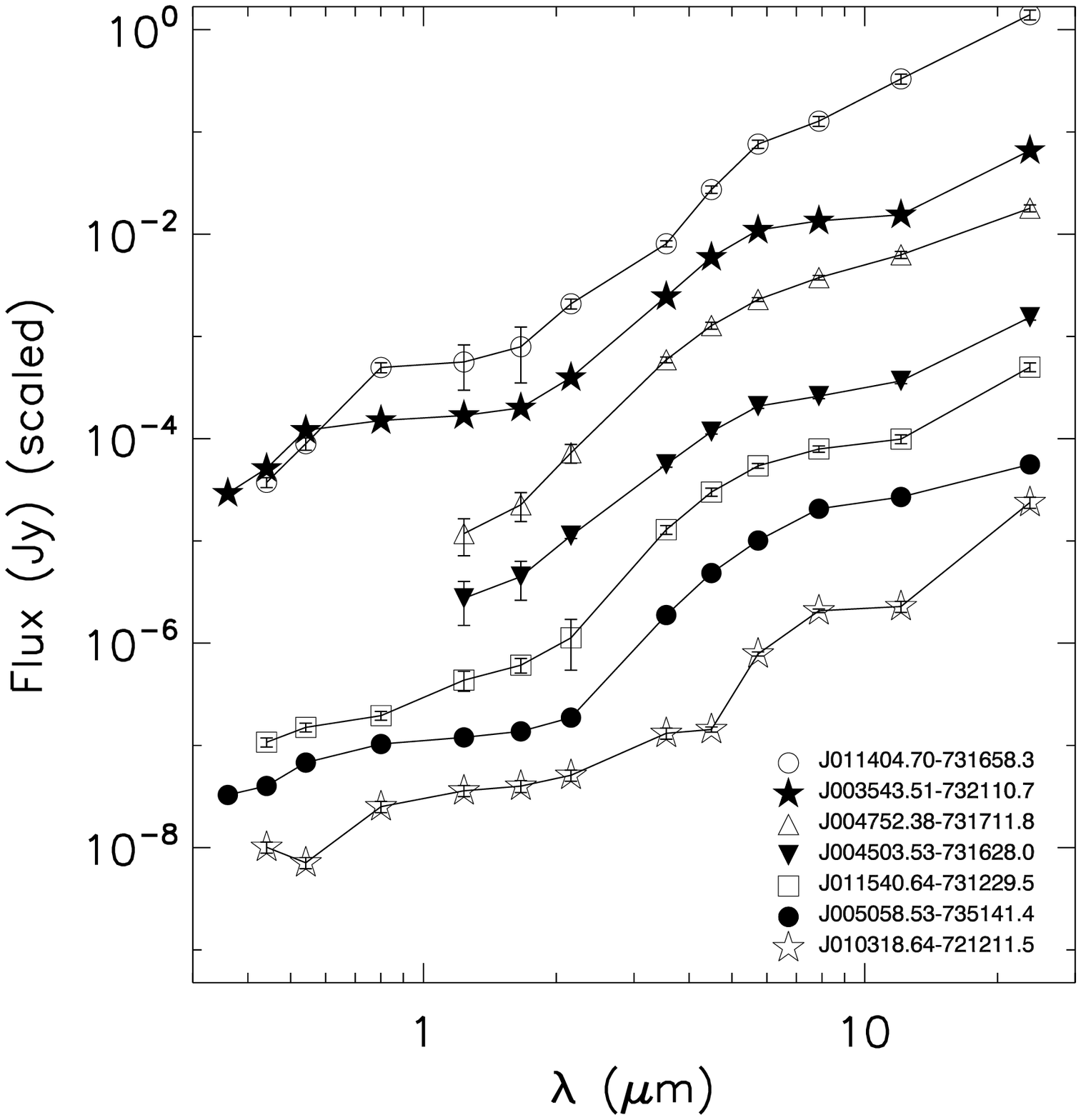}
\caption{The Group 7 FIR objects all show a flat SED in the optical/near-IR, then a sharp rise towards 24 \mic. Image inspection shows all the sources to either be associated with diffuse star formation or suffer from blending issues. The SEDs are progressively scaled down by a factor of five for clarity.\label{fig:FIR_group7}}
\end{figure}

\item {\it SED shapes, spectroscopic identifications and optical/mid-IR variability:} We compile the spectroscopic and variability information for the surviving \NvetFIR\ FIR candidates in Table \ref{tab:FIRcategories}. We split these objects into seven groups based on their SED shapes, as shown in Figs. \ref{fig:FIR_group1}--\ref{fig:FIR_group7}. Group 1 includes all the sources that have been classified by \citet{Ruffleetal2015} as AGB or RSG based on their mid-IR spectra  (Objects 1--5 and 7 in Table \ref{tab:FIRcategories}). SSTISAGEMA~J010041.61--723800.7 (Object 33 in Table \ref{tab:FIRcategories}), a newly-discovered mid-IR variable \citep{Riebeletal2015} is placed in its own group (Group 2). While Groups 3 (Objects 6, 8, 9, 10, 11, and 13 in Table \ref{tab:FIRcategories}) and 4 (Objects 14, 16, 19, and 23 in Table \ref{tab:FIRcategories}) both include objects previously classified as AGB stars based on optical spectra, some Group 4 sources are also found to vary in the mid-IR \citep{Polsdoferetal2015,Riebeletal2015}. The SEDs in Group 3 show a peak in the near-IR followed by a rise beyond 5.8 \mic\ that is comparable in flux to the near-IR peak, while those in Group 4 have much larger fluxes in the mid-IR beyond 5.8 \mic. Based on the similar SED shapes within each group, we treat all the members as possible AGB/RSG candidates. Group 4 SEDs are affected by contamination from diffuse mid-IR emission (see Table \ref{tab:FIRcategories}), so that not all the flux at longer wavelengths may be due to a single source.\\
The best fits for Group 3 (see Section \ref{subsec:FIRfitsdiscussion}) show prominent silicate emission. These sources are perhaps the fainter counterparts of the Group 1 sources, which were preferentially selected for spectroscopic followup based on their brightness (mid-IR flux $>$ 0.1 Jy).\\

Groups 5 (Objects 17 and 18 in Table \ref{tab:FIRcategories}), 6 (Objects 15, 20, 21, 22, 24, and 25 in Table \ref{tab:FIRcategories}), and 7 (Objects 26--32 in Table \ref{tab:FIRcategories}) contain sources that are likely post-AGB/planetary nebulae, background galaxies or YSOs. While some of these objects suffer contamination from diffuse emission or blending of multiple sources at longer wavelengths, they are also in the vicinity of star-forming regions (see Table \ref{tab:FIRcategories}). In our discussion of the dust budget, we will only include the contribution from sources in Groups 1--4.

\vspace{1cm}

Table \ref{tab:sourcecounts} shows the source counts for our dataset after the filtering employed in this section.  Our final list has 154 fewer sources than the original (see Table \ref{tab:sourcecounts}), resulting from reductions in the number of x--AGB and RSG stars, and FIR objects. Photometry for the vetted source list is available online (see Table \ref{tab:metatable_vettedphot_fitresults} for a description of the columns in the online table). In the next section, we describe the procedure used to fit these SEDs and to determine their dust-production rates.

\begin{table*}
\scriptsize
\centering
\begin{minipage}{180mm}
\caption{FIR objects classified based on spectroscopic identification, optical/mid-IR variability, and SED/image inspection.\label{tab:FIRcategories}}
\begin{tabular}{@{}lllllllll@{}}
\hline
No. & Designation\footnote{SAGE-SMC IRAC mosaic archive designation; should be preceded by `SSTISAGEMA '.} & SIMBAD ID & \multicolumn{2}{c}{Spectroscopic\footnote{Identification of the photospheric/dust chemistry from optical/mid-IR spectra.}} & OGLE\footnote{Counterpart in the OGLE-III Catalog of Variable Stars. Should be preceded by `OGLE--SMC--LPV--'.} & mid-IR\footnote{ID in the \citet{Polsdoferetal2015} variable list. IDs followed by `(R)' are also in the \citet{Riebeletal2015} list of mid-IR variables.} & SED group\footnote{Grouping based on SED shape; see Section \ref{subsubsec:FIRobjects} for details.} & Notes\footnote{Notes from image inspection.}\\
 & &\multicolumn{2}{c}{identification} & ID & variable & &\\
 \cline{3-4}
& & & \multicolumn{1}{c}{Opt.\footnote{O(--rich) or C(--rich); references: [1] = \citet{GroenewegenBlommaert1998}, [2] = \citet{vanLoonetal2010}, [3] = \citet*{Rebeirotetal1993}.}} & \multicolumn{1}{c}{mid-IR\footnote{Identifications from \citet{Ruffleetal2015} based on {\it Spitzer} IRS spectra. The `SMC IRS' designations used in the paper are included in parenthesis.}} & & ID & &\\
\hline
1&  J010810.28-731552.5 & IRAS F01066-7332           & O [1]     & O-AGB(178)   & 17253 & 725 & Group 1 & -\\
2&  J005212.91-730853.0 & BMB-B  75                  & O [2]     & O-AGB(177)   & 09518 & 338(R) & Group 1 & -\\
3&  J004631.56-732846.3 & OGLE J004631.61-732846.0   & -           & O-AGB(121)   & 05887 & 148 & Group 1 & -\\
4&  J010902.23-712410.2 & SV* HV 12956               & O [1]     & O-AGB(277)   & 17525 & 744 & Group 1 & -\\
5&  J004452.53-731825.9 & OGLE J004452.59-731825.4   & -           & O-AGB(175)   & 04992 & - & Group 1& -\\
6&  J005402.70-725330.4 & RAW  839                   & C [3]     & -            & 10720 & - & Group 3 & -\\
7& J005044.38-723739.0 & RAW  631                   & C [3]     & O-AGB(113)   & 08525 & - & Group 1 & -\\
8&  J011336.29-714242.8 & 2MASS J01133627-7142428    & -           & -            & 18434 & - & Group 3  & -\\
9&  J005219.97-721046.2 & 2MASS J00521995-7210461    & -           & -            & 09590 & - & Group 3    & -\\
10&  J003953.94-725124.5 & 2MASS J00395393-7251246    & -           & -            & 02960 & - & Group 3    & -\\
11&  J004618.13-732332.7 & OGLE SMC-LPV-5748          & -           & -            & 05748 & - & Group 3    & -\\
12&  J011358.09-731747.4 & [MA93] 1780                & -           & -            & 18490 & - & Group 6 & Diffuse emission/star-forming\\
  & & & & & & & & region (SFR); possible YSO?\\
13&  J004248.73-734522.3 & RAW  171                   & C [3]     & -            & 04054 & - & Group 3 & -\\
14&  J004641.37-730613.5 & 2MASS J00464140-7306135    & -           & -            &   -   & 151(R) & Group 4& Diffuse emission.\\
15&  J012237.00-731016.6 & 2MASS J01223695-7310165    & -           & -            &   -   & - & Group 6 & -\\
16&  J005442.68-740450.4 & IRAS F00530-7421           & C [1]     & -            & 11159 & - & Group 4 & -\\
17&  J005803.13-732245.2 &   OGLE SMC-LPV-13055       &   -       &   -       & 13055 &   -   &  Group 5    & -\\
18&  J004114.10-741130.2 &   OGLE SMC-LPV-3409        &   -       &   -       & 03409 &   -   &  Group 5    & -\\
19&  J004805.77-731743.5 & 2MASS J00480580-7317435    & -           & -            &   -   & 203 &  Group 4 & Diffuse emission; possible blend\\ 
  & & & & & & & & at 24 $\mu$m.\\
20&  J010309.60-715354.3 & 2MASS J01030962-7153541    & -           & -            &   -   & - &   Group 6  & -\\
21&  J011405.98-731705.6 & 2MASS J01140597-7317054    & -           & -            &   -   & - &   Group 6  & SFR; YSO?\\
22&  J012651.46-712640.8 & 2MASS J01265140-7126407    & -           & -            &   -   & - &   Group 6  & -\\
23&  J010307.18-720629.5 & 2MASS J01030720-7206295    & -           & -            &   -   & 629 &   Group 4 & Blend at 24 $\mu$m.\\
24&  J005930.20-713121.9 & 2MASS J00593021-7131220    & -           & -            &   -   & - &   Group 6  & -\\
25&  J004124.35-730127.4 & 2MASS J00412434-7301274    & -           & -            &   -   & - &   Group 6  & -\\
26&  J011404.70-731658.3 &   2MASS J01140467-7316584  &   -       &   -       &   -   &   784 &   Group 7 & SFR; YSO?\\
27&  J003543.51-732110.7 & 2MASS J00354347-7321106    & -           & -            &   -   & - &   Group 7  & 2 sources at 24 $\mu$m.\\
28&  J004752.38-731711.8 & 2MASS J00475238-7317116    & -           & -            &   -   & - &   Group 7  & Blending with diffuse emission;\\
  & & & & & & & & SFR; possible YSO?\\
29&  J004503.53-731628.0 & 2MASS J00450354-7316280    & -           & -            &   -   & - &   Group 7  & SFR; possible YSO?\\
30&  J011540.64-731229.5 & IRAS 01016-7228            & -           & -            &   -   & - &   Group 7  & Diffuse emission/SFR; possible\\
    & & & & & & & & YSO?\\
31&  J005058.53-735141.4 & 2MASS J01154060-7312293    & -           & -            &   -   & - &   Group 7  & -\\
32&  J010318.64-721211.5 & 2MASS J00505852-7351417    & -           & -            &   -   & - &   Group 7  & Some diffuse emission at 24 micron\\
33&  J010041.61-723800.7 & 2MASS J01154060-7312293    & -           & -            &   -   & (R) &   Group 2 & -\\
\hline

\end{tabular}
\end{minipage}
\end{table*}
\end{enumerate}

\begin{table*}
\scriptsize
\centering
\begin{minipage}{180mm}
\caption{Source counts, by colour class, before and after the filtering in Section \ref{sec:vetting} is applied.\label{tab:sourcecounts}}
\begin{tabular}{@{}llllllllllllll@{}}
\hline
Colour & Number & Number & \multicolumn{11}{c}{Number of counterparts}\\
class\footnote{Classification using the same scheme as in \citet{B2011}} & (\citetalias{B2011})\footnote{Numbers from the \citet{B2011} study.} & (this paper) & \multicolumn{3}{c}{IRAC Archive Epoch\footnote{For the {\it Spitzer} data, Epoch 0 denotes the S$^3$MC photometry, while Epochs 1 and 2 are from SAGE-SMC.}} & \multicolumn{3}{c}{MIPS24 Epoch\footnote{MIPS~24~\mic\ photometry from the Catalog List (see Section \ref{subsec:iracmipssourcelist}).}} & OGLE & 2MASS & IRSF & AKARI & WISE\\
&&&1&2&0&1&2&0&&&&&\\
\hline
O--AGB & 2\,478 & \NrawO  & 2472 & 2448 & 1133 & 105 & 112 & 105 & 1991 & 2472 & 2054 & 110 & 2385\\
~~~(filtered) & & \NvetO  & 2471 & 2447 & 1132 & 105 & 112 & 105 & 1990 & 2471 & 2053 & 110 & 2384\\
a--AGB & 1\,244 & \NrawaO  & 1710 & 1703 & 798 & 462 & 524 & 388 & 1543 & 1698 & 1425 & 68 & 1685\\
~~~(filtered) & & \NvetaO  & 1710 & 1703 & 798 & 462 & 524 & 388 & 1543 & 1698 & 1425 & 68 & 1685\\
C--AGB & 1\,729 & \NrawC  & 352 & 349 & 149 & 317 & 323 & 129 & 286 & 343 & 289 & 15 & 348\\
~~~(filtered) & & \NvetC  & 341 & 337 & 142 & 309 & 314 & 125 & 283 & 333 & 278 & 15 & 337\\
X--AGB & 349 & \NrawX  & 1195 & 1184 & 524 & 18 & 18 & 19 & 1017 & 1196 & 983 & 34 & 1164\\
~~~(filtered) & & \NvetX  & 1195 & 1184 & 524 & 18 & 18 & 19 & 1017 & 1196 & 983 & 34 & 1164\\
RSG & 3\,325 & \NrawRSG  & 3848 & 3810 & 1545 & 307 & 335 & 259 & 201 & 3906 & 2812 & 123 & 3799\\
~~~(filtered) & & \NvetRSG  & 3765 & 3730 & 1495 & 293 & 320 & 252 & 199 & 3823 & 2745 & 119 & 3724\\
FIR & 360 & \NrawFIR  & 95 & 95 & 58 & 70 & 76 & 49 & 36 & 91 & 85 & 2 & 90\\
~~~(filtered) & & \NvetFIR  & 33 & 33 & 19 & 31 & 32 & 19 & 16 & 33 & 30 & 0 & 33\\
\hline
Total & 9\,539 & \Nrawev & 9672 & 9589 & 4207 & 1279 & 1388 & 949 & 5074 & 9706 & 7648 & 352 & 9471 \\
~~~(filtered) & & \Nvetev & 9515 & 9434 & 4110 & 1218 & 1320 & 908 & 5048 & 9554 & 7514 & 346 & 9327 \\
\hline
\end{tabular}
\end{minipage}
\end{table*}

\section{Fitting and fit refinement}
\label{sec:fitting}
Our procedure to compute best fits and subsequently improve these fits is similar to that of Section 2.3 in \citetalias{R2012}. We first fit GRAMS models of both chemical types (O--rich\footnote{Note that the GRAMS models do not distinguish O--rich AGB stars from RSGs; the latter is a category derived from the \citetalias{B2011} colour classification and variability information.} and C--rich) to each SED in our list. By default, we fit the optical, near-infrared and {\it Spitzer} photometry. Where available, we also fit the {\it AKARI} {\it S11} and {\it L15} photometry, as well as the {\it WISE} {\it W3} band. GRAMS is a pre-computed grid of models for circumstellar dust shells around hydrostatic photosphere models with two fixed prescriptions for the dust properties, one each for O--rich and carbon dust. For more details on the models, we refer the reader to \citet{Sargentetal2011} and \citet{Srinivasanetal2011}. The fitting procedure involves comparing each source to a set of model templates. In such a situation, it is not straightforward to identify the number of degrees of freedom in the problem. Moreover, the reduced \chisq\ is not a meaningful statistic given the non-linear parameter dependence \citep{Ye1998}. For this reason, we quantify the fit quality using the \chisq\ per data point (\chisq\ divided by the number of bands with valid flux measurements). This method has also been used in fitting pre-computed YSO models to data \citep{Robitailleetal2007}.\\

The model with the lowest \chisq\ per data point (hereafter abbreviated as \chisq) determines the best-fit parameters -- the chemical type (`GRAMS class'), luminosity and dust-production rate being the most important. The reliability of the chemical classification depends on the relative difference between the O--rich and C--rich best-fit \chisq\ values; this issue is addressed in Section \ref{subsec:chemicalclassification}. For each chemical type, we also define a range of `acceptable' fits using the hundred models with the lowest \chisq\ values. We set the uncertainty in each best-fit parameter to the median absolute deviation of the acceptable values from their median \citepalias[MADM; see][]{R2012}.\\

For dusty sources, the optical part of the SED is much fainter than the mid-IR, and the luminosity and dust-production rate are most sensitive to the mid-IR fluxes. Moreover, the optical range is most affected by variability due to the presence of deep molecular absorption bands \citep{Bladhetal2013}. In such cases, it is not desirable to assign equal weights to the optical and mid-IR fluxes. Following \citetalias{R2012}, we disregard the optical photometry for sources with $I-J > 1.4$ mag. Of the sources in our filtered dataset (Table \ref{tab:sourcecounts}) we ignore optical data for a total of \Nvetignoreoptical\ sources -- about 25--30\% of the sample. This fraction includes dusty sources, as well as those in the high-\chisq\ group (see Section \ref{subsubsec:highchisq}). \citetalias{R2012} disregarded optical fluxes for a similar fraction of objects in their LMC sample.

As mentioned in \ref{subsubsec:FIRobjects}, it is possible for a blue source to appear dusty at 24 \mic\ due to a bright IR neighbour. We adjust for this by lowering the weight assigned to the 24 \mic\ photometry for our entire sample -- the relative uncertainty at 24 \mic\ is set to be at least equal to the largest relative uncertainty among the IRAC bands.\\

The GRAMS dust-production rates are computed assuming a constant expansion speed $v_{\rm exp} = 10$ km s$^{-1}$. In practice, the expansion speed varies from star to star. For consistency with \citetalias{B2012}, we scale our expansion speeds using the \citet{vanLoon2006} scaling relation:
\begin{equation}
\label{eqn:vexpscaling}
{v_{\rm exp}\over 10~{\rm km}~{\rm s}^{-1}} = \left({L\over 3\times 10^4~{\rm L}_\odot}\right)^{1/4}\left({\Psi_{\rm SMC}\over 200}\right)^{-1/2}
\end{equation}
The gas:dust ratio $\Psi$ is not well constrained for Magellanic Cloud stars; the gas:dust ratio is expected to increase at lower metallicity. However, the dependence of the gas:dust ratio on metallicity is much weaker for carbon stars. In this paper, we therefore use a gas:dust ratio of 500 (1000) for O--rich stars in the LMC (SMC) and a ratio of 200 for carbon stars in both Clouds. The DPR is directly proportionate to the expansion speed, so it also scales according to Equation \ref{eqn:vexpscaling}. We use these scaled DPRs throughout the rest of the paper, except when comparing our results to those of \citet{Groenewegenetal2009} (Fig. \ref{fig:G2009DPRvsGRAMSDPR}).\\

\begin{figure*}
 \includegraphics[width=84mm]{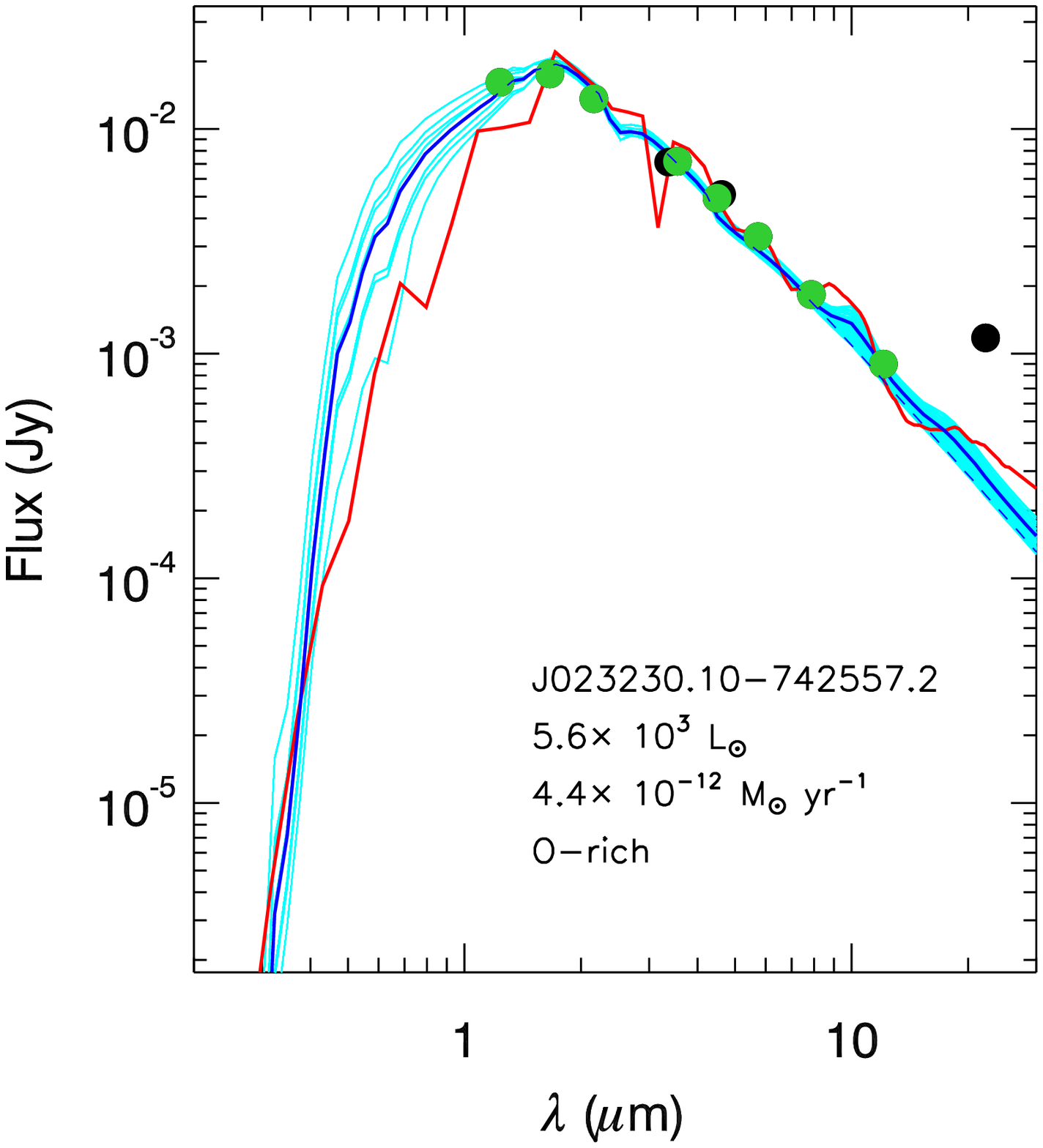} \includegraphics[width=84mm]{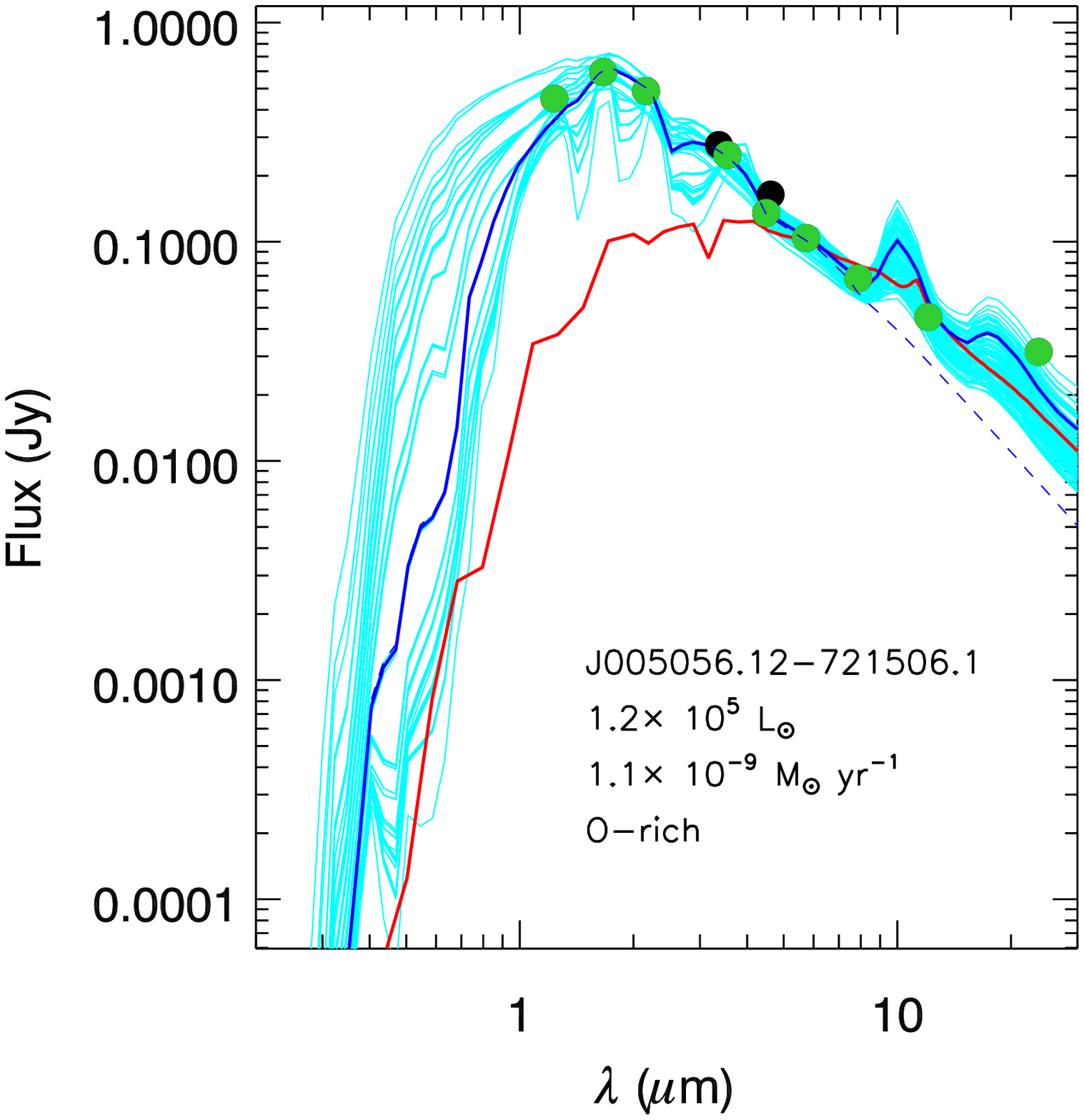}\\
 \includegraphics[width=84mm]{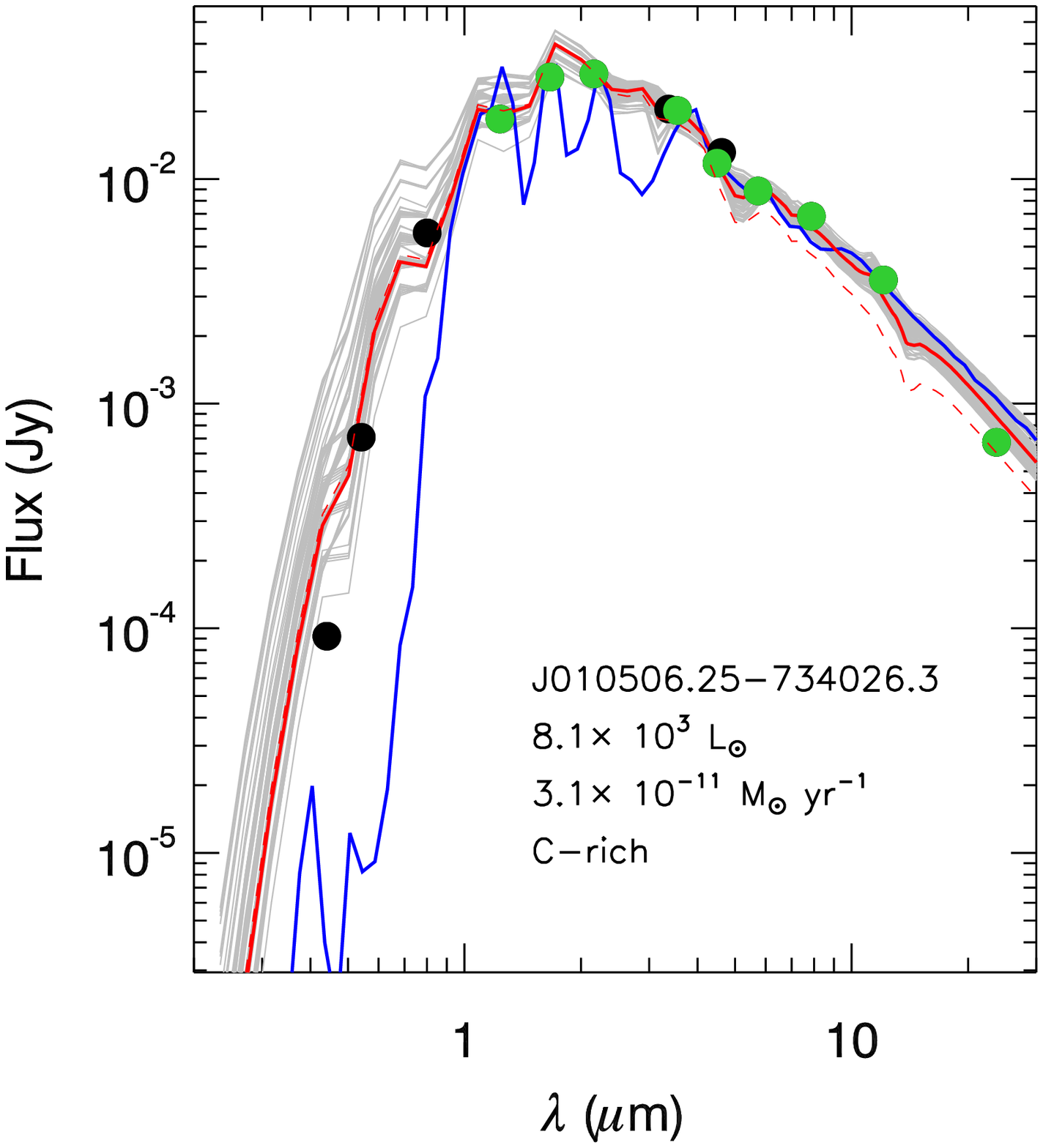} \includegraphics[width=84mm]{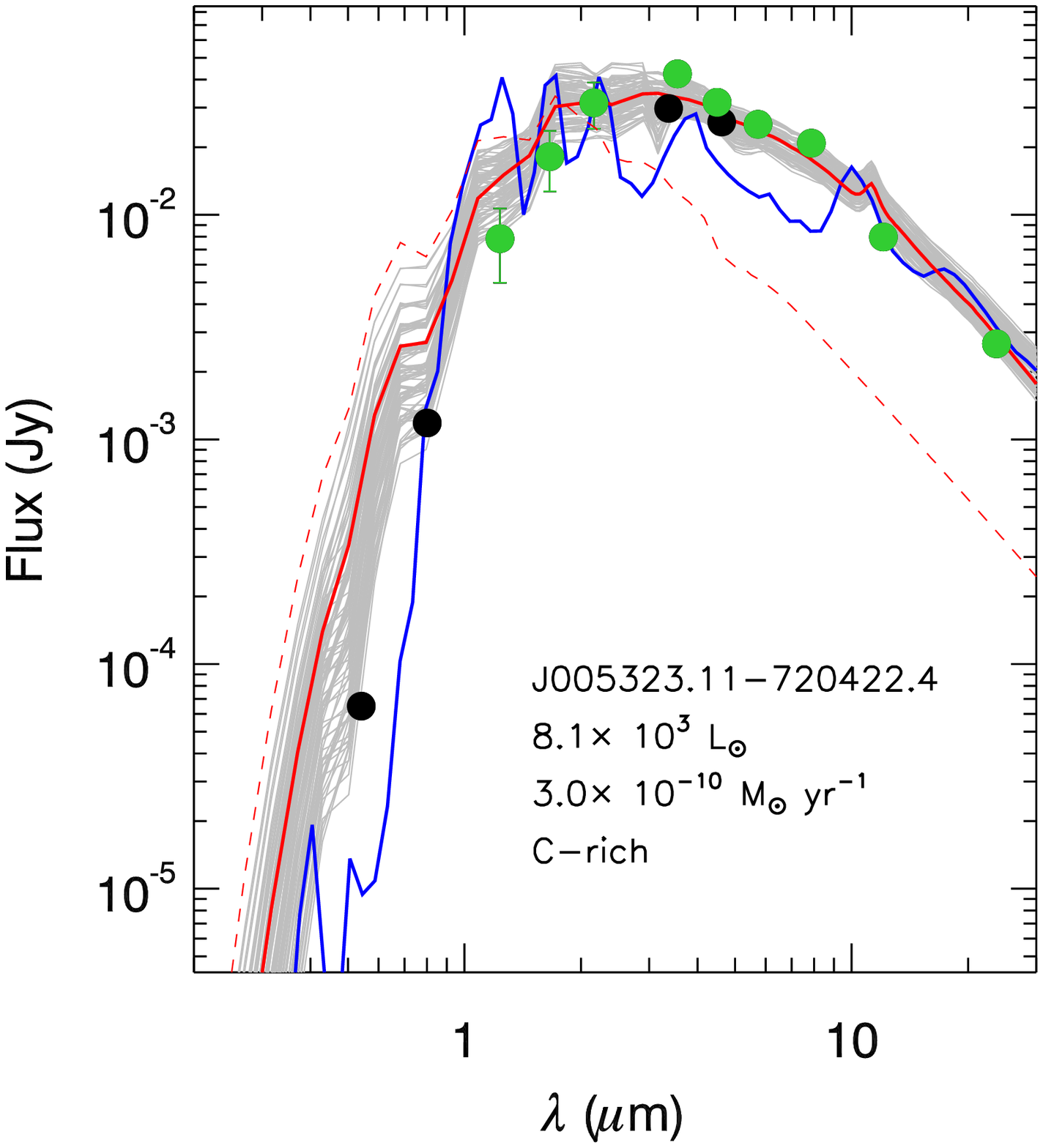}
 \caption{Typical best-fits for sources of both chemical types. Data points (filled circles) are coloured based on whether they are used to compute the fit (green: used, black: ignored). Each plot shows the best-fit O--rich/C--rich model (solid blue/red curve) and the corresponding best-fit photosphere (dashed curve). The best-fit model is overlaid on the 100 acceptable models, colour coded by chemical type (O--rich: cyan, C--rich: grey). For instance, SSTISAGEMA~J010308.56--731718.0 is classified as O--rich; the best-fit model is shown by the blue curve, and the cyan curves are the 100 acceptable O--rich models with the lowest \chisq. The best-fit C--rich model (red curve) is also shown. For each source, the figure also displays the luminosity and DPR estimates corresponding to the best-fit chemical type. \label{fig:typicalfits}}
\end{figure*}

Fig. \ref{fig:typicalfits} shows two SEDs of each chemical type, along with their best-fit model spectra. These fits are `typical' in the sense that their \chisq\ values are close to the median for sources of that chemical type. In each case, the best-fit model of each type is also shown, as well as the hundred acceptable models of the best-fit chemical type. This `envelope' of acceptable models demonstrates the range that results in reasonable fits to the overall SED. For comparison, the best-fit bare photosphere model is also shown in each case.\\

\subsection{Fit quality}
\subsubsection{High-\chisq\ and high-DPR fits}
\label{subsubsec:highchisq}
We inspect the fits in the 95th percentile of \chisq\ values and find that, in many cases, fits without the optical do a noticeably better job of reproducing the mid-IR SED (which strongly influences the DPR). As the optical data are not obtained concurrently with the {\it Spitzer} data, large variations are expected. For these high-\chisq\ sources, therefore, we ignore the optical fluxes if doing so reduces the \chisq. In all, we disregard the optical data for \Nvetignoreopticalhighchisq\ of the $\approx$500 sources in the high-\chisq\ sample.\\
 
\begin{figure*}
 \includegraphics[width=84mm]{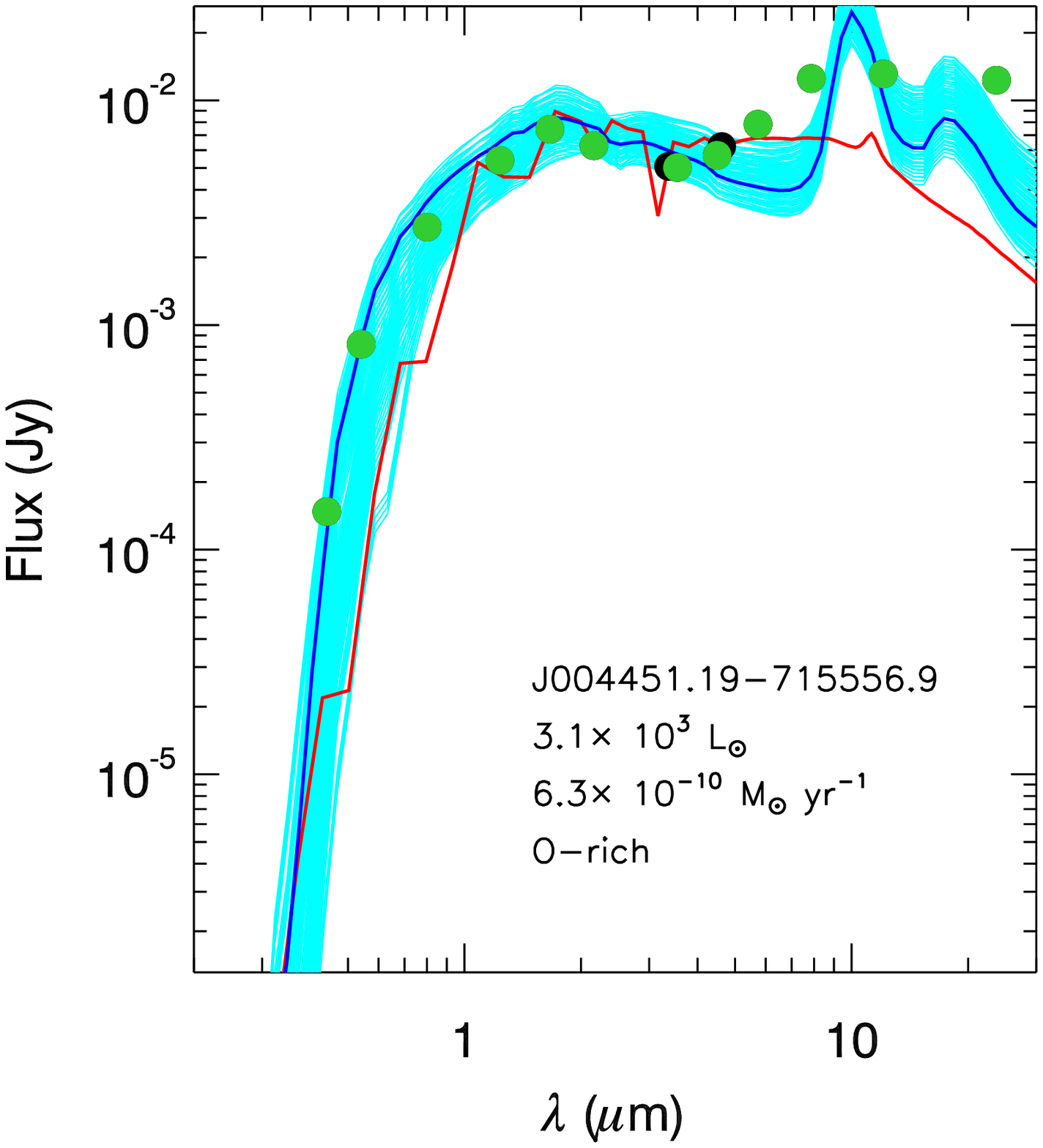} \includegraphics[width=84mm]{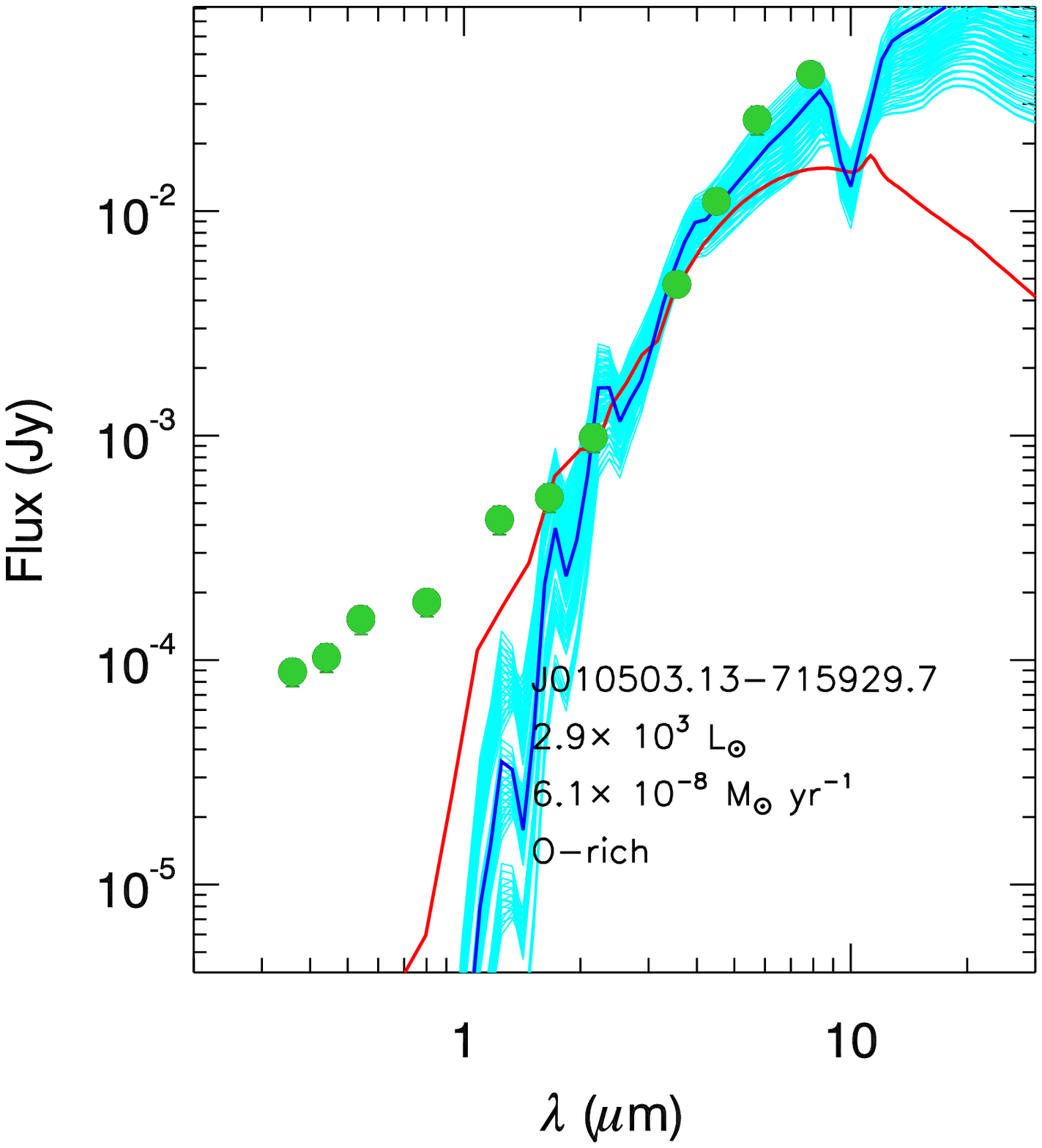}\\
 \includegraphics[width=84mm]{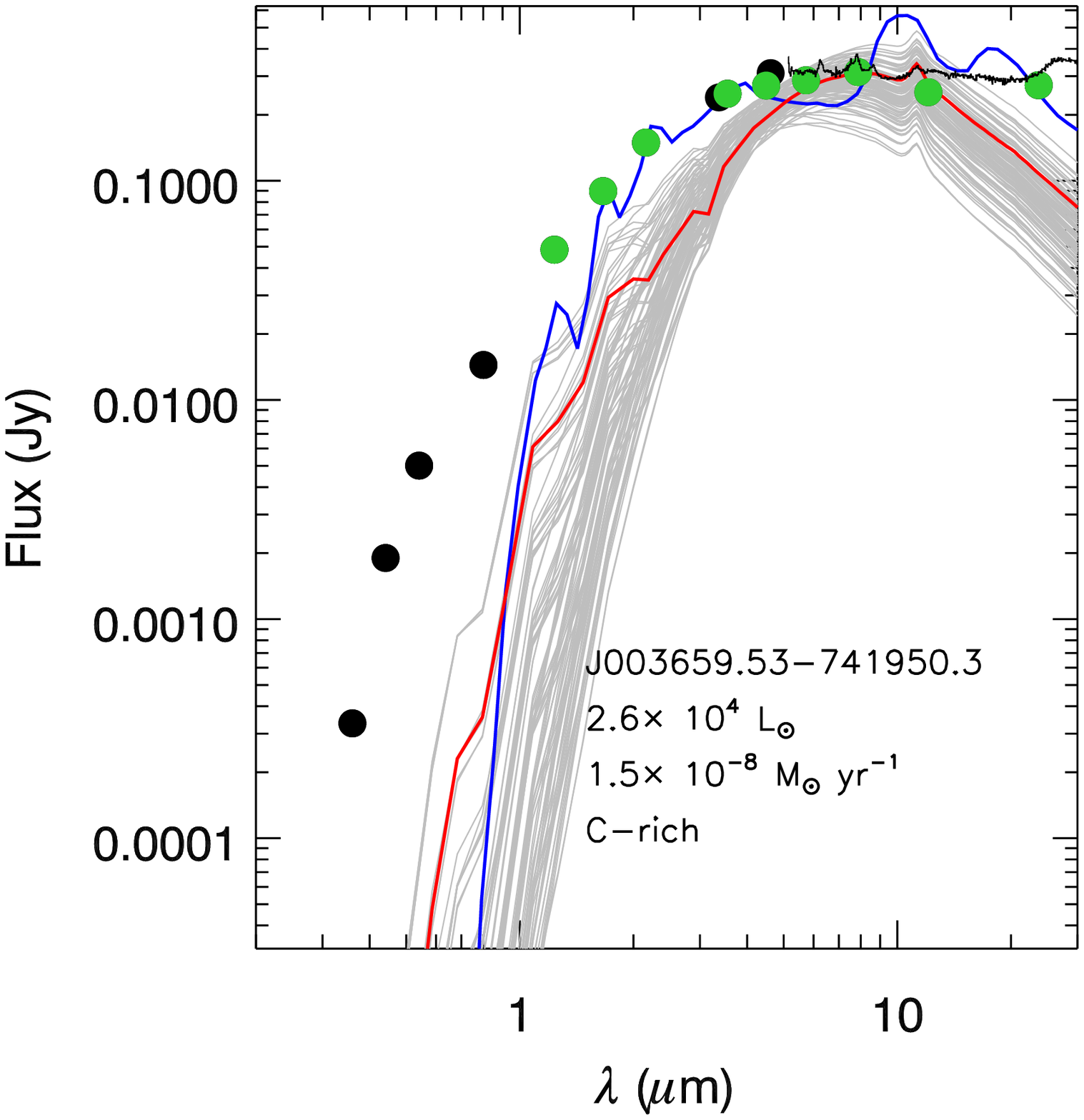} \includegraphics[width=84mm]{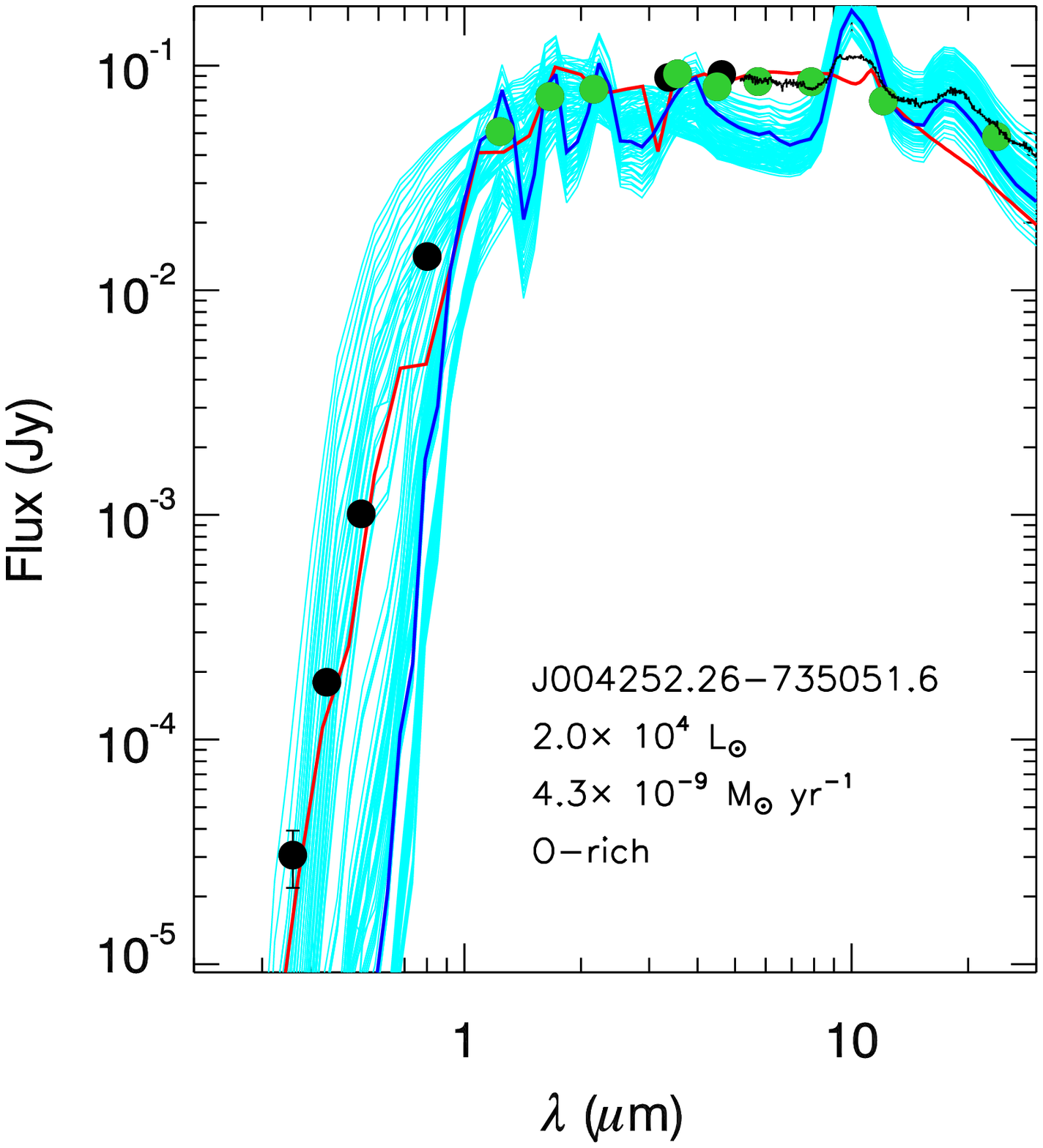}
 \caption{\emph{Top}: Examples of bad fits -- a double-peaked SED (\emph{left}) and a rising SED that could be a YSO or background galaxy (\emph{right}), the latter with a high dust-production rate fit. \emph{Bottom}: Examples of fits whose chemical types are manually set. The source on the left is IRAS~00350--7436. The chemical type was modified from O--rich (blue line) to C--rich to agree with the carbon-rich features in its IRS spectrum. The source on the right shows a 10 \mic\ silicate feature in its IRS spectrum, but was originally classified as C--rich (red line). See text for details.\label{fig:himlr}}
\end{figure*}

To ensure an accurate determination of the global dust budget, we examine the SEDs and fits to the 200 non-FIR sources with the highest DPRs. We find that \NvetnonFIRflaggedinvalid\ sources have either rising SEDs (likely YSOs or background galaxies) or double-peaked SEDs (disc-like or detached shell sources, including post-AGB stars). The top panel of Fig. \ref{fig:himlr} shows an example of each. While the best-fit DPR for the rising SED is very high, the fit quality is quite poor. We flag all these SEDs as bad/invalid fits.\\

For \Nvetchangechemtype\ sources, we find either that the C--rich model is a better fit to the mid-IR SED than the O--rich one, or that the GRAMS class does not agree with the spectroscopic identification. As we are interested in comparing the dust contribution from both chemical types, it is important to correct any such misclassification. In such cases, we set the alternate model as the best-fit. The bottom panel of Fig. \ref{fig:himlr} shows two examples. On the left is IRAS~00350--7436, the most luminous carbon-rich object in the SMC. It has variously been classified as a carbon star \citep{Ruffleetal2015}, a post-AGB \citep[e.g.,][]{Matsuuraetal2005}, and a symbiotic star \citep{Whitelocketal1989}. Our fitting routine classified it as O--rich, and we modify this to be consistent with the carbonaceous features in its IRS spectrum. It is not surprising that the fit quality remains poor, given the unique SED shape. The SED in the bottom right panel of the same figure was classified as C--rich; however, its IRS spectrum shows the 10 and 18 \mic\ silicate features, so we modify the best-fit chemical type accordingly.\\

\subsubsection{Confidence in chemical classification}
\label{subsubsec:chemclassconfidence}
We determine the chemical classification for each source by comparing the \chisq\ values of the best-fit O--rich and C--rich models. The lower of the two is referred to as the `best' model, and the other as the `alternate' model. The best model sets the chemical classification as well as the values of the other fit parameters, while the alternate model is used to determine the quality of the classification. 

\begin{figure}
 \includegraphics[width=84mm]{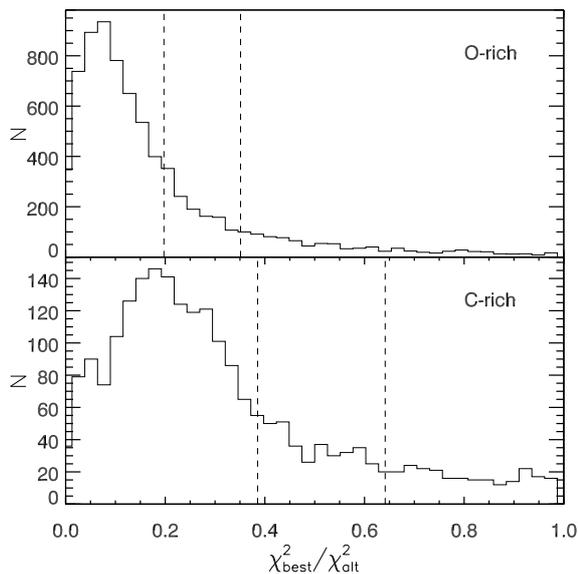}
 \caption{The ratio of the best-fit model \chisq\ to that of the alternate model for O--rich (top) and C--rich (bottom) chemical types. A low value of $\chi^2_{\rm best}/\chi^2_{\rm alt}$ corresponds to a high-confidence chemical classification. For each chemical type, the vertical lines differentiate the high-, medium-, and low-confidence regimes.\label{fig:chemtypeconfidence}}
\end{figure}

Fig. \ref{fig:chemtypeconfidence} shows the distribution of ratios of the \chisq\ of the best-fit model to that of the alternate model for both chemical classes. These distributions are qualitatively similar to fig. 10 in \citetalias{R2012}, but the O--rich distribution for the SMC data is broader than its LMC counterpart. As in \citetalias{R2012}, we define high-, medium-, and low-confidence intervals. We deem fits with best-to-alternate ratio under \chiratiolowo\ (\chiratiolowc) to have high-confidence O--rich (C--rich) classifications, and fits with ratios over \chiratiohio\ (\chiratiohic) to have low-confidence O--rich (C--rich) classifications. Classifications for intermediate values are considered medium-confidence. We choose the boundaries for the medium- and low- confidence levels such that their best-to-alternate ratios are approximately twice and three times the value at the peaks of the distributions. The vertical lines in Fig. \ref{fig:chemtypeconfidence} demarcate the three confidence intervals for each chemical type. Based on these definitions, we find \Nhicono\ (\Nhiconc) high-confidence O--rich (C--rich) classifications for our filtered sample with valid fits and reliable DPR estimates. These fractions (\percenthicono\% and \percenthiconc\% for O-- and C--rich classifications respectively) are similar to those obtained for the LMC data \citepalias[Table 6 of ][]{R2012}.

\subsubsection{Sensitivity of DPR determination}
\begin{figure}
\includegraphics[height=90mm]{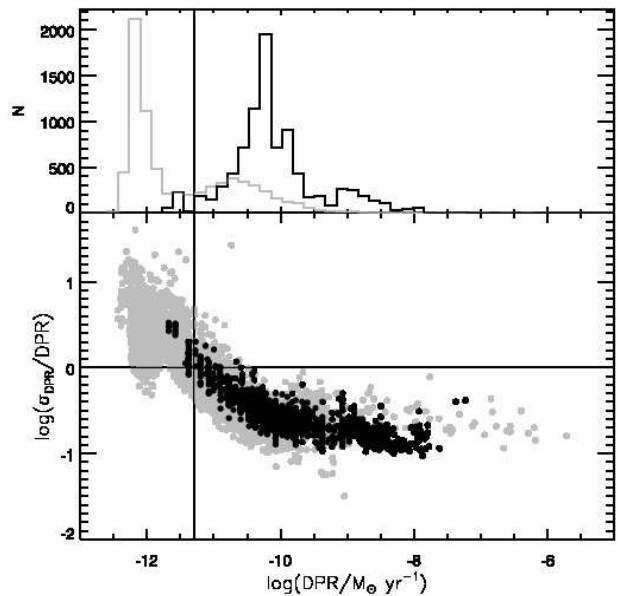}
\caption{{\it Top}: the distribution of best-fit dust-production rates for O--rich (grey) and C--rich (black) chemical types. The C--rich distribution has been multiplied by 4 for clarity. Most of the O--rich fits have DPRs less than $10^{-11.3}$ \msunperyr\ (vertical line). {\it Bottom:} the relative uncertainty in the DPR for both chemical types falls below unity only for DPRs greater than this threshold.\label{fig:MLRquality}}
\end{figure}

Fig. \ref{fig:MLRquality} shows the distribution of best-fit DPRs for each chemical type, as well as the relative uncertainty in the DPR as a function of the DPR. As in \citetalias{R2012}, we set a lower limit on the relative uncertainty in DPR determination at the 1-$\sigma$ level, shown by the horizontal line in the lower panel. The top panel shows that the O--rich sources are split into two populations; a much larger fraction of these objects are located at very low DPRs ($< 10^{-11}$ \msunperyr). This bimodal distribution of O--rich DPRs is similar to that seen in fig. 16 of \citet{R2012}. As in that paper, we set the sensitivity threshold for GRAMS fits to `detect' dust production at $10^{-11.3}$ \msunperyr, shown by the vertical line Fig. \ref{fig:MLRquality}. All of the fits with DPRs $\ga 10^{-10.5}$ \msunperyr\ are above the 1-$\sigma$ limit. We only use DPRs with relative uncertainties above this cutoff to compute the dust budget.

The fitting procedure and refinement described above results in fits to between 6 and 13 (with a median of 12) valid data points with finite fluxes for each source.

\begin{figure*}
\includegraphics[width=55mm]{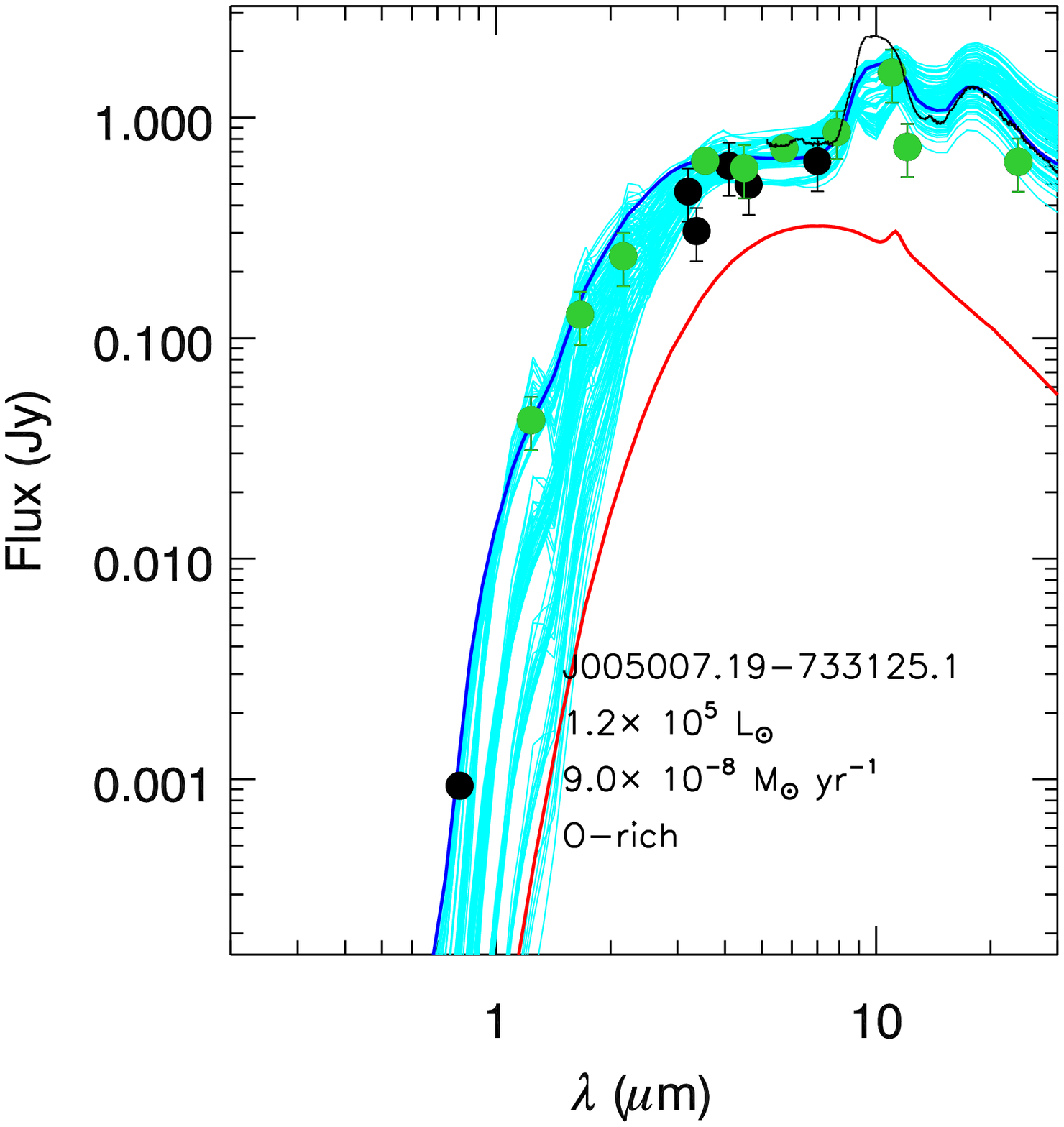} \includegraphics[width=55mm]{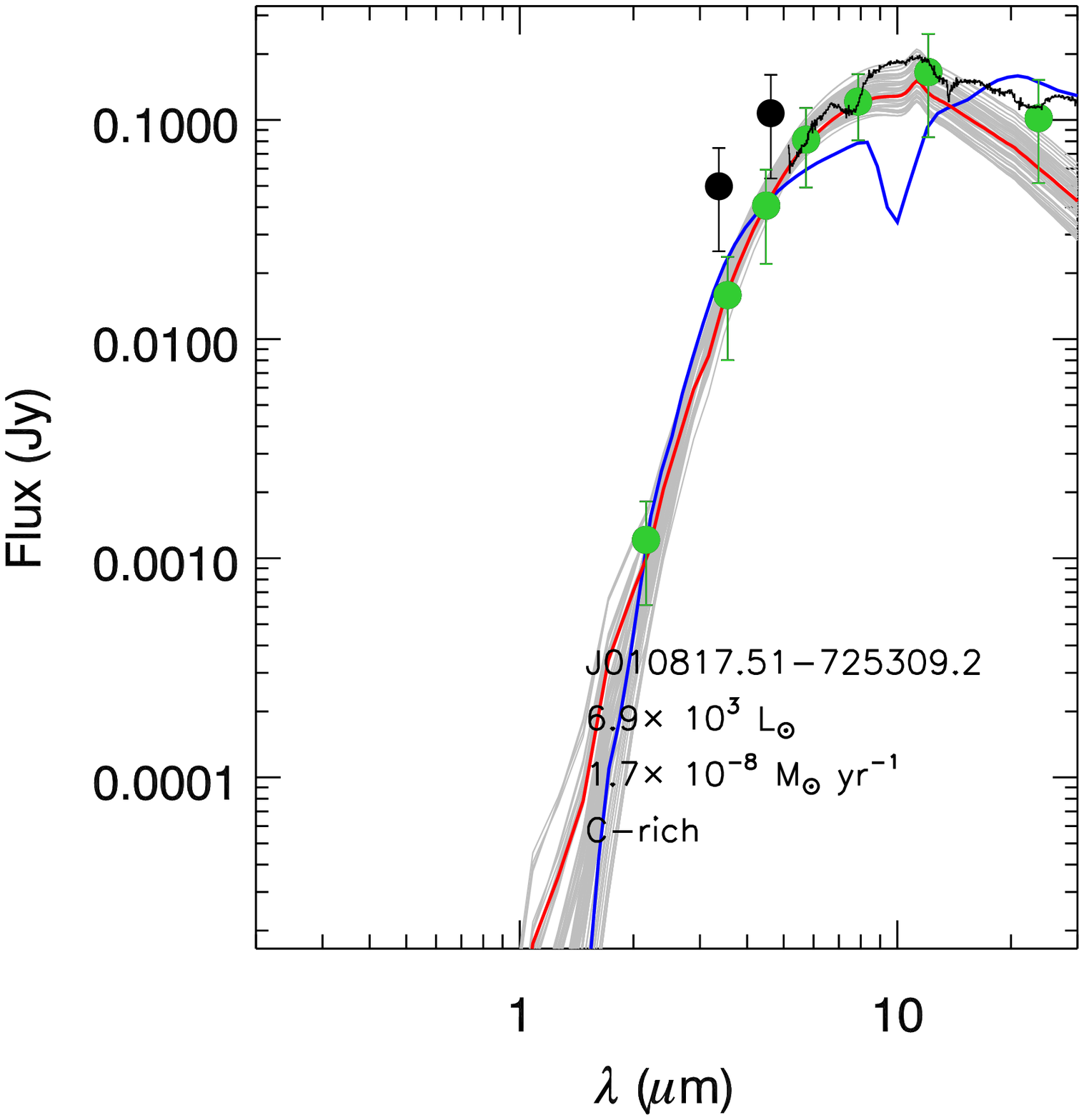} \includegraphics[width=55mm]{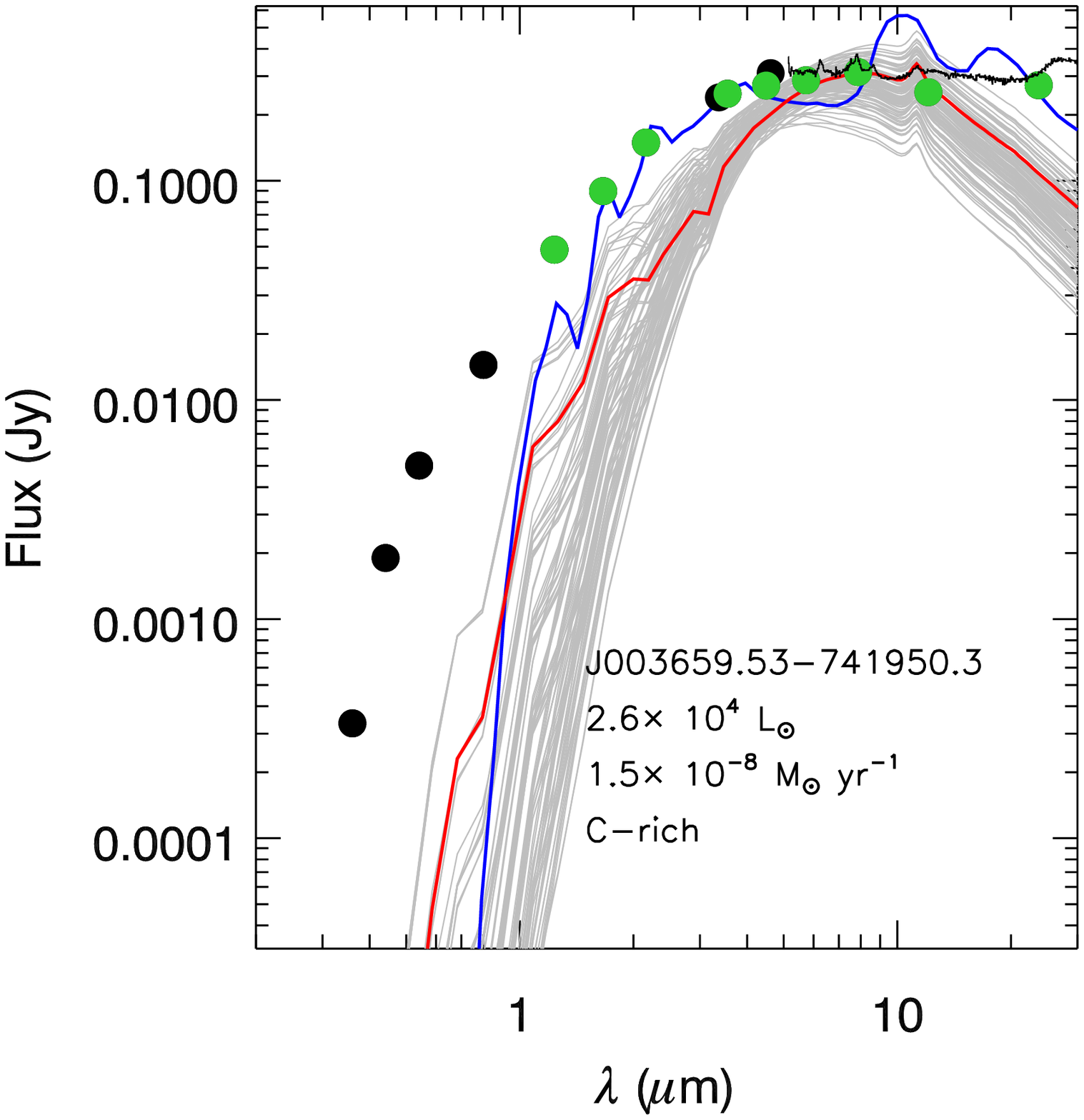}\\
\includegraphics[width=55mm]{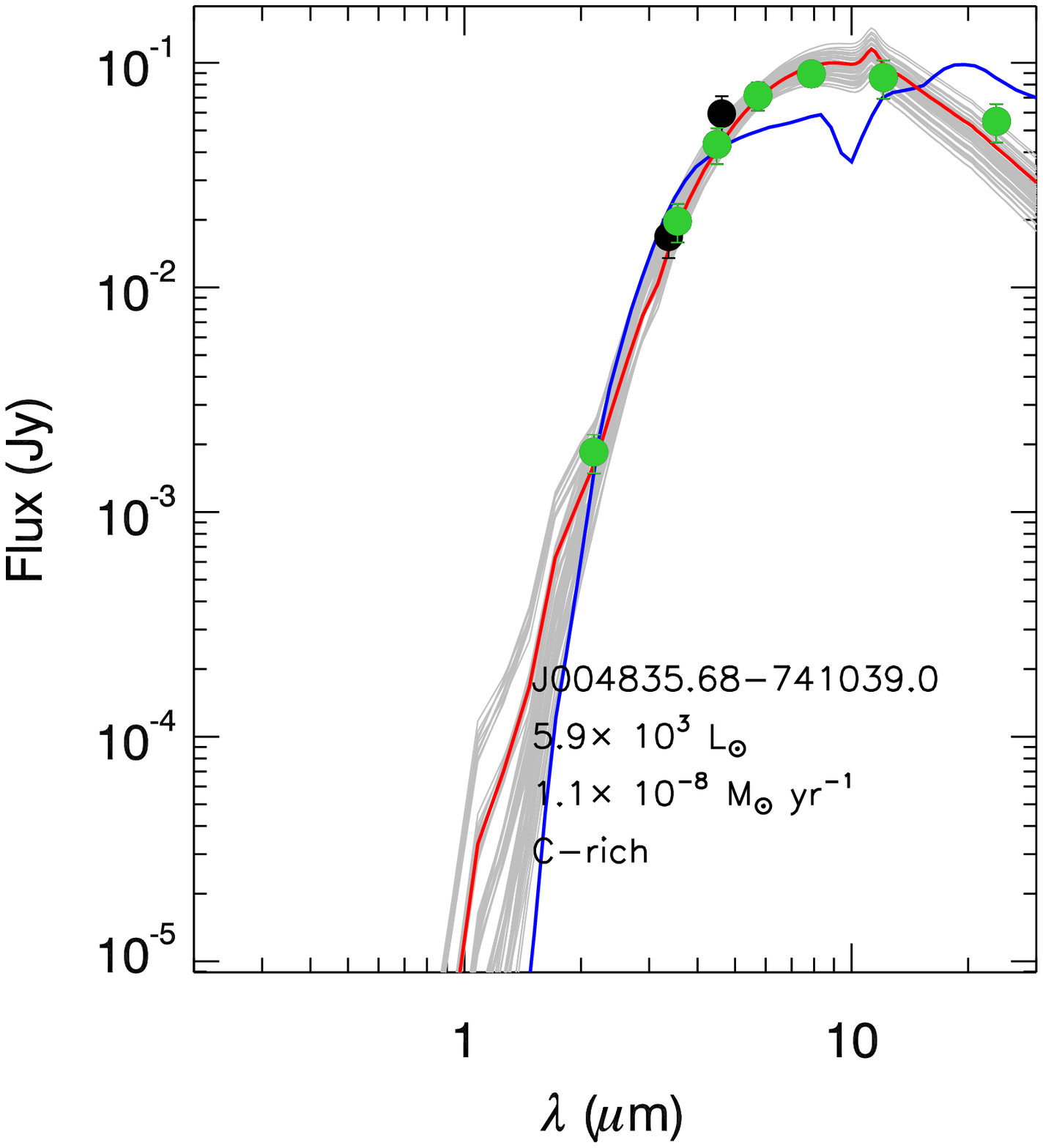} \includegraphics[width=55mm]{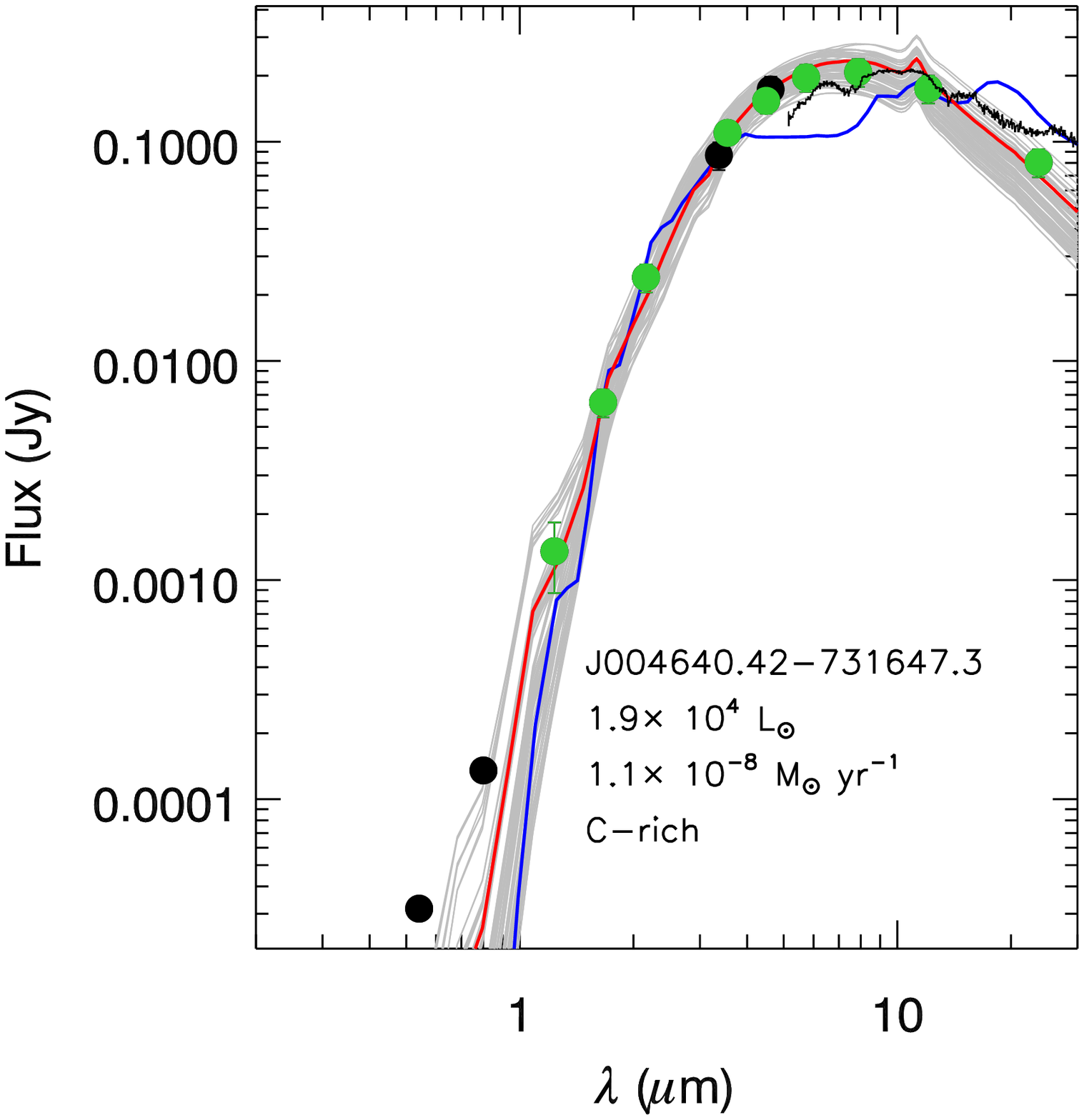} \includegraphics[width=55mm]{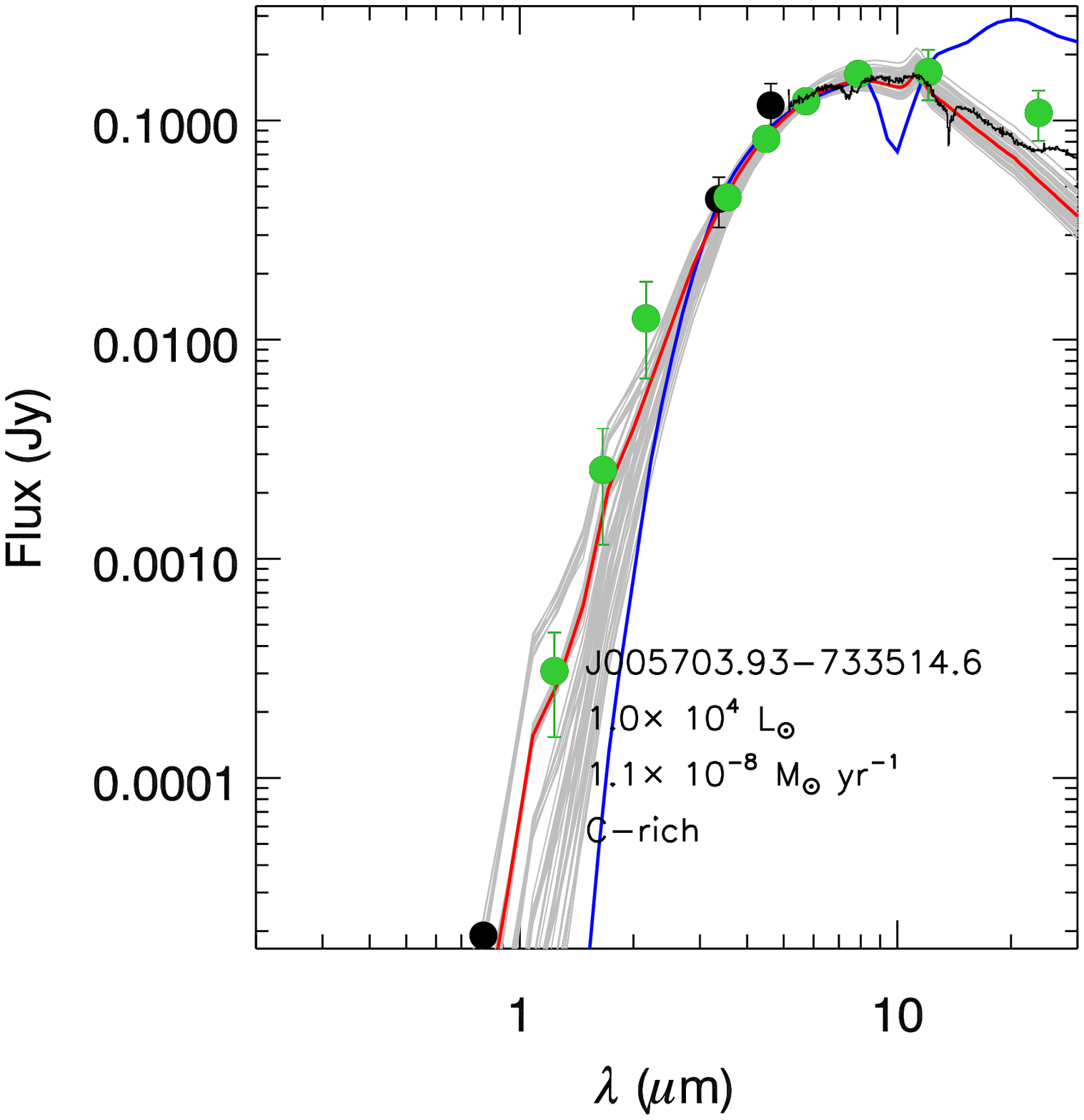}\\
\includegraphics[width=55mm]{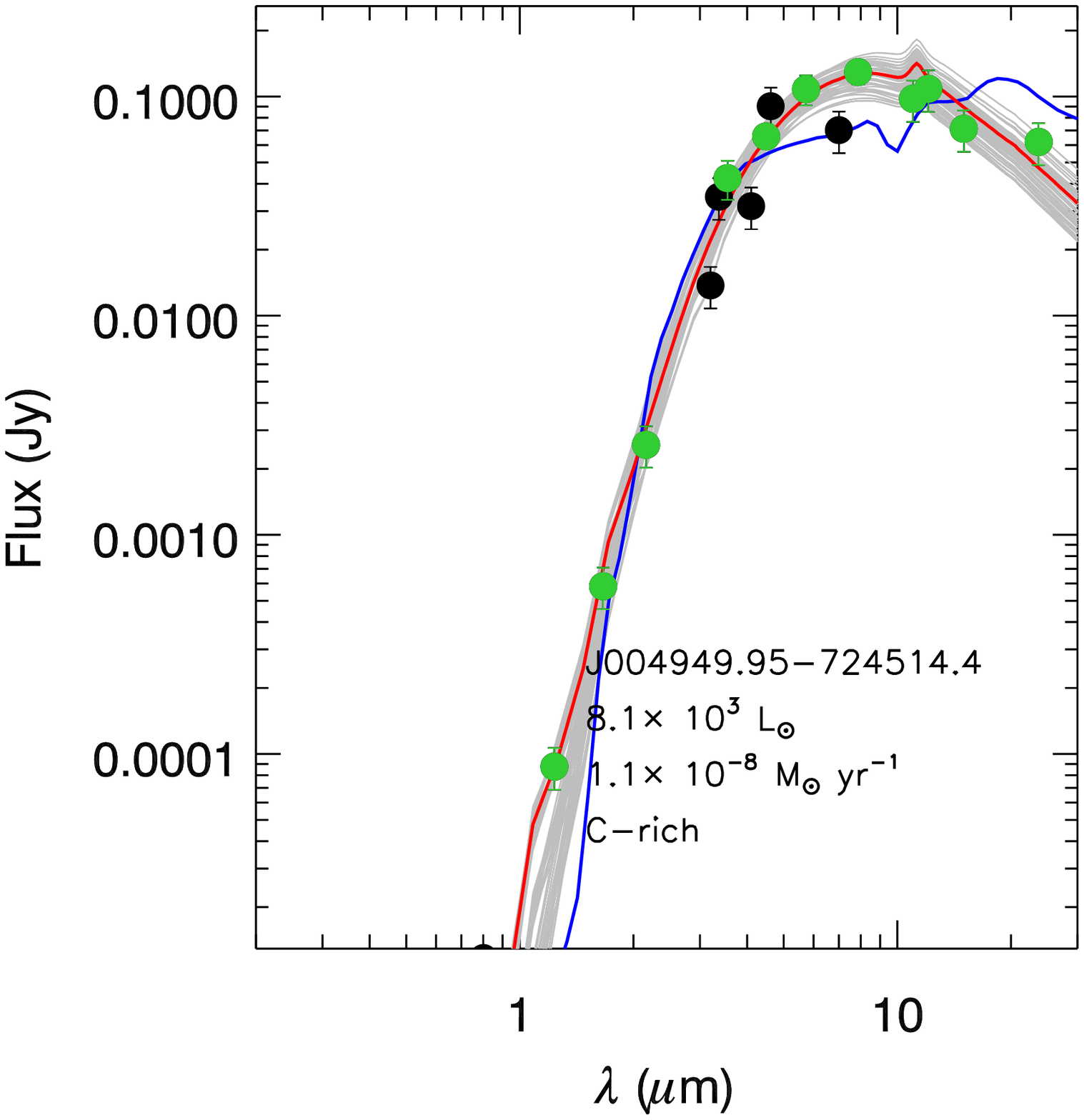} \includegraphics[width=55mm]{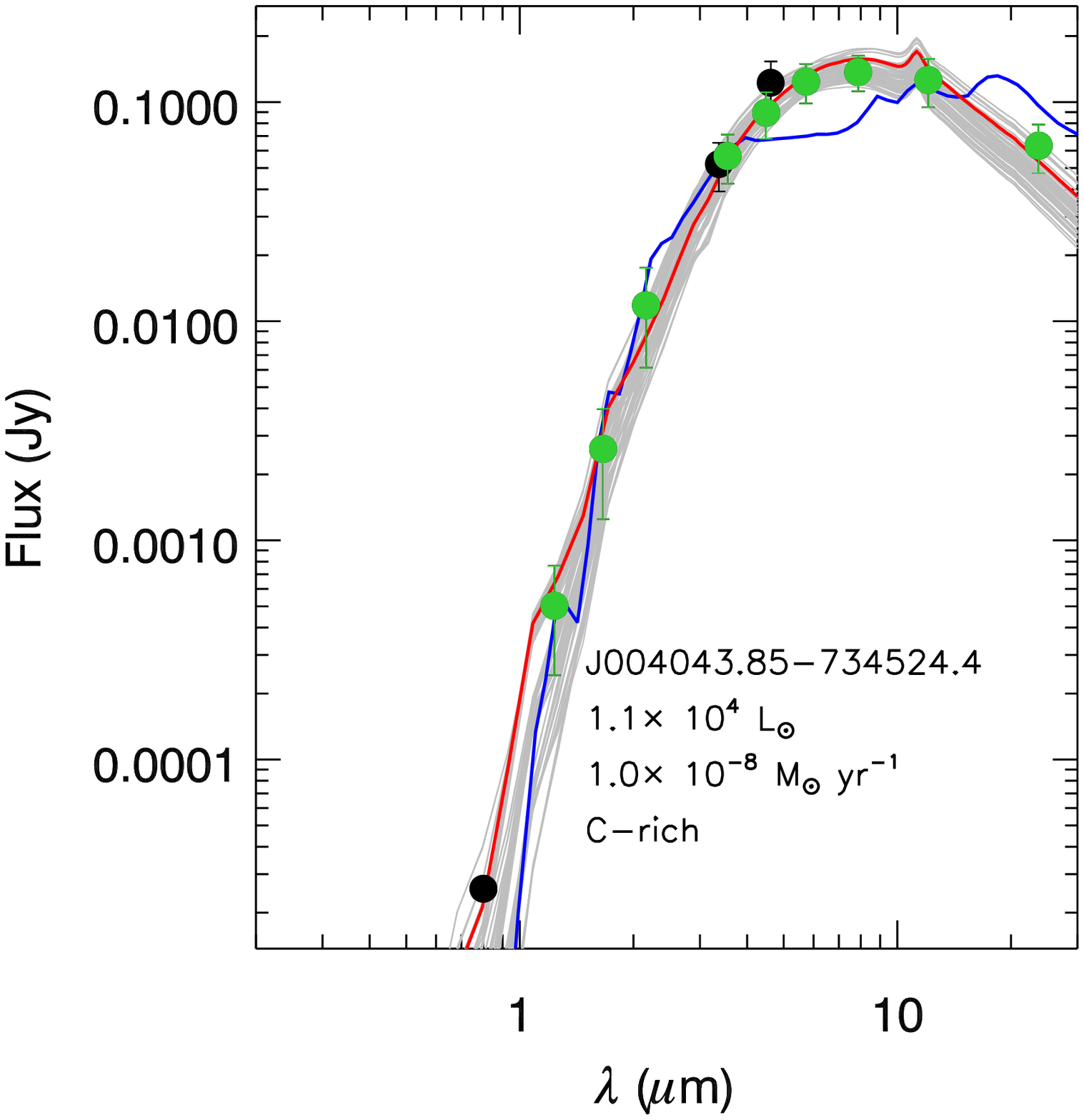} \includegraphics[width=55mm]{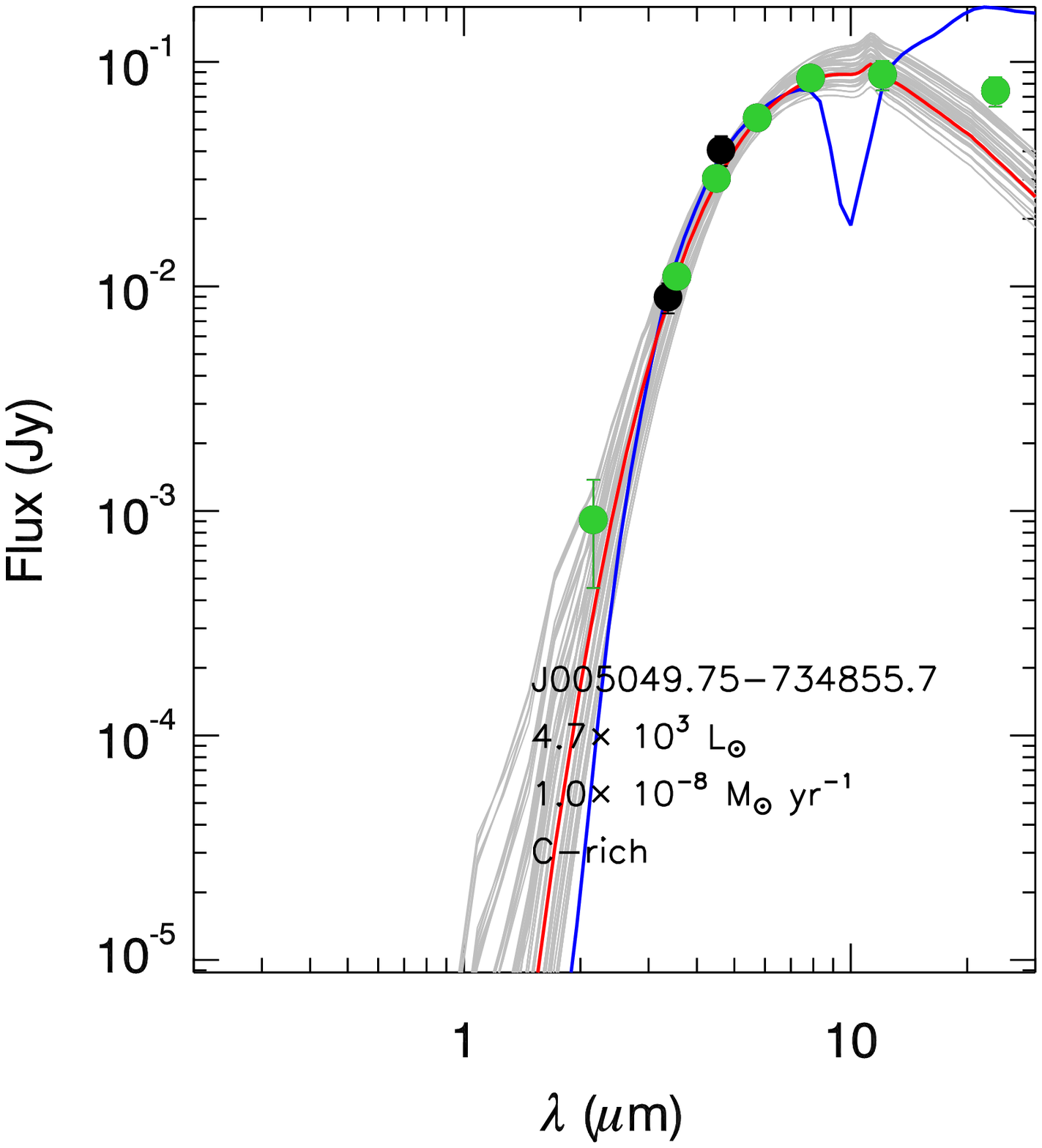}
\caption{Fits to the nine sources in the non-FIR sample with the highest estimated DPRs. The best-fit O--rich (C--rich) model is shown in blue (red), and all the acceptable fits are shown in cyan (grey). \label{fig:vetted_himlr}}
\end{figure*}

\section{Results}
\label{sec:results}
\subsection{Chemical classification}
\label{subsec:chemicalclassification}
\begin{figure*}
\includegraphics[width=55mm]{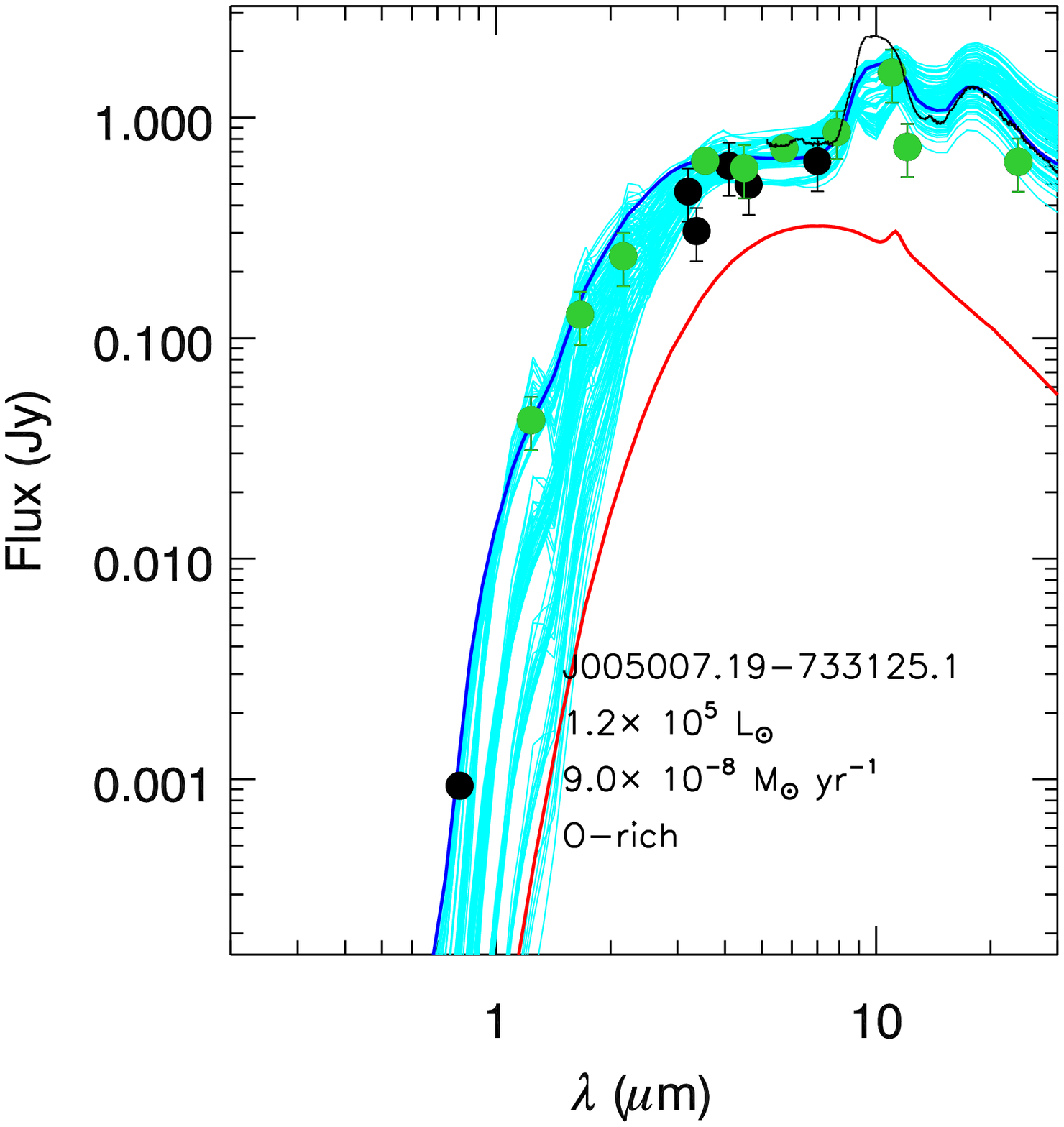} \includegraphics[width=55mm]{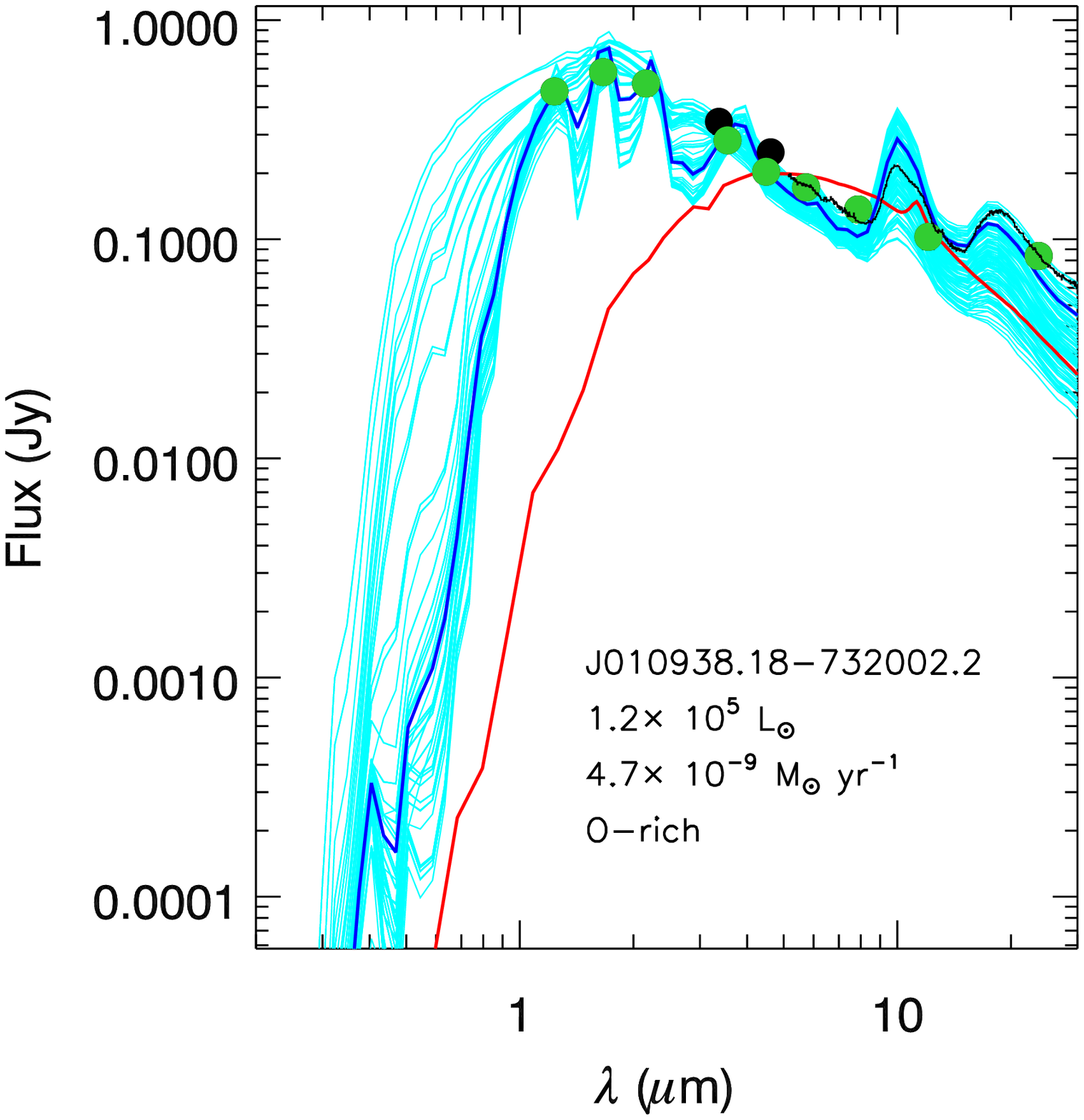} \includegraphics[width=55mm]{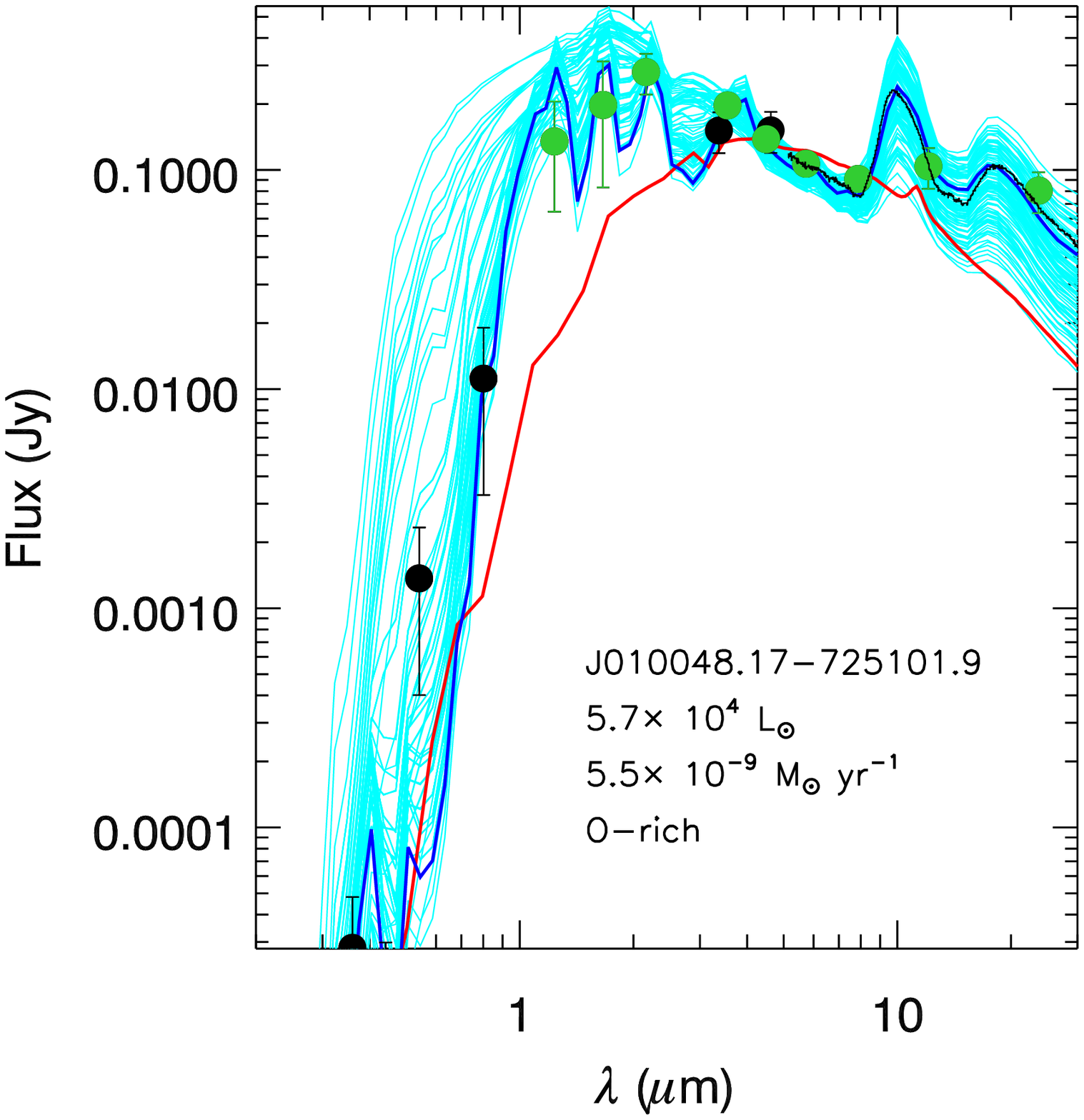}\\
\includegraphics[width=55mm]{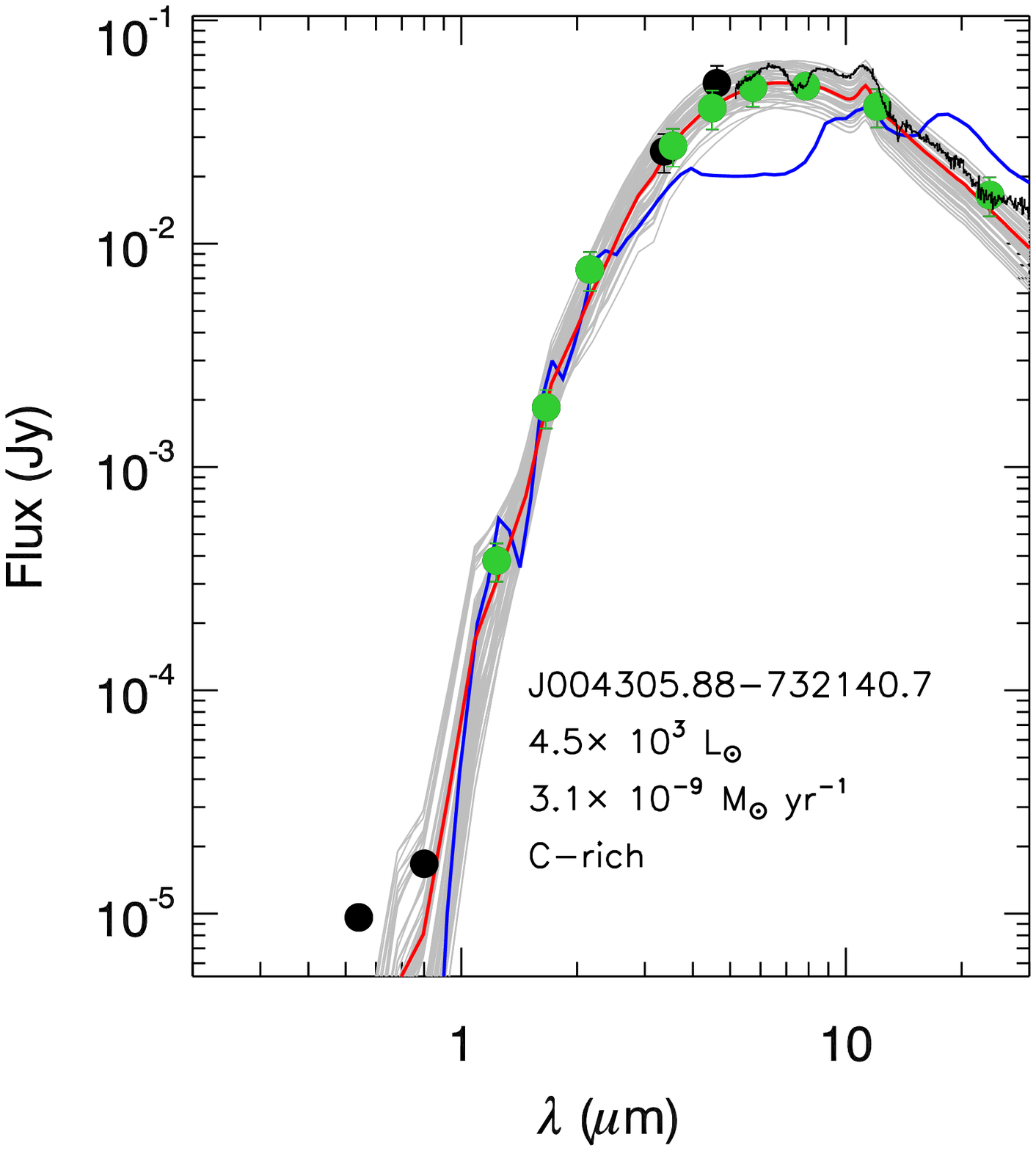} \includegraphics[width=55mm]{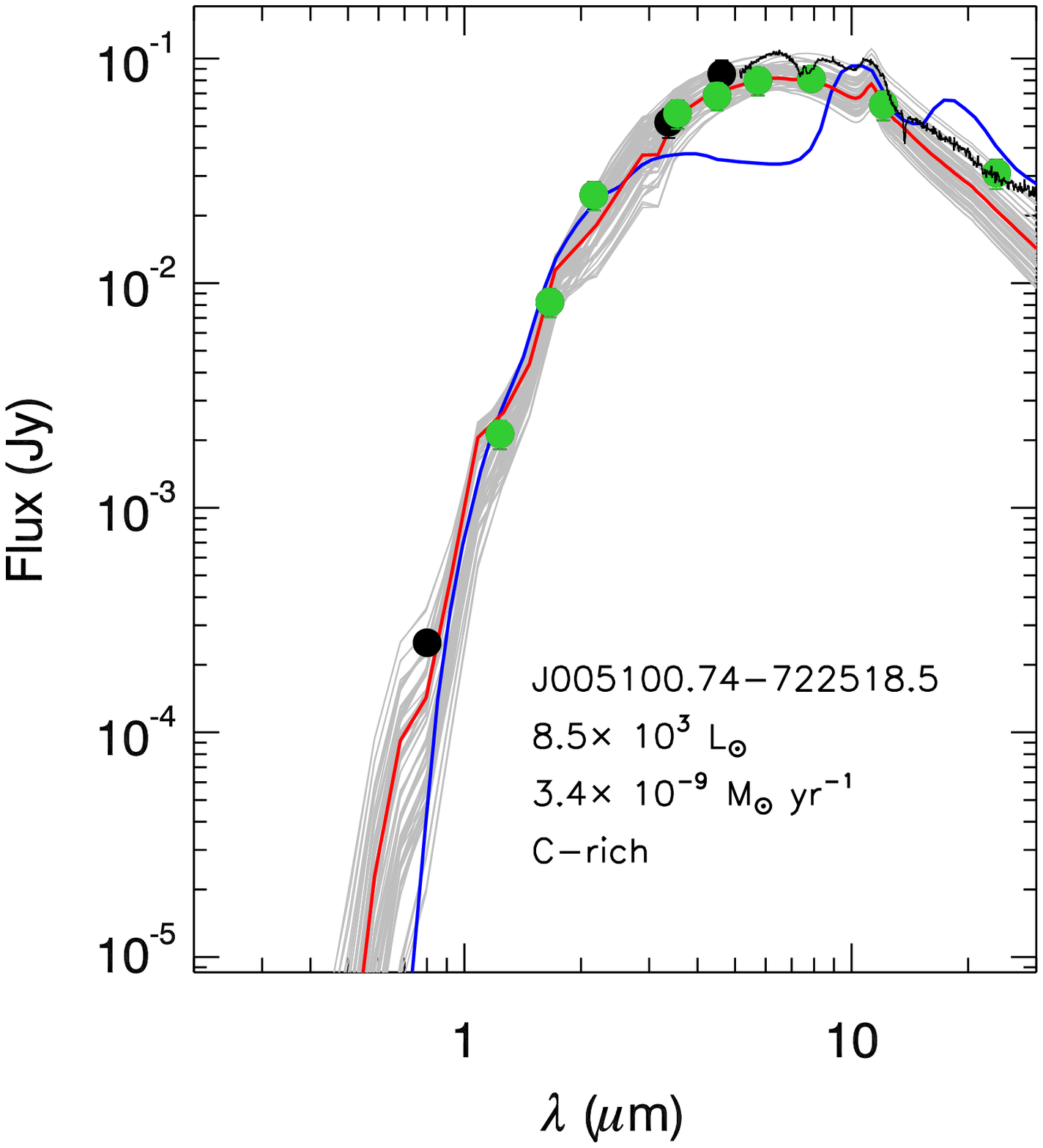}
 \caption{Examples of good fits to the SED as well as IRS spectrum. The top panel shows three good O--rich fits, and the bottom shows two good C--rich fits. All O--rich (C--rich) fits reproduce the silicate (SiC) feature strengths well.\label{fig:goodfits}}
\end{figure*}

Our fit-based chemical classifications are in excellent agreement with the spectroscopic identifications and colour classes used in this paper. Fig. \ref{fig:goodfits} shows some examples of good fits that are able to reproduce the strengths of the silicate or SiC feature in the IRS spectra in addition to the overall SED. For the sample with IRS spectra, our classification differs from that of \citet{Ruffleetal2015} in only two cases -- we classify a pair of O--EAGB stars as C--rich. However, both are low-confidence classifications (the O--rich \chisq\ is about only about 20\% higher). We classify the S star BF~1 as C-rich.\\

Table \ref{tab:colourclassvsgramsclass} demonstrates the excellent agreement between the GRAMS chemical type and our colour-based classifications, and also shows the percentage of low-confidence classifications for each colour class. It is interesting to note that, among the non-FIR sources, the anomalous AGB stars have the largest fraction of low-confidence classifications. This is consistent with the \citet{Boyeretal2015} conclusion, based in part on optical spectra, that their initial masses are very close to the limit where the third dredge-up is sufficient to form a carbon star.\\

We classify all but one of the extreme AGB candidates as C--rich, the exception being IRAS~F00483--7347, an evolved O--rich star classified as an RSG by \citet{Ruffleetal2015} based on its IRS spectrum and luminosity. Only about 2\% of the x--AGB classifications are low-confidence. This is an important result, because 235 of the top 300 non-FIR dust producers are extreme AGB stars, and our classification implies that the mass loss is dominated by carbon stars. This situation, however, is complicated by the results for the FIR objects. While we identify only one FIR object as carbon-rich, over half of the FIR fits have low-confidence classifications, and in many cases the 24 \mic\ data may be contaminated by off-source emission. We discuss these issues in Section \ref{subsec:FIRfitsdiscussion}.

\begin{table}
\scriptsize
\centering
\begin{minipage}{180mm}
\caption{Colour class {\it vs.} GRAMS class.\label{tab:colourclassvsgramsclass}}
\begin{tabular}{@{}lrrr@{}}
\hline
Colour class & \multicolumn{2}{c}{GRAMS class} & \% Low\footnote{Percentage of low-confidence classifications.}\\
& O--rich & C--rich & \\
\hline
O--AGB &802& 82 & 13\\
aO/O &537& 86 & 31\\
aO/C &20& 0 & 20\\
C--AGB &100& 1552 & 12\\
x--AGB &1& 336 & 1.8\\
RSG &1384& 26 & 20\\
FIR &22& 1 & 52\\
\end{tabular}
\end{minipage}
\end{table}

\subsection{Luminosity functions}
\begin{figure}
 \includegraphics[width=84mm,height=84mm]{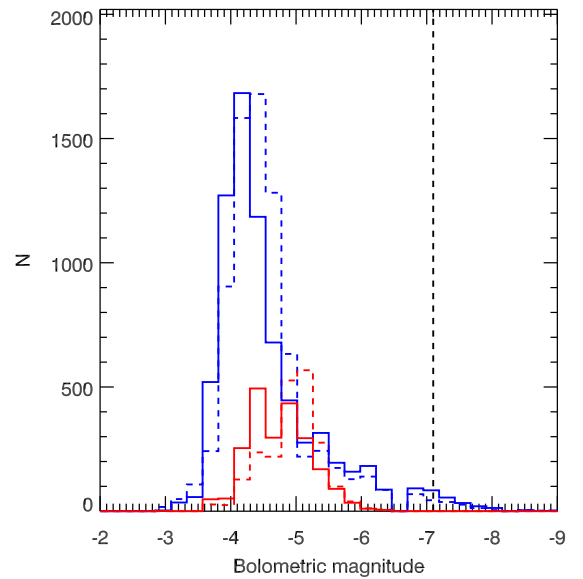}
 \caption{Distribution of luminosities for sources identified by the fits as O--rich (solid blue) and C--rich (solid red). The LMC distributions are shown as dashed lines, and have been scaled down by a factor of 3.3 for direct comparison. The vertical line at $M_{\rm bol}=-7.1$ mag is the classical AGB luminosity limit.
 \label{fig:LF}}
\end{figure}

As in \citetalias{R2012}, we present the luminosity functions and \ks--band bolometric corrections for our sample using the fit-based luminosity estimates. Fig. \ref{fig:LF} compares the luminosity functions derived from the fits for both chemical types with the corresponding LMC distributions from \citetalias{R2012}. Overall, there is good agreement in the range of luminosities as well as the shapes of the luminosity functions. The O--rich distributions are very similar; the SMC O--AGB LF peaks at $-4.2$ mag, about 0.1 mag fainter than for the LMC. The difference in the peaks of the carbon-star LFs is significant -- the SMC distribution peaks at $M_{\rm bol}^{\rm peak}$$\approx -$4.5 mag, about half a magnitude fainter than its LMC counterpart. This is probably due to the SMC's lower mean metallicity, which lowers the mass limit for stars to turn C--rich.

The most luminous carbon stars in our sample are SSTISAGEMA~J010245.16--741257.4 and IRAS~00350--7436, with luminosities around 25\,000 L$_\odot$. The fit quality for the latter source is very poor. Previous studies estimate a luminosity of about 40\,000 L$_\odot$ for IRAS~00350. A more detailed analysis of this source is out of the scope of the current work, but will be the focus of an upcoming paper. We only note here that we do not find any carbon stars above the classical AGB luminosity limit. There are \NOrichaboveLAGB\ sources chemically classified as O--rich above this limit, \NRSGaboveLAGB\ of which are classified as RSGs based on their colours. None of these sources have matching OGLE information. Three of the remaining four -- designated as RSGs by \citet{Ruffleetal2015} -- are OGLE long-period variables, with periods $>$ 500 d, and are thus likely to be massive O--rich AGB stars. Only one of these sources, an extreme AGB, is significantly more luminous than the classical limit; it is the well-known long-period Mira \citep[1749~d,][]{Groenewegenetal2009} MSX SMC 55. Our luminosity estimate of $124\, 000$ L$_\odot$ ($M_{\rm bol}=-8.0$ mag) for this star is identical to the value derived by \citet{Groenewegenetal2009}, who noted it to be the only possible super-AGB candidate in their SMC sample. The RSG classification in, e.g., \citet{Ruffleetal2015} is primarily based on the classical AGB luminosity limit as defined by \citet{VassiliadisWood1993}. Their definition in turn was based on the identification of a small sample of RSG candidates, quite a few of which have since shown abundance anomalies pointing to an AGB nature. Based on our luminosities for long-period variables, there could be a larger number of sources above the classical limit undergoing hot-bottom burning than previously expected.\\

\subsubsection{Bolometric correction for the $K_{\rm s}$ band}
\begin{figure}
 \includegraphics[width=84mm,height=84mm]{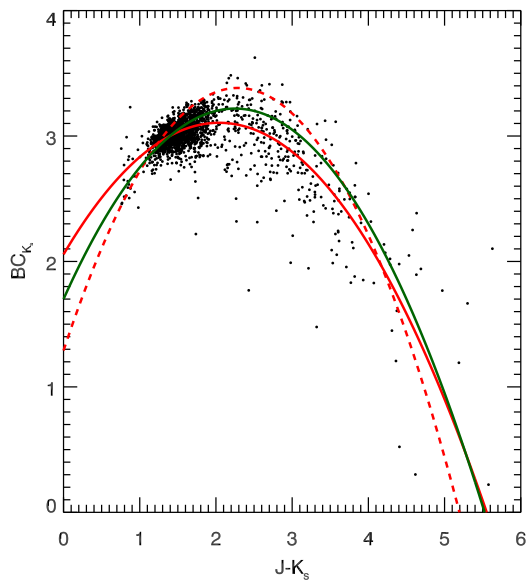}
 \caption{The $K_{\rm s}$ band bolometric correction as a function of the $J-K_{\rm s}$ colour for all sources in our list identified by our fits as carbon-rich. The quadratic fit to these data (solid red) is compared to those of \citetalias{R2012} (dashed red) and \citet{Kerschbaumetal2010} (solid green). \label{fig:bolcorr_Ks}}
\end{figure}

Fig. \ref{fig:bolcorr_Ks} shows the bolometric correction (BC) in the $K_{\rm s}$ band as a function of the $J-K_{\rm s}$ colour. We compute a quadratic fit of the form
\begin{equation}
BC_{K_{\rm s}}=a_0+a_1(J-K_{\rm s})+a_2(J-K_{\rm s})^2
\end{equation}
and obtain $a_0=$\bolcorra, $a_1=$\bolcorrb, and $a_2=$\bolcorrc. Fig. \ref{fig:bolcorr_Ks} compares our fit with two other quadratic fits. The \citet*{Kerschbaumetal2010} relation, derived for nearby field stars, shows good agreement with our result. The \citetalias{R2012} fit is also consistent with ours within the range of variation of the data, with differences arising mainly due to the smaller number of SMC sources at redder colours. The discrepancy in BC is at most about 0.5 mag, at intermediate $J-K_{\rm s}$ colours. At these colours, the \citetalias{R2012} relation prediction for a given $J-K_{\rm s}$ colour and $K_{\rm s}$ magnitude under-predicts the luminosity by about 60\%. The \citet{Kerschbaumetal2010} relation differs, at worst, by about 20\% in this range.\\

\subsection{Dust-production rates}
Fig. \ref{fig:vetted_himlr} shows the fits to the nine highest-DPR sources (excluding the FIR objects). The source with the highest DPR is SSTISAGEMA J005007.19--733125.1 (IRAS 00483--7347), a red supergiant \citep{Ruffleetal2015}. We estimate its DPR to be $(1.4\pm 0.6)\times 10^{-7}$ \msunperyr, 5--6 times higher than the second-highest DPR. We are also able to reproduce the silicate emission features in its spectrum (black curve) reasonably well. Only one more of the top nine dust producers is classified O--rich. All the sources in the top nine classified as carbon-rich have DPRs over $10^{-8}$ \msunperyr. They are all very red, and very faint in the near-infrared. Except for IRAS~00350--7436, they all have similar SED shapes. Two of these carbon stars have IRS spectra that show weak SiC emission along with strong 13.7 \mic\ C$_2$H$_2$ absorption and perhaps a weak 30 \mic\ feature.
\\

Including the FIR objects, there are \NhiDPR\ sources with DPRs $>$$10^{-9}$ \msunperyr, \NhiDPRO\ of which are classified by GRAMS as O--rich. The total DPR from O--rich and C--rich sources in this high-DPR subsample is \totalDPRhiDPRO\ and \totalDPRhiDPRC\ \msunperyr\ respectively, with mean DPRs of \meanDPRhiDPRO\ and \meanDPRhiDPRC\ \msunperyr\ respectively. The high-DPR subsample contains \NhiDPRxAGB\ x--AGB stars, only one of which is classified by GRAMS as O--rich.\\

\begin{table*}
\scriptsize
\centering
\begin{minipage}{180mm}
\caption{Comparison of total dust-production rates by colour class between \citet[][LMC]{R2012}, \citet[][SMC]{B2012}, and this paper.
\label{tab:compareDPRs_B2012_R2012}}
\begin{tabular}{@{}llllllllll@{}}
\hline
Colour class & \multicolumn{3}{c}{\citetalias{R2012} (LMC)} & \multicolumn{3}{c}{\citetalias{B2012} (SMC)} & \multicolumn{3}{c}{This work (SMC)}\\
& Number & Total DPR\footnote{The LMC DPRs have been scaled using Equation \ref{eqn:vexpscaling} with a gas:dust ratio of 500 for O--rich stars and 200 for C--rich stars.} & Mean DPR & Number & Total DPR & Mean DPR & Number & Total DPR & Mean DPR\\
\cline{3-4}\cline{6-7}\cline{9-10}\\
& & \multicolumn{2}{c}{(\msunperyr)}& & \multicolumn{2}{c}{(\msunperyr)}\\ 
\hline
x--AGB & 1347 & $1.2\times 10^{-5}$ & $8.6\times 10^{-9}$ & 313 & 6.3$\times 10^{-7}$ & 2.0$\times 10^{-9}$ & 337 & $6.8\times 10^{-7}$ & $2.0\times 10^{-9}$\\
C--AGB & 6662 & $7.6\times 10^{-7}$ & $1.1\times 10^{-10}$ & 1\,559 & 1.2$\times 10^{-7}$ & 7.8$\times 10^{-11}$ & 1652 & $1.2\times 10^{-7}$ & $7.1\times 10^{-11}$\\
O--AGB & 12031 & $9.7\times 10^{-7}$ & $8.1\times 10^{-11}$ & 1\,851 & 5.2$\times 10^{-8}$ & 2.8$\times 10^{-11}$ & 884 & $3.3\times 10^{-8}$ & $3.7\times 10^{-11}$\\
aO--AGB &  -- & -- & -- & 1\,243 & 2.6$\times 10^{-8}$ & 2.1$\times 10^{-11}$ & 643 & $1.6\times 10^{-8}$ & $2.4\times 10^{-11}$\\
RSG & 3589 & $1.4\times 10^{-6}$ & $4.0\times 10^{-10}$ & 2\,611 & 3.1$\times 10^{-8}$ & 1.2$\times 10^{-11}$ & 1410 & $4.6\times 10^{-8}$ & $3.3\times 10^{-11}$\\
FIR\footnote{The FIR contribution from this paper includes only sources from Groups 1--4.} &  -- & -- & -- & 50 & 9.6$\times 10^{-8}$ & 1.9$\times 10^{-9}$ & 17 & $4.5\times 10^{-7}$ & $2.6\times 10^{-8}$\\
Total & 23629 & $1.5\times 10^{-5}$ & $6.2\times 10^{-10}$ & 7\,627 & 9.5$\times 10^{-7}$ & 1.2$\times 10^{-10}$ & 4943 & $1.3\times 10^{-6}$ & $2.7\times 10^{-10}$\\
Total, no FIR & 23629 & $1.5\times 10^{-5}$ & $6.2\times 10^{-10}$ & 7\,577 & 8.6$\times 10^{-7}$ & 1.1$\times 10^{-10}$ &  4926 & $8.9\times 10^{-7}$ & $1.8\times 10^{-10}$\\
\end{tabular}
\end{minipage}
\end{table*}

Table \ref{tab:compareDPRs_B2012_R2012} shows the cumulative DPRs for the different colour classes, and compares the results from the current paper to those of \citetalias{B2012} for the SMC, which were derived using excess--DPR relations, and of \citetalias{R2012} for the LMC, who used a procedure similar to ours. For direct comparison, we have scaled the \citetalias{R2012} DPRs using Equation 1, with gas:dust ratios of 500 and 200 for O--rich and C--rich stars respectively. It is interesting to note that the mean DPR in every colour class is consistently lower for the SMC than for the LMC. The lower average dust-production rate in the SMC could be a result of the lower mean metallicity of the galaxy. The deficit is most pronounced in the case of the RSGs; this can be due to the lack of luminous dusty RSGs, as a large fraction of RSG dust is contributed by the most luminous stars. Red supergiants contribute only about 2.5\% of the total dust input to the SMC (including the FIR objects). In this paper, we increased the number of RSG candidates by adding $\approx$80 sources occupying the `gap' between the RSG and O--AGB colour classes of \citetalias{B2011} (see Section \ref{subsec:candidateselection}). While the actual nature (RSG or O--AGB) of these sources is unclear, they do not contribute significantly to the dust budget -- we find that this new population accounts for less than 4\% of the dust input from RSG stars.\\

The value of the total rate is very sensitive to the DPRs determined for the FIR sources. Most of the FIR sources in our original list were rejected based on CMDs or spectroscopic data which pointed to a non-AGB nature. The best-fit DPRs for these objects are, on average, higher than for the rest of the sample. We only take sources from Groups 1--4 into account when computing the dust budget. When the FIR sources are included in the estimate, we find a cumulative DPR of \globaldustbudget\ \msunperyr.\\

Our fit results for the filtered dataset are available as an online table. This table includes the best-fit parameters (luminosity, dust-production rate, optical depth, inner radius of the dust shell, stellar effective temperature) as well as their uncertainties. The columns in this table are described in Table \ref{tab:metatable_vettedphot_fitresults}. The GRAMS grid offers only coarse sampling in some parameters such as the effective temperature of the central star. In cases where all 100 best-fit models correspond to the same value of such a parameter, we cannot compute a parameter uncertainty.

\begin{table*}
  \caption{Names and descriptions for the columns in the online table of photometry and fit results.}
\label{tab:metatable_vettedphot_fitresults}
\begin{tabular}{lll}
  \hline
  Column(s) & Name & Description \\
  \hline
  1 & IRAC\_{}DESIGNATION & The SAGE-SMC IRAC Mosaic (if unavailable, Epoch 1)\\
  & & Archive identifier; should be preceded by `SSTISAGE'.\\
  2 & COLOUR\_{}CLASS & Colour classification using the scheme described in \\
  & & \citet{B2011} and this paper.\\
  3 & OGLE3\_{}ID & Identifier in the OGLE-III Catalog of Variable Stars;\\
  & & should be preceded by `OGLE-SMC-LPV-'.\\
  4 & AKARI\_{}ID & Identifier in the AKARI All-Sky Point-Source Catalog.\\
  5 & WISE\_{}ID & Identifier in the WISE All-Sky Point-Source Catalog.\\
  6--13 & FLUXU, DFLUXU, FLUXB, DFLUXB, FLUXV, DFLUXV,& Matching photometry from the MCPS survey in the \\
  &  FLUXI, DFLUXI & $U$, $B$, $V$, and $I$ bands, and the associated errors. Where\\
  & &  possible, the $V$- and $I$-band fluxes are replaced with the\\
  & &  MACHO/OGLE-III mean measurements. Accordingly,\\
  & &  the errors are inflated using the MACHO/OGLE-III\\
  & &  $V$-/$I$-band amplitude. See Section \ref{subsec:variability} for details.\\
  14--19 & FLUXJ, DFLUXJ, FLUXH, DFLUXH, FLUXK\_{}S, DFLUXK\_{}S & Matching near-IR photometry and the related uncertainties.\\
  & & These data consist of matching 2MASS and/or IRSF\\
  & & information. If both 2MASS and IRSF fluxes are available,\\
  & &  these fields contain the mean fluxes and variability-adjusted\\
  & &  uncertainties in each band. See Section \ref{subsec:variability} for details.\\
  20--27 & FLUX3\_{}6, DFLUX3\_{}6, FLUX4\_{}5, DFLUX4\_{}5, FLUX5\_{}8, & Mean IRAC fluxes and associated uncertainties computed from\\
  & DFLUX5\_{}8, FLUX8\_{}0, DFLUX8\_{}0 &  two epochs of SAGE-SMC and one epoch of matching S$^3$MC\\
  & &  data, if available. See Section \ref{subsec:variability} for details.\\
  28--29 & FLUX24, DFLUX24 & Mean MIPS24 flux and associated uncertainty computed from \\
  & & two epochs of SAGE-SMC and one epoch of matching S$^3$MC\\
  & &  data, if available. The uncertainty is also inflated to reduce the\\
  & &  weight on the 24 \mic\ band when computing the fit. See\\
  & & Sections \ref{subsec:variability} and \ref{sec:fitting} for details.\\
  30--33 & FLUXN11, DFLUXN11, FLUX{\it L15}, DFLUX{\it L15} & Matching photometry from the AKARI survey for the $N11$ \\
  & & and ${\it L15}$ bands. The uncertainties have been inflated to account\\
  & & for variability according to Section \ref{subsec:variability}.\\
  34--35 & FLUX{\it W3}, DFLUX{\it W3} & Matching photometry from the WISE survey for the ${\it W3}$ band. \\
  & & The uncertainty has been inflated to account\\
  & &  for variability according to Section \ref{subsec:variability}.\\
  36--43 & MAGU, DMAGU, MAGB, DMAGB, MAGV, DMAGV, & Magnitudes and uncertainties computed from the fluxes and \\
  & MAGI, DMAGI& uncertainties in fields 6--13.\\
  44--49 & MAGJ, DMAGJ, MAGH, DMAGH, MAGK\_{}S, DMAGK\_{}S & Magnitudes and uncertainties computed from the fluxes and \\
  & & uncertainties in fields 14--19.\\
  50--57 & MAG3\_{}6, DMAG3\_{}6, MAG4\_{}5, DMAG4\_{}5, MAG5\_{}8,& Magnitudes and uncertainties computed from the fluxes and \\
  &  DMAG5\_{}8, MAG8\_{}0, DMAG8\_{}0  & uncertainties in fields 20--27.\\
  58--59 & MAG24, DMAG24 & Magnitude and uncertainty computed from fields 28 and 29.\\
  60--63 & MAGN11, DMAGN11, MAG{\it L15}, DMAG{\it L15} & Magnitudes and uncertainties computed from the fluxes and \\
  & & uncertainties in fields 30--33.\\
  64--65 & MAG{\it W3}, DMAG{\it W3} & Magnitudes and uncertainties computed from the fluxes and \\
  & & uncertainties in fields 34--35.\\
  66 & FIR\_{}GROUP & Group number for FIR objects (see Section \ref{subsubsec:FIRobjects}).\\
  67 & SMC\_{}IRS & SMC IRS identification number from \citet{Ruffleetal2015}.\\
  68 & SAGE\_{}SPEC\_{}CLASS & Spectroscopic classification from \citet{Ruffleetal2015}.\\
  69 & GRAMS\_{}CLASS & Best-fit chemical classification.\\
  70 & CLASS\_{}CONFIDENCE & Confidence of classification (see Section \ref{subsubsec:chemclassconfidence}).\\
  71--72 & CHISQ\_{}BEST, CHISQ\_{}ALT & Lowest \chisq\ values for the best-fit and alternate chemical classes.\\
  73--76 & LUM, DLUM, DPR, DDPR & Best-fit luminosities and dust-production rates (DPRs), and\\
  & & associated uncertainties. Parameter uncertainties are computed\\
  & &  using all 100 acceptable models (see Section \ref{sec:fitting}).\\
  & & Note: the DPRs in this table have been scaled according to \\
  & & Equation \ref{eqn:vexpscaling}.\\
  77--82 & TAU, DTAU, RIN, DRIN, TIN, DTIN & Best-fit optical depth, dust shell inner radius, and temperature\\
  & &  at this radius in K. The optical depth is measured at \\
  & & 10 \mic\ (11.3 \mic) for O--rich (C--rich) dust, and the inner radius\\
  & & is in units of stellar radii.\\
  83--86 & TEFF, DTEFF, LOGG, DLOGG & Effective temperature and surface gravity ($\log{[{\rm g}/{\rm cm~s}^{-1}]}$) of \\
  & & best-fit photosphere model. See \citet{Sargentetal2011} and \\
  & & \citet{Srinivasanetal2011} for details.\\
  87 & FITCOMMENT & Notes that the optical data is ignored (Section \ref{sec:fitting}), or that the\\
  & &  best-fit chemistry is changed (Section \ref{subsubsec:highchisq}).\\
\hline
\end{tabular}
\end{table*}

\section{Discussion}
\label{sec:discussion}

\subsection{Fits for FIR objects}
\label{subsec:FIRfitsdiscussion}

\begin{figure*}
\includegraphics[width=50mm]{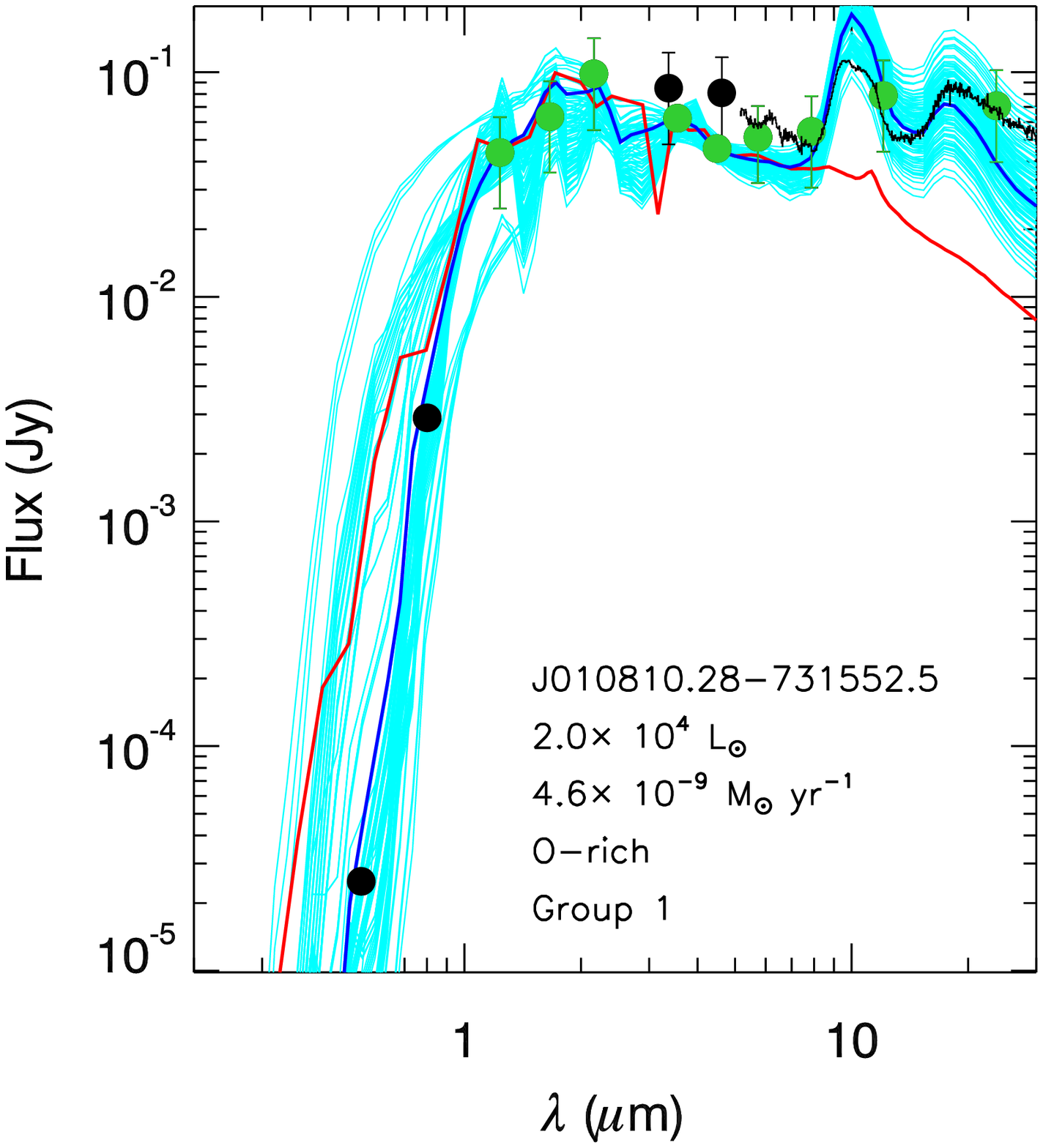} \includegraphics[width=50mm]{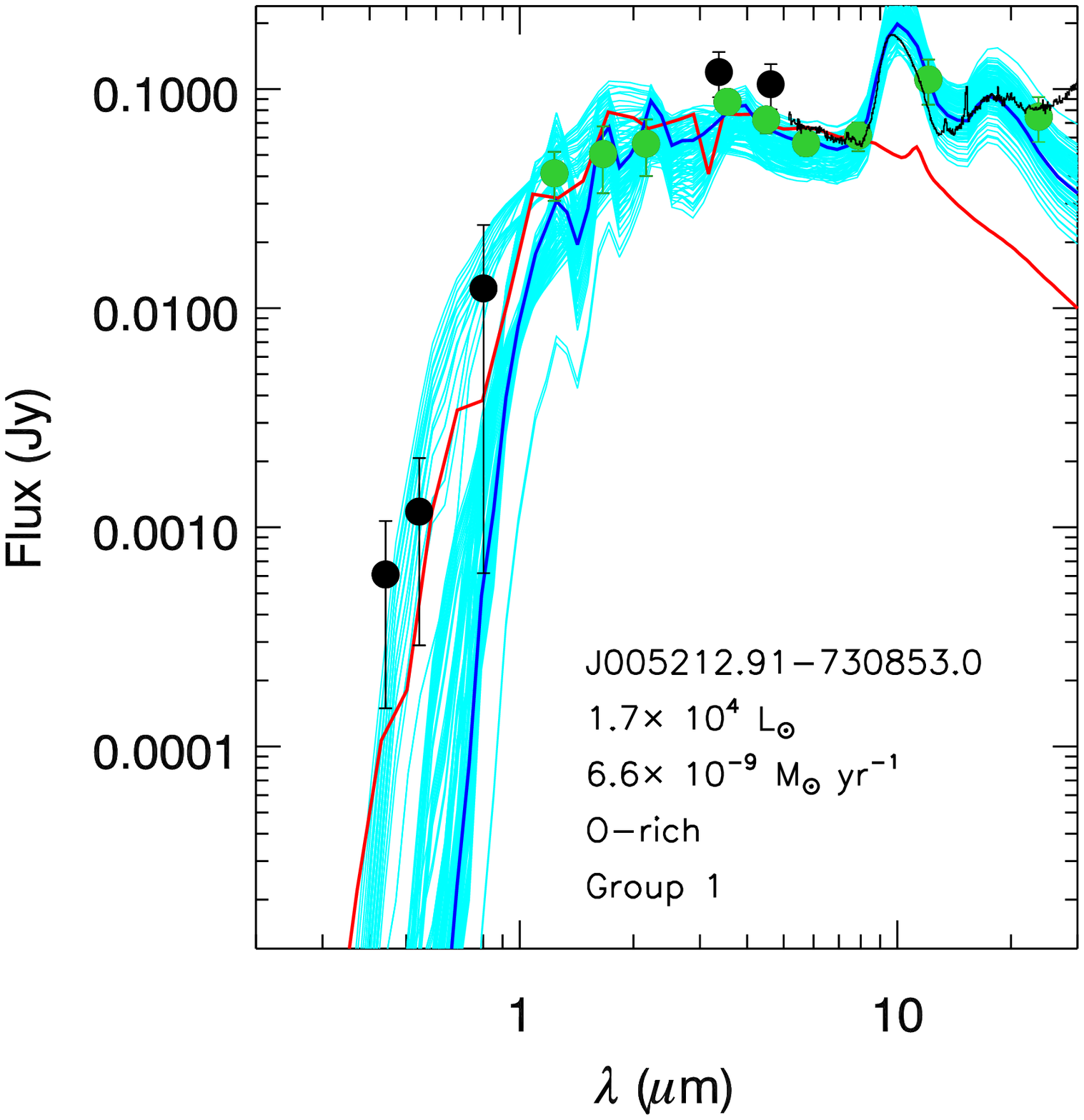} \includegraphics[width=50mm]{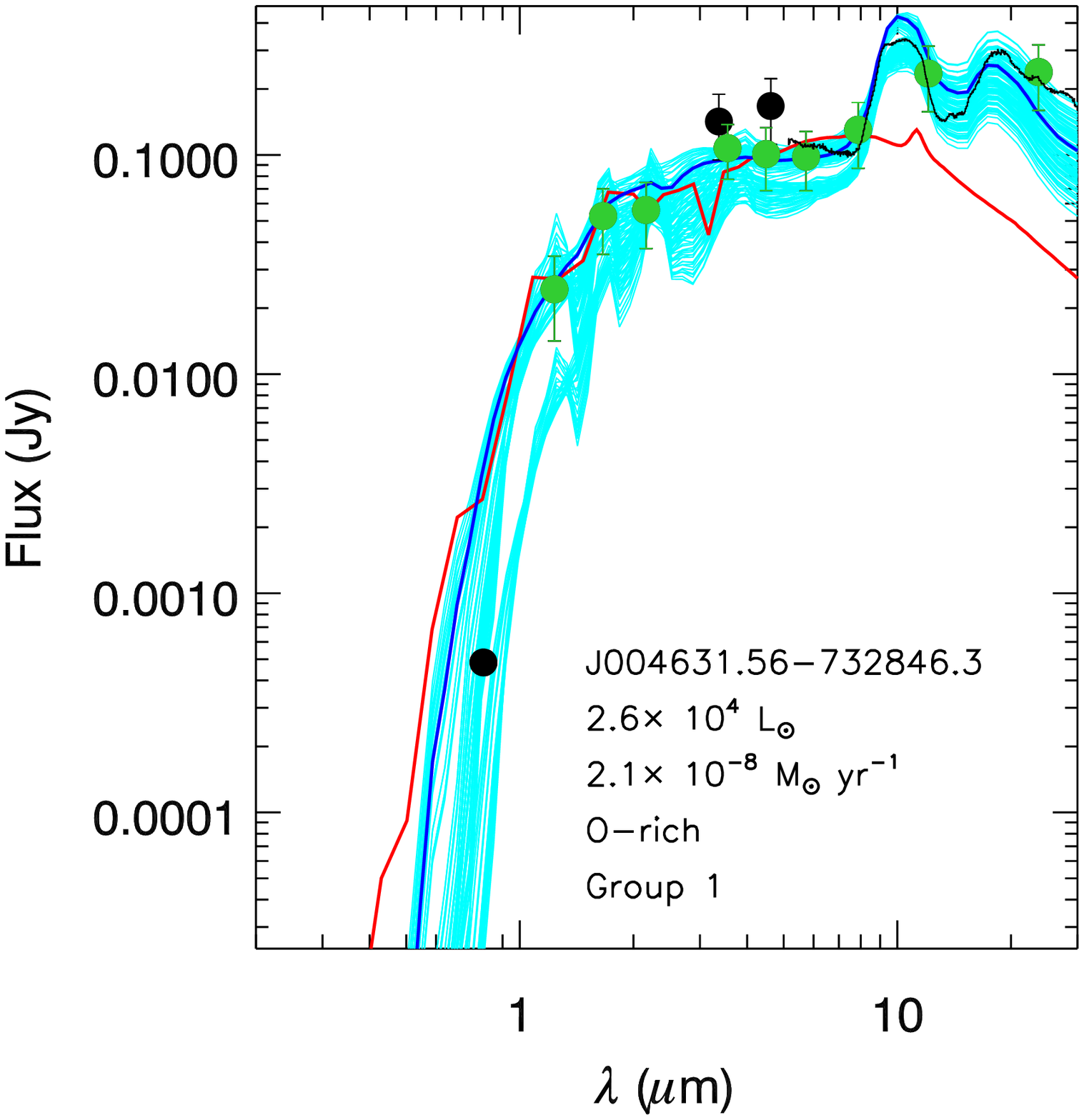}\\
\includegraphics[width=50mm]{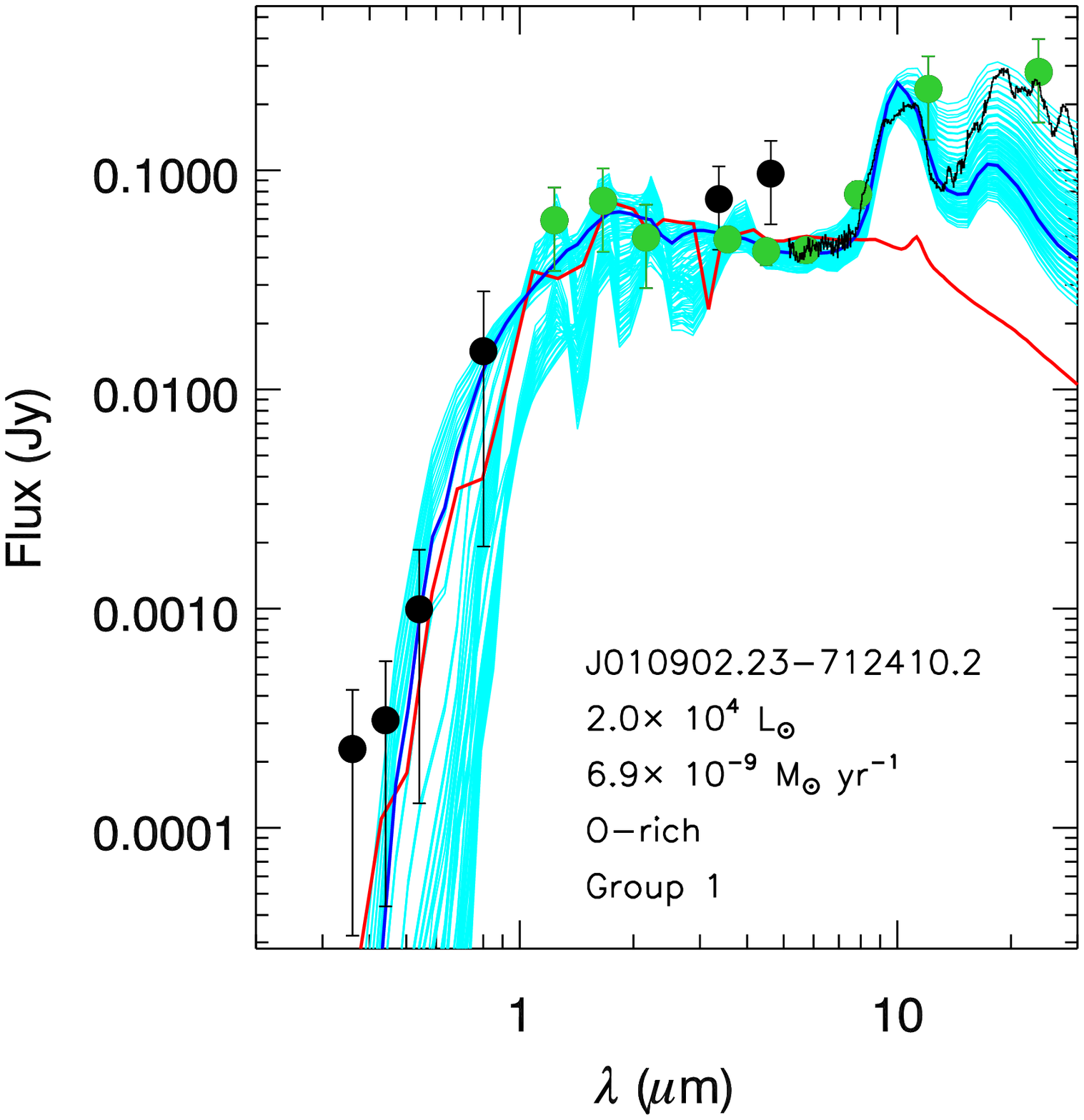}\includegraphics[width=50mm]{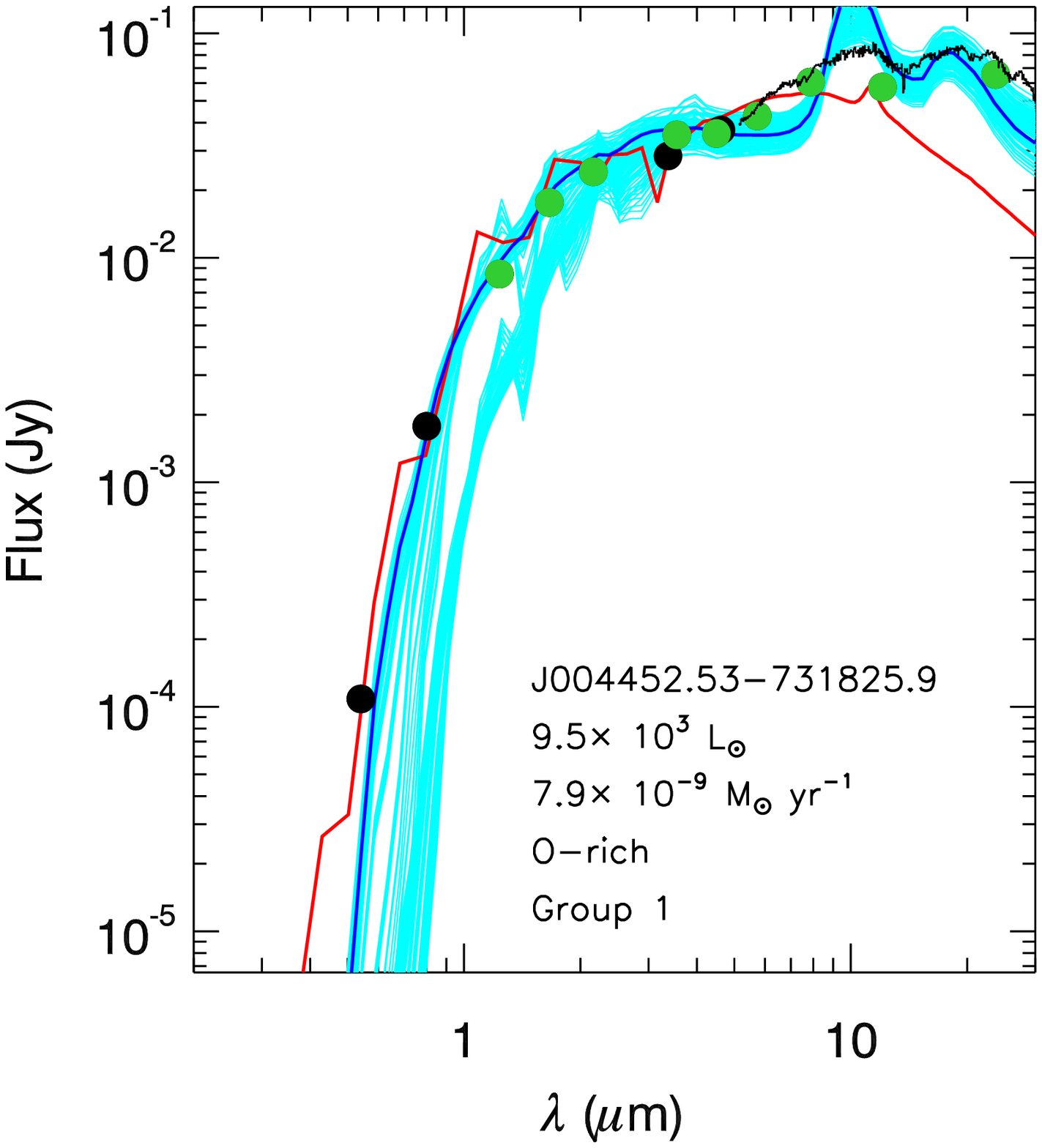}\includegraphics[width=50mm]{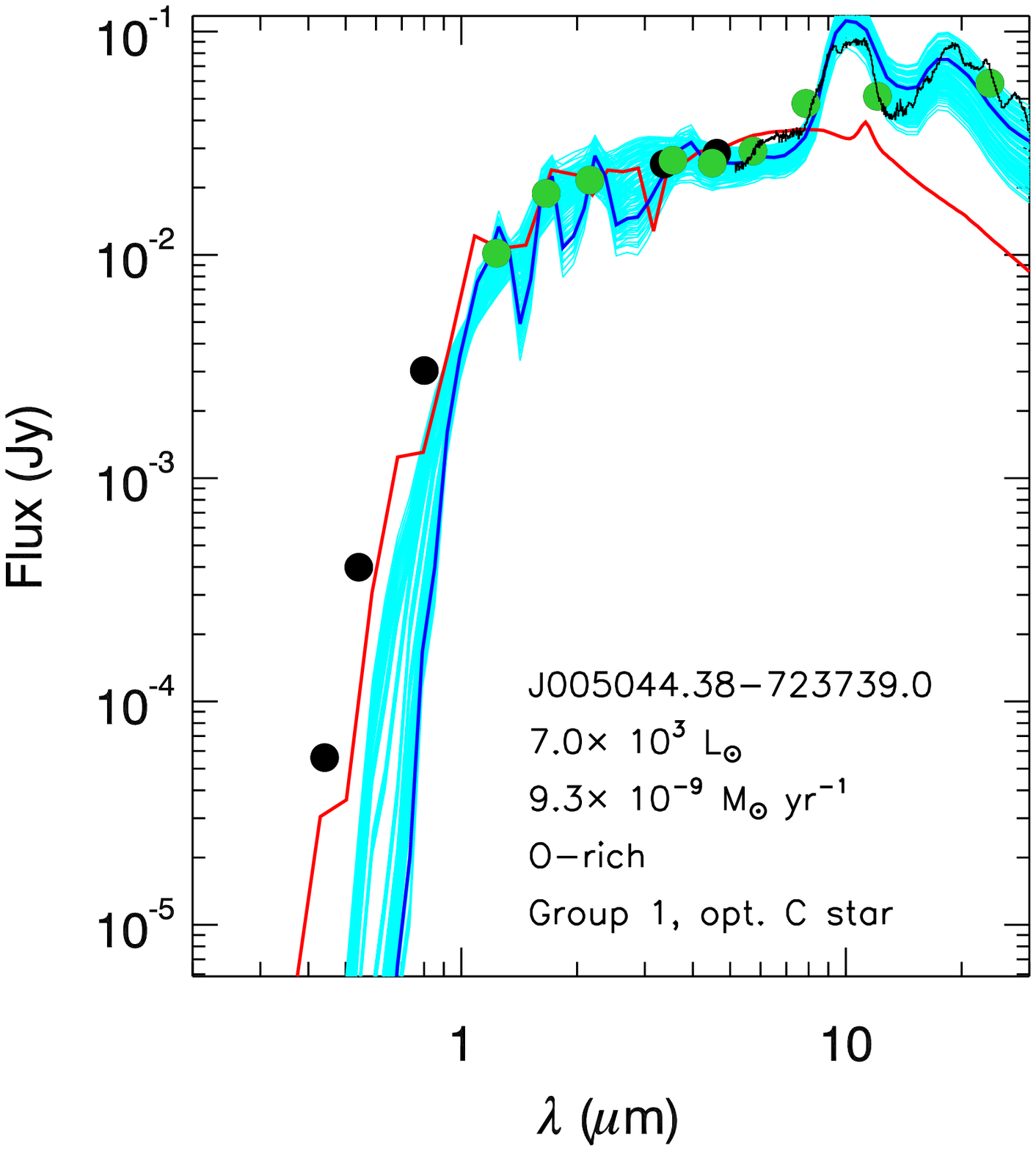}\\
\includegraphics[width=50mm]{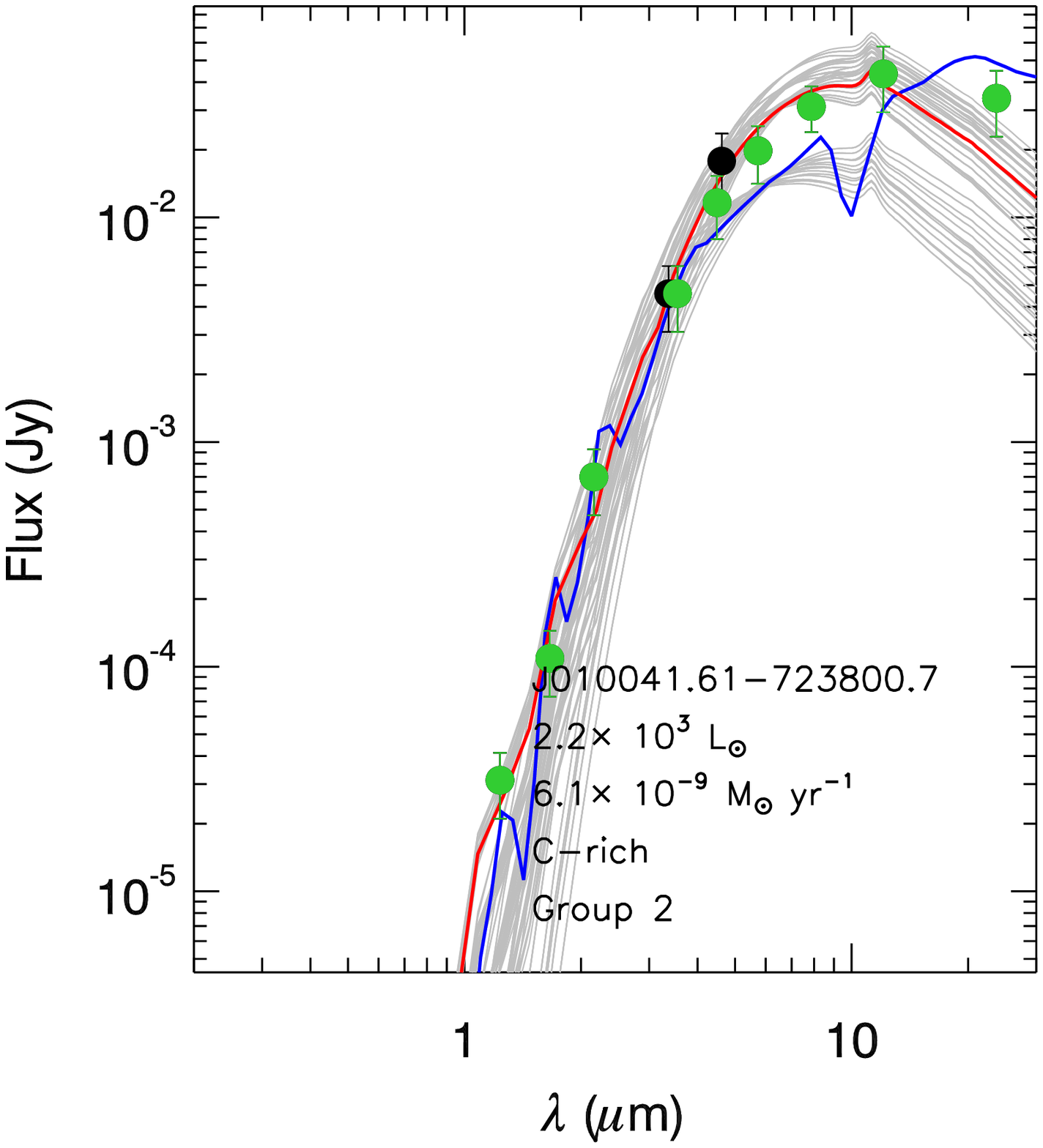}\includegraphics[width=50mm]{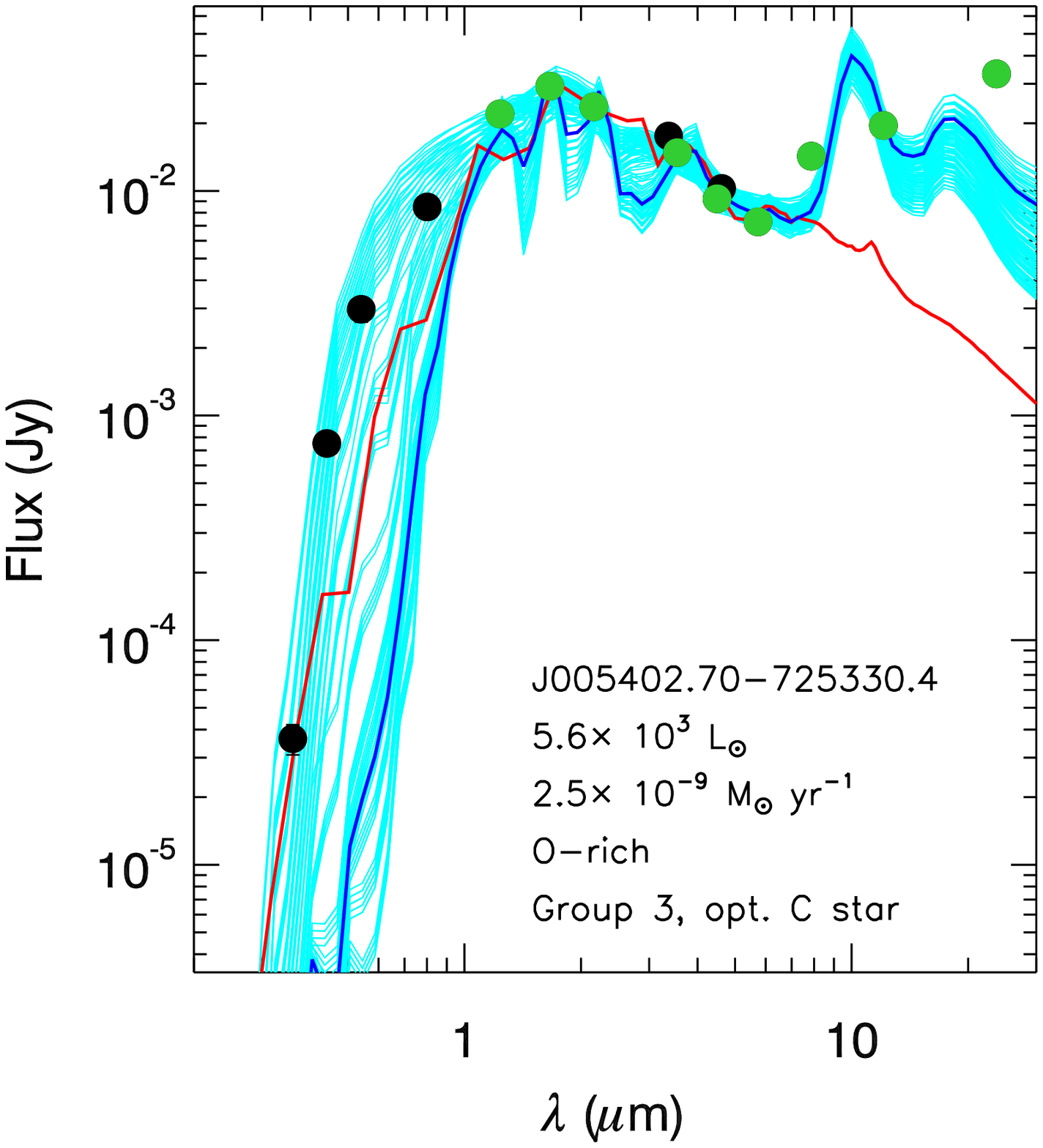}\includegraphics[width=50mm]{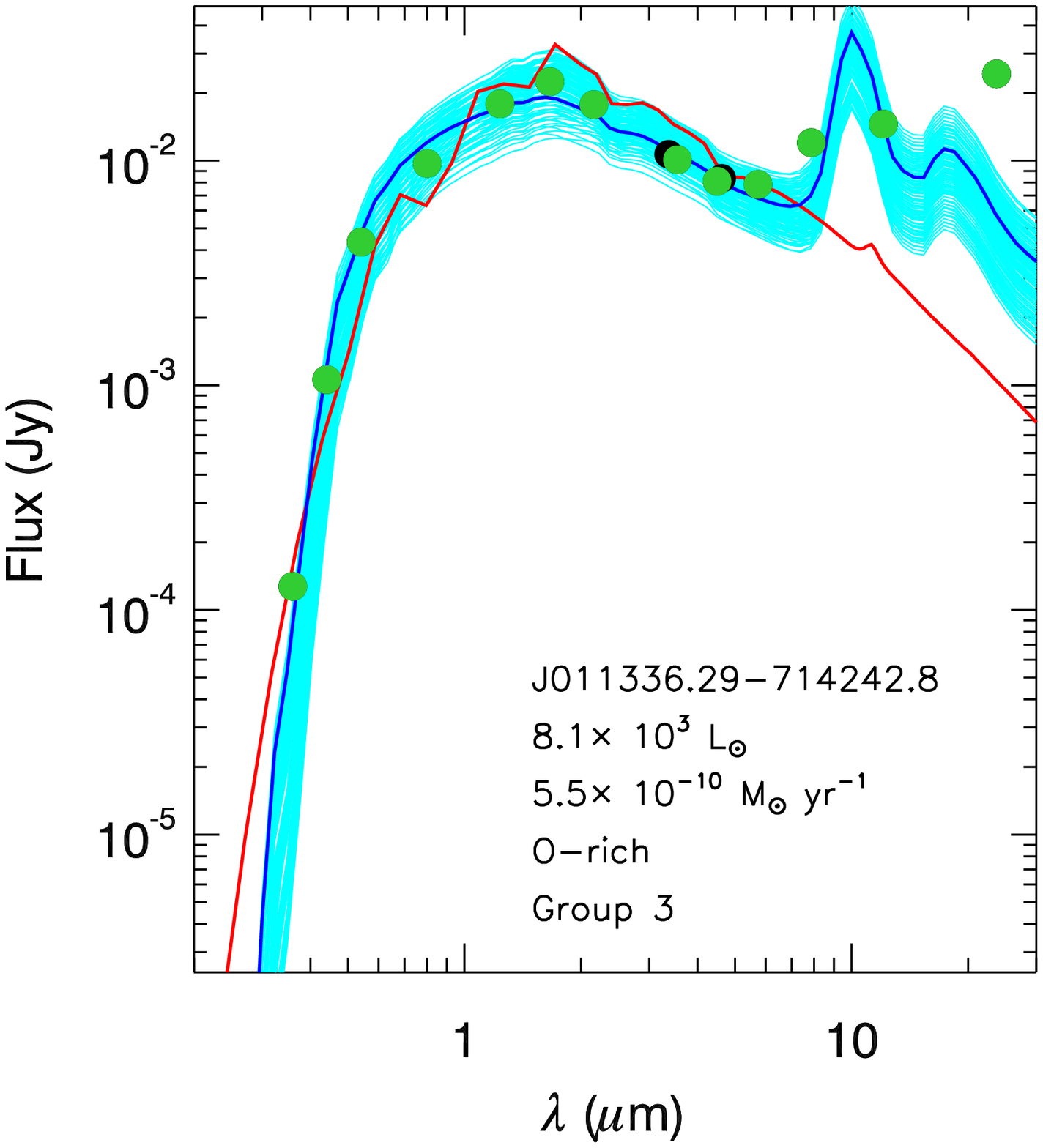}\\
\includegraphics[width=50mm]{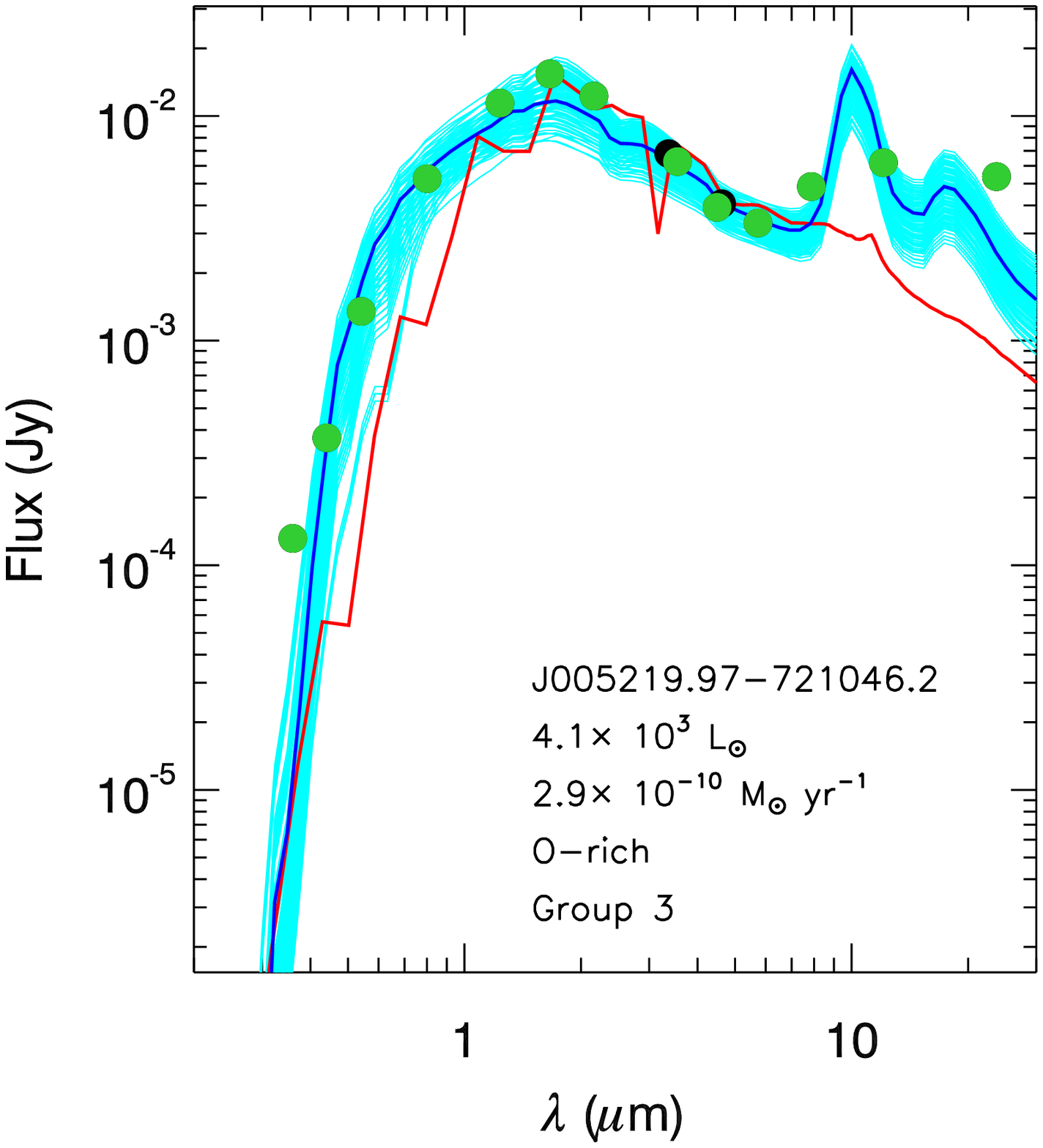}\includegraphics[width=50mm]{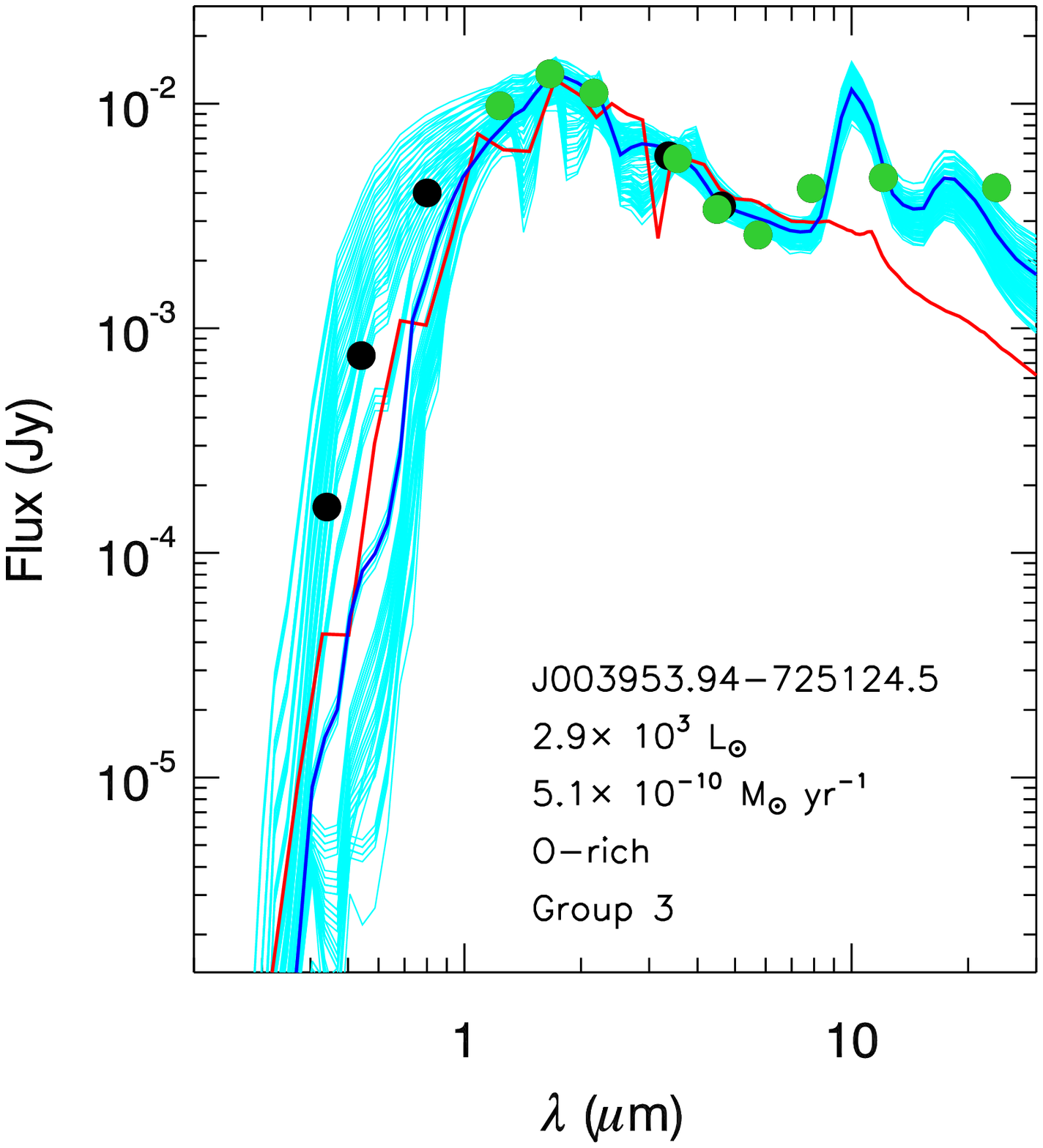}\includegraphics[width=50mm]{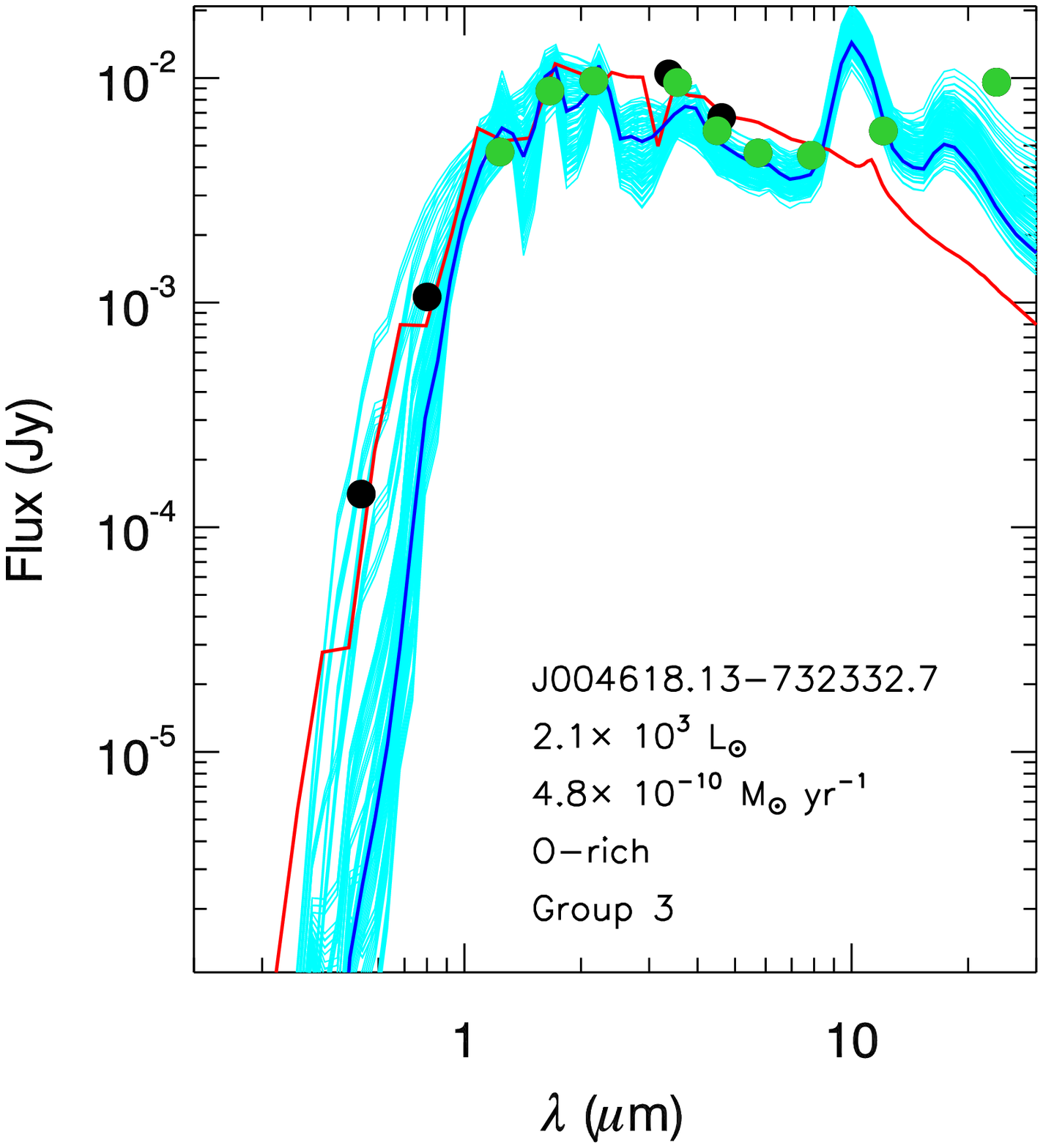}\\
\caption{FIR objects with valid fits (best-fit O--rich: blue, best-fit C--rich: red). The 100 acceptable models of the best-fit chemical type are also shown (cyan: O--rich, grey C--rich). Optical carbon stars are also indicated.\label{fig:validFIRfits}}
\end{figure*}

\renewcommand{\thefigure}{\arabic{figure} (Cont.)}
\addtocounter{figure}{-1}

\begin{figure*}
\includegraphics[width=50mm]{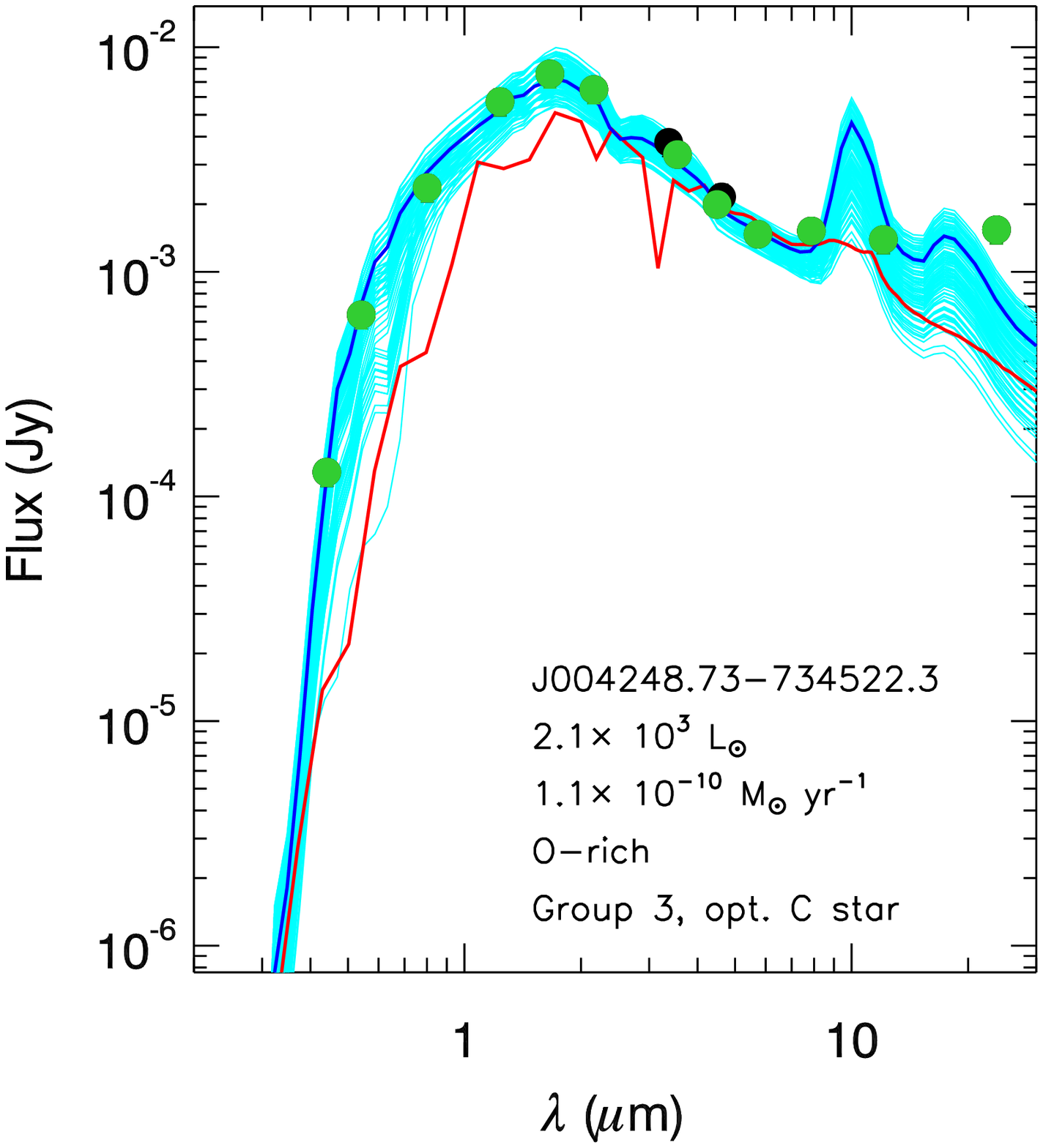}\includegraphics[width=50mm]{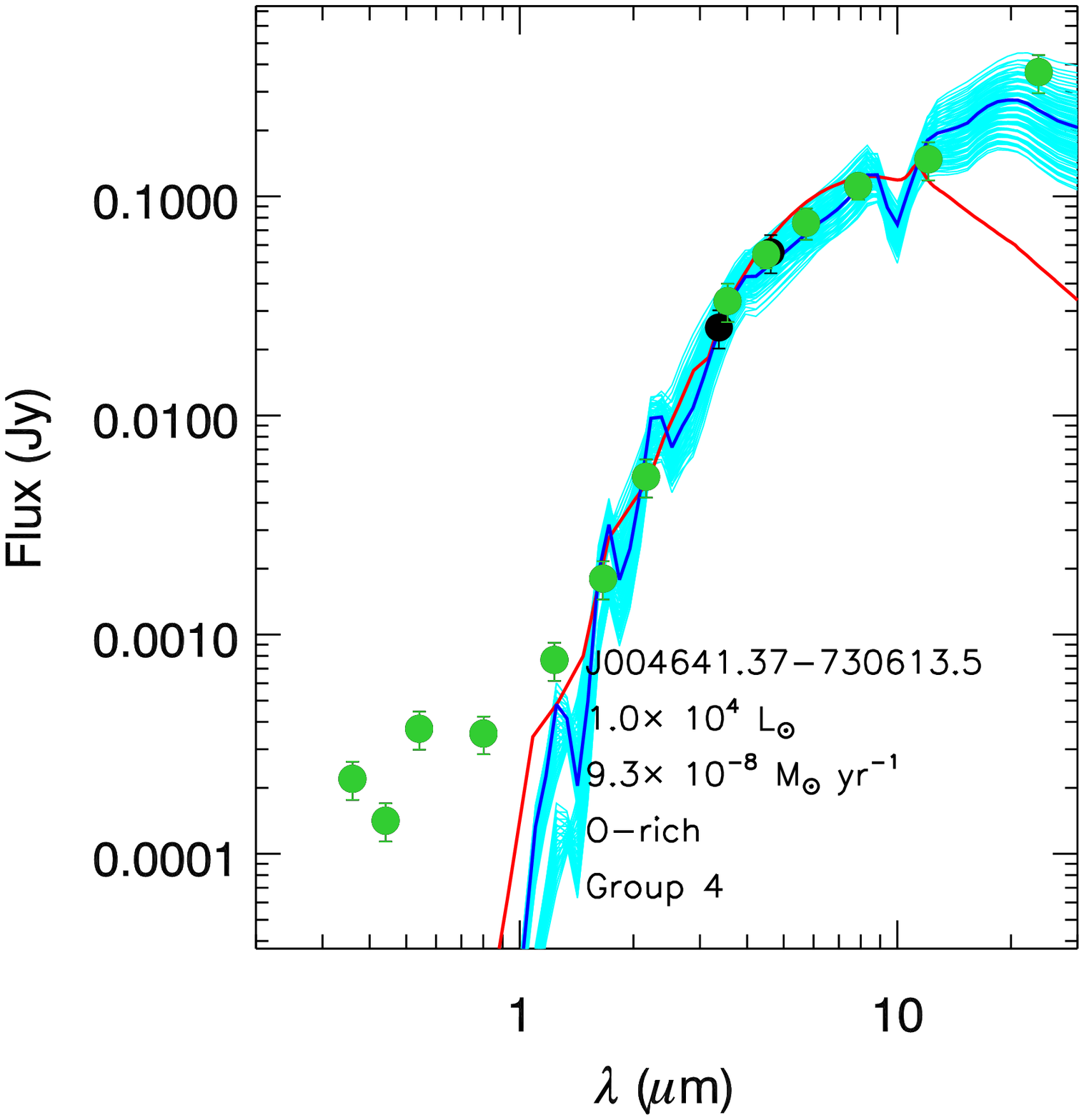}\includegraphics[width=50mm]{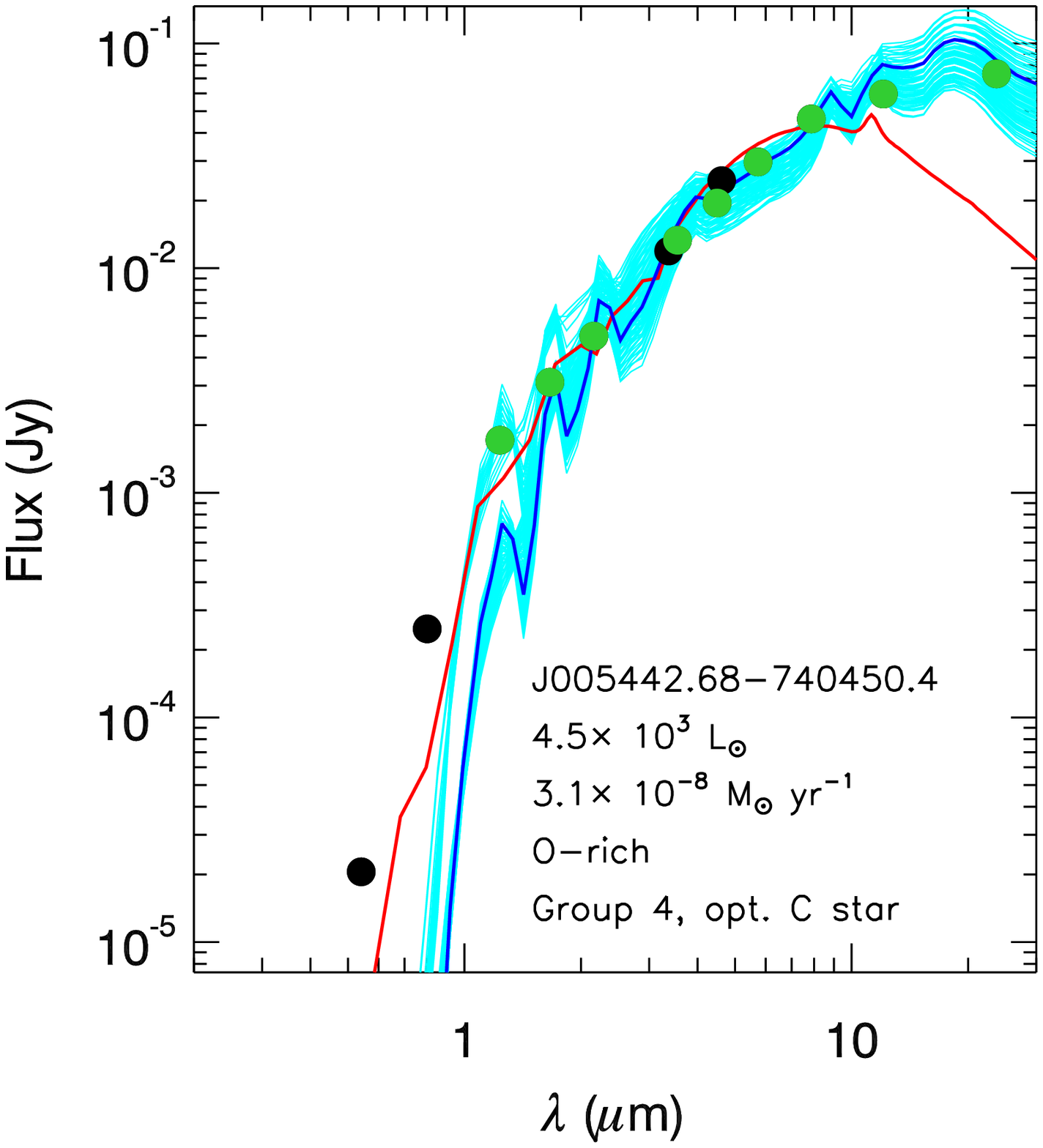}\\
\includegraphics[width=50mm]{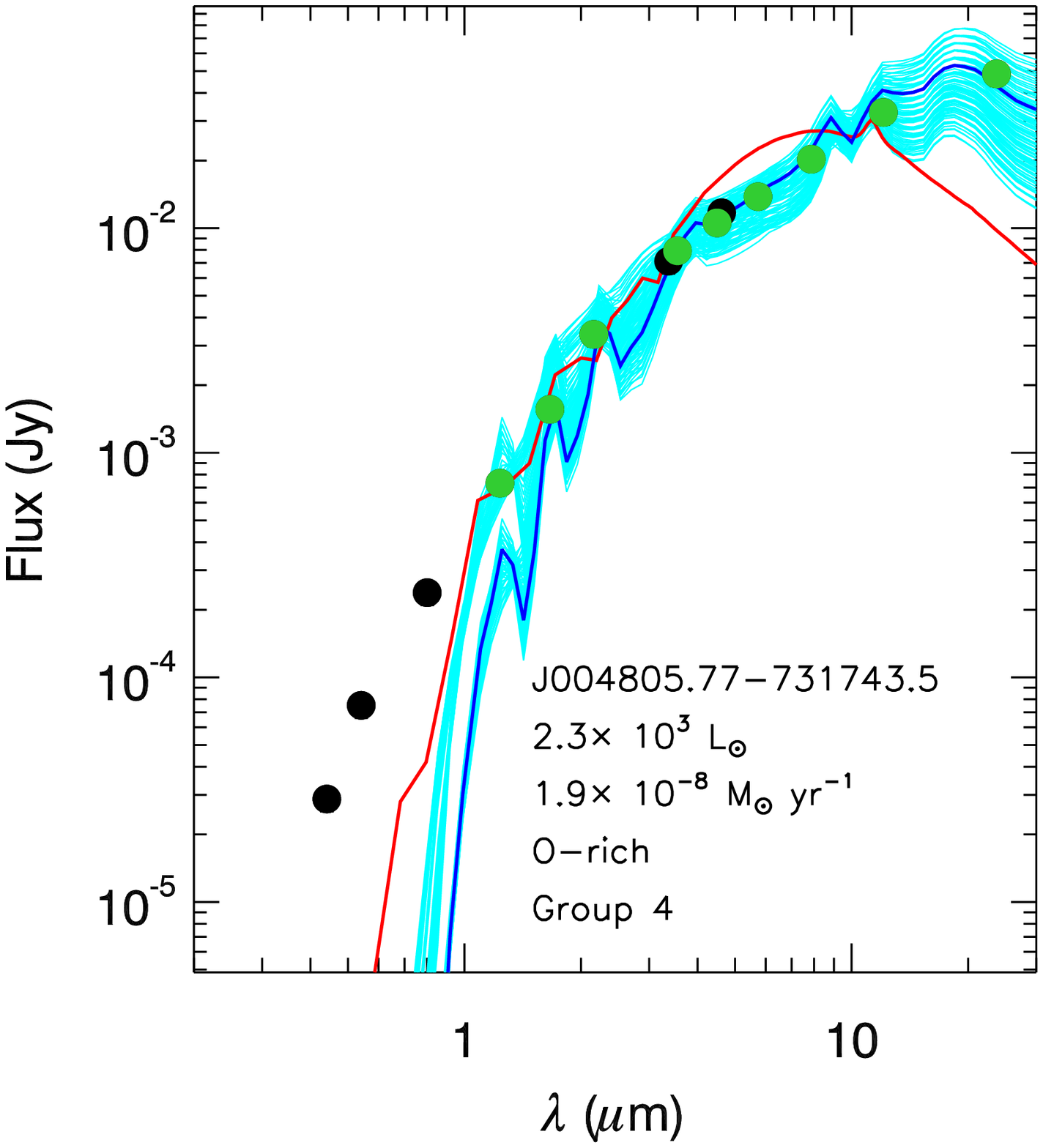}\includegraphics[width=50mm]{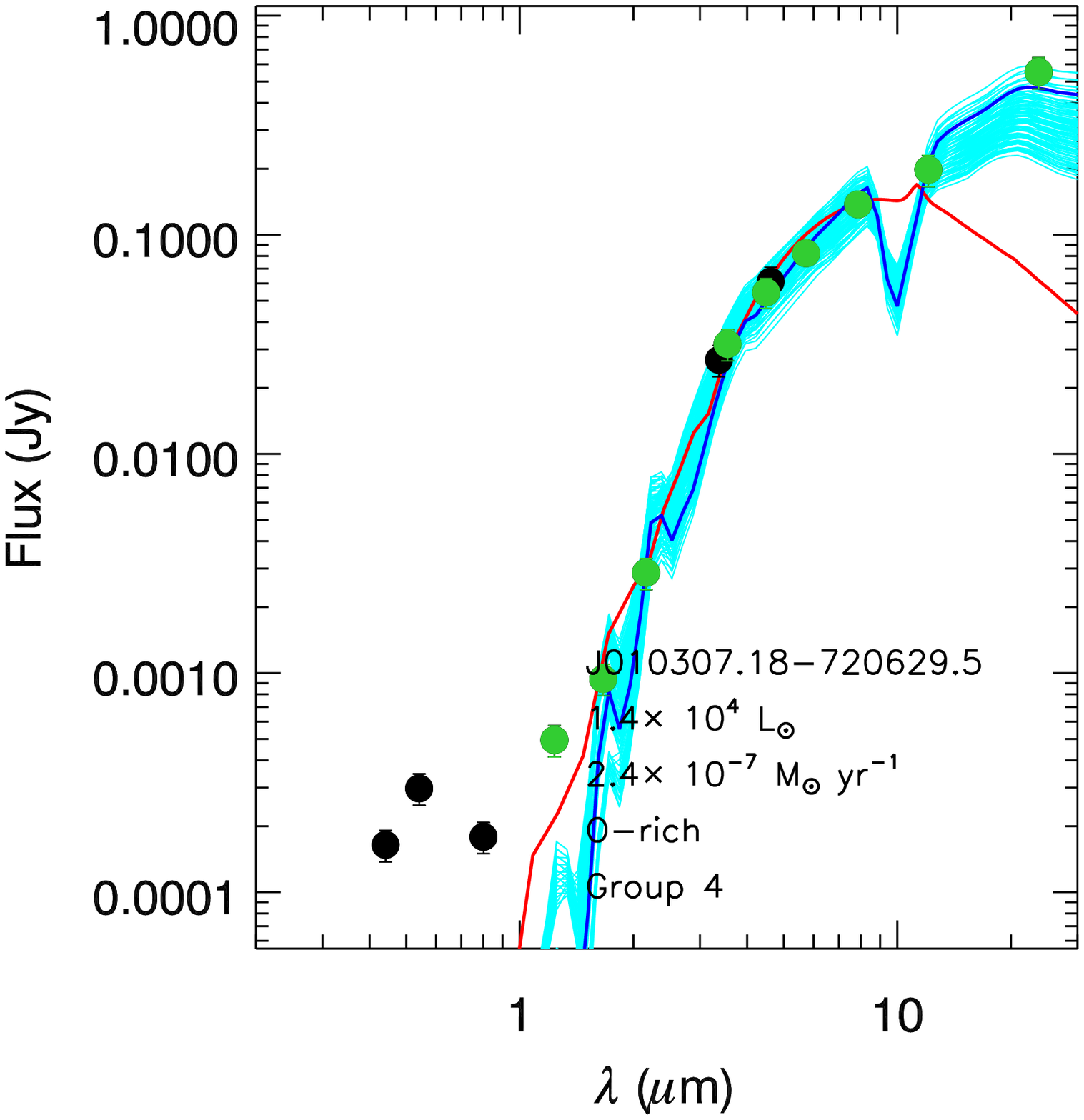}
\caption{}
\end{figure*}

\renewcommand{\thefigure}{\arabic{figure}}

Table \ref{tab:totaldpr_FIRcategory} displays the total dust-production rate for each FIR object category. The DPRs are considerably higher than our previous excess-based estimates \citepalias{B2012}, which is probably because of the cooler dust around FIR objects that may not contribute significantly to the 8 \mic\ excess. However, we did not find any genuine {\it Herschel} counterparts to these sources \citep{Jonesetal2015}, which indicates the lack of a large dust reservoir beyond 24 \mic.\\

Fig. \ref{fig:validFIRfits} shows the fits for objects in FIR Groups 1--4 (the rest of the fits are shown in Fig. \ref{fig:invalidFIRfits}). Our fitting procedure classifies almost all \NvetFIR\ FIR objects as O--rich, including the optical carbon stars RAW~631, RAW~171, and IRAS~F00530--7421. However, the GRAMS classifications for the six FIR objects with IRS spectra agree with the \citet{Ruffleetal2015} determinations. Such a `mixed chemistry' effect could also explain our classifications for the three optical carbon stars without IRS spectra; for instance, \citet{GroenewegenBlommaert1998} suggested that IRAS F00530--7421 is a post-AGB star or has a binary companion that is responsible for its peculiar optical spectrum and atypical infrared colours. It is interesting to note that -- in the case of RAW 631, for example -- the alternate best-fit (in this case, C--rich) model is able to reproduce the shape of the SED in the optical through IRAC bands, consistent with a C--rich photosphere. In all the cases shown in Fig. \ref{fig:validFIRfits}, the best-fit model can reproduce the near- and mid-IR SED. The exception is that the observed 24 \mic\ flux is sometimes higher than the model prediction. As discussed before, the 24 \mic\ emission may not necessarily be associated with a single source at the SMC distance (see notes in Table \ref{tab:FIRcategories}). If the 24 \mic\ flux is indeed associated with the IRAC source, the GRAMS best-fit DPR is an underestimate of its dust content. In the current work, we consider all the sources in Groups 1--4 as mass-losing evolved stars and include their contribution in the global dust budget.\\

SSTISAGEMA~J010041.61--723800.7 (Object 33 in Table \ref{tab:FIRcategories}) is the only FIR candidate with a valid best fit that is C--rich. This source is one of six newly-discovered mid-infrared variables \citep{Riebeletal2015}. While our fit procedure classifies IRAS~F00530 (Object 16 in Table \ref{tab:FIRcategories}) as O--rich, the carbon-star model is a good fit to the SED at wavelengths below 24 \mic. The same holds true for almost all the remaining Group 3 and 4 sources. If all Group 3 and 4 sources are O--rich, and the 24 \mic\ emission is associated with the source, the total DPR from Groups 3 and 4 is \GroupIIIandIVtotalDPRifallO\ \msunperyr. If, instead, they are all C--rich, the total DPR is much lower: \GroupIIIandIVtotalDPRifallC\ \msunperyr.\\

The total DPR from all valid FIR objects (i.e., Groups 1--4) is thus in the range \GroupItoIVtotalDPR\ \msunperyr. Fits to Groups 5--7 predict a total DPR for that sample of about \GroupVtoVIItotalDPR\ \msunperyr. By careful elimination of contaminants to the mass-losing evolved star sample, we have avoided an overestimate of the global dust budget! \emph{The uncertain chemical nature of a small number of FIR objects with valid fits is thus the largest source of uncertainty in the total dust budget}.\\

\begin{table}
\scriptsize
\centering
\begin{minipage}{180mm}
\caption{Total FIR dust-production rate by group.\label{tab:totaldpr_FIRcategory}}
\begin{tabular}{@{}llll@{}}
\hline
Group\footnote{Only the objects in Groups 1--4 are likely AGB/RSG stars.} & Number & Total DPR & Mean DPR\\
\cline{3-4}\\
      &        & \multicolumn{2}{c}{(\msunperyr)}\\
\hline
1 & 6 & $(5.6\pm 0.5) \times 10^{-8}$ & $(9.3\pm 0.8) \times 10^{-9}$\\
2 & 1 & $(6.1\pm 0.6) \times 10^{-9}$ & $(6.1\pm 0.6) \times 10^{-9}$\\
3 & 6 & $(4.4\pm 0.3) \times 10^{-9}$ & $(7.4\pm 0.4) \times 10^{-10}$\\
4 & 4 & $(3.8\pm 0.07) \times 10^{-7}$ & $(9.6\pm 0.2) \times 10^{-8}$\\
5 & 2 & $(3.1\pm 0.02) \times 10^{-8}$ & $(1.6\pm 0.02) \times 10^{-8}$\\
6 & 7 & $(1.9\pm 0.04) \times 10^{-7}$ & $(2.6\pm 0.06) \times 10^{-8}$\\
7 & 7 & $(1.3\pm 0.02) \times 10^{-6}$ & $(1.8\pm 0.04) \times 10^{-7}$\\
\end{tabular}
\end{minipage}
\end{table}

\subsection{Uncertainties in dust-production rates}
\label{subsec:dpruncertainties}
Before we compare our results to previous efforts and discuss the SMC dust budget, we discuss the various sources of uncertainty associated with the DPR estimates.\\

The effect of scaling our DPRs using Equation \ref{eqn:vexpscaling} is an overall reduction in DPR. None of the sources classified by our fitting technique as carbon-rich have luminosities above 30\,000 L$_\odot$, and the DPRs of the oxygen-rich sources brighter than this value are suppressed due to the dependence on the gas:dust ratio. The cumulative DPR with and without FIR objects in Table \ref{tab:compareDPRs_B2012_R2012} are $1.8\times 10^{-6}$ and $8.9\times 10^{-7}$ \msunperyr\ respectively. If we use the unscaled DPRs, these values are 2.3 and 1.6 times higher respectively. This factor of $\sim$2 difference due to the lack of knowledge of expansion speeds is already a significant contribution to the uncertainty in the absolute value of the global dust budget, and it is important to consider this issue when comparing our results with those from other studies.\\

\begin{figure}
\includegraphics[width=85mm]{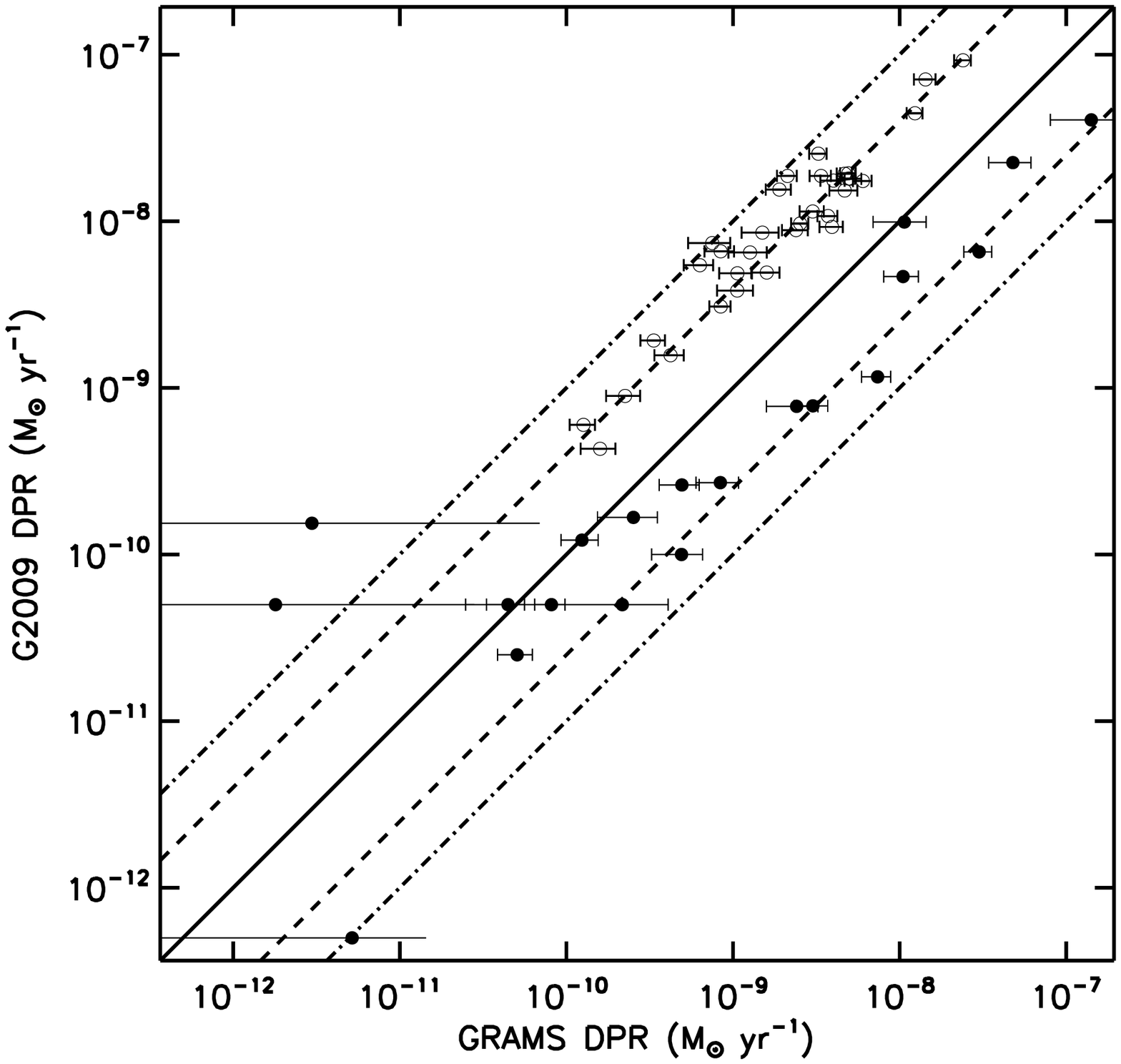}
 \caption{Comparison of DPR estimates for the \citet[][G2009]{Groenewegenetal2009} M stars (filled circles) and carbon stars (open circles). The solid line represents 1:1 agreement, while the dashed and dot-dashed lines correspond to a 4$\times$ and 10$\times$ discrepancy respectively. The GRAMS DPRs are consistently higher (lower) than the \citet{Groenewegenetal2009} values for the O--rich (C--rich) stars. \label{fig:G2009DPRvsGRAMSDPR}}
\end{figure}

Perhaps the most important source of systematic uncertainty in the DPR is the choice of dust optical constants when computing the radiative transfer models. In Fig. \ref{fig:G2009DPRvsGRAMSDPR}, we compare our unscaled DPRs with those of \citet{Groenewegenetal2009} for their M star and carbon star samples. Our C--rich DPRs are consistently about 4--10 times lower; this is consistent with the discussion in \citet{Srinivasanetal2011} for the LMC sample, and can be explained by our differing choice of optical constants for amorphous carbon. At DPRs above $10^{-10}$ \msunperyr, our O--rich DPRs are higher than the corresponding \citet{Groenewegenetal2009} values for all but two stars. The reason for the larger variation is unclear \citep[see Section 4.2.2 in][]{Sargentetal2011}, although it is likely due to the larger range of O--rich dust properties explored in that paper. This (4--10)$\times$ discrepancy only worsens when the scaling relation (Equation \ref{eqn:vexpscaling}) is incorporated. This discrepancy should be taken into account when comparing our results with other recent studies \citep[e.g.,][]{Matsuuraetal2013} that rely on the \citet{Groenewegenetal2009} DPRs, as our scaled DPRs can be {\em up to 20$\times$ higher (lower) for O--rich (C--rich) stars in comparison.}\\

We can place an upper bound on the total DPR by combining the two effects above. If all the stars have an expansion speed of 10 km s$^{-1}$, and if the set of optical constants used by \citet{Groenewegenetal2009} is appropriate for SMC evolved stars, then the total DPR can be up to $(3-13)\times$ higher (The scaled C--rich and O--rich DPR contributions are almost equal, and the systematic uncertainties mentioned above suppress the O--rich DPRs) than the total value quoted in Table \ref{tab:compareDPRs_B2012_R2012}. We will use this information in Section \ref{subsec:dustbudgetcrisis}.\\

The relative contributions of O--rich and C--rich dust are also uncertain due to (a) the low-confidence classifications as well as (b) the uncertain chemistry of the FIR objects. We verified that the contribution to the dust budget due to low-confidence classifications is $\la 10\%$. As discussed in Section \ref{subsec:FIRfitsdiscussion}, (b) dominates the uncertainty in relative contributions.\\

\subsection{Dust budget}
\begin{figure*}
\includegraphics[width=85mm]{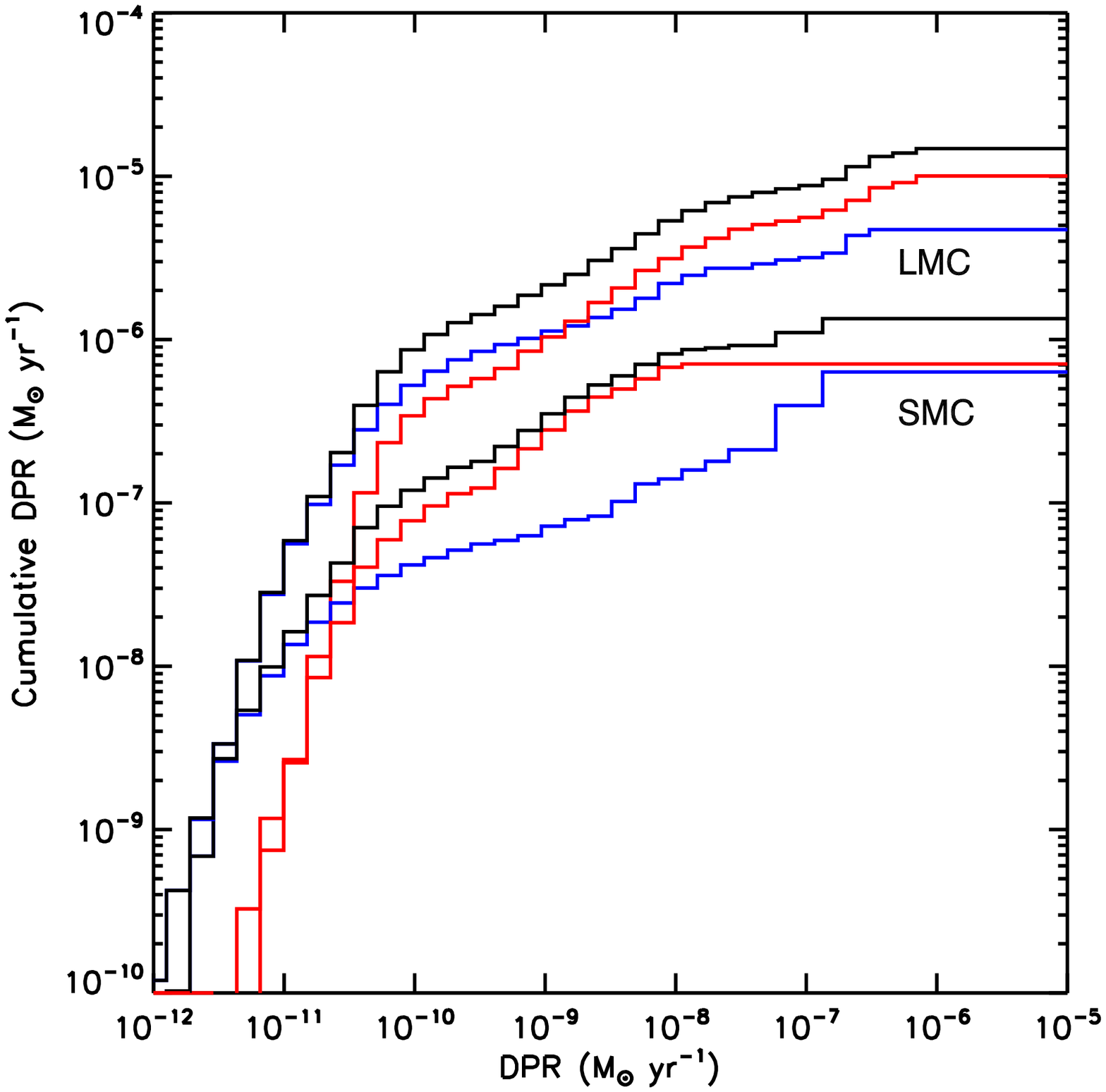} \includegraphics[width=85mm]{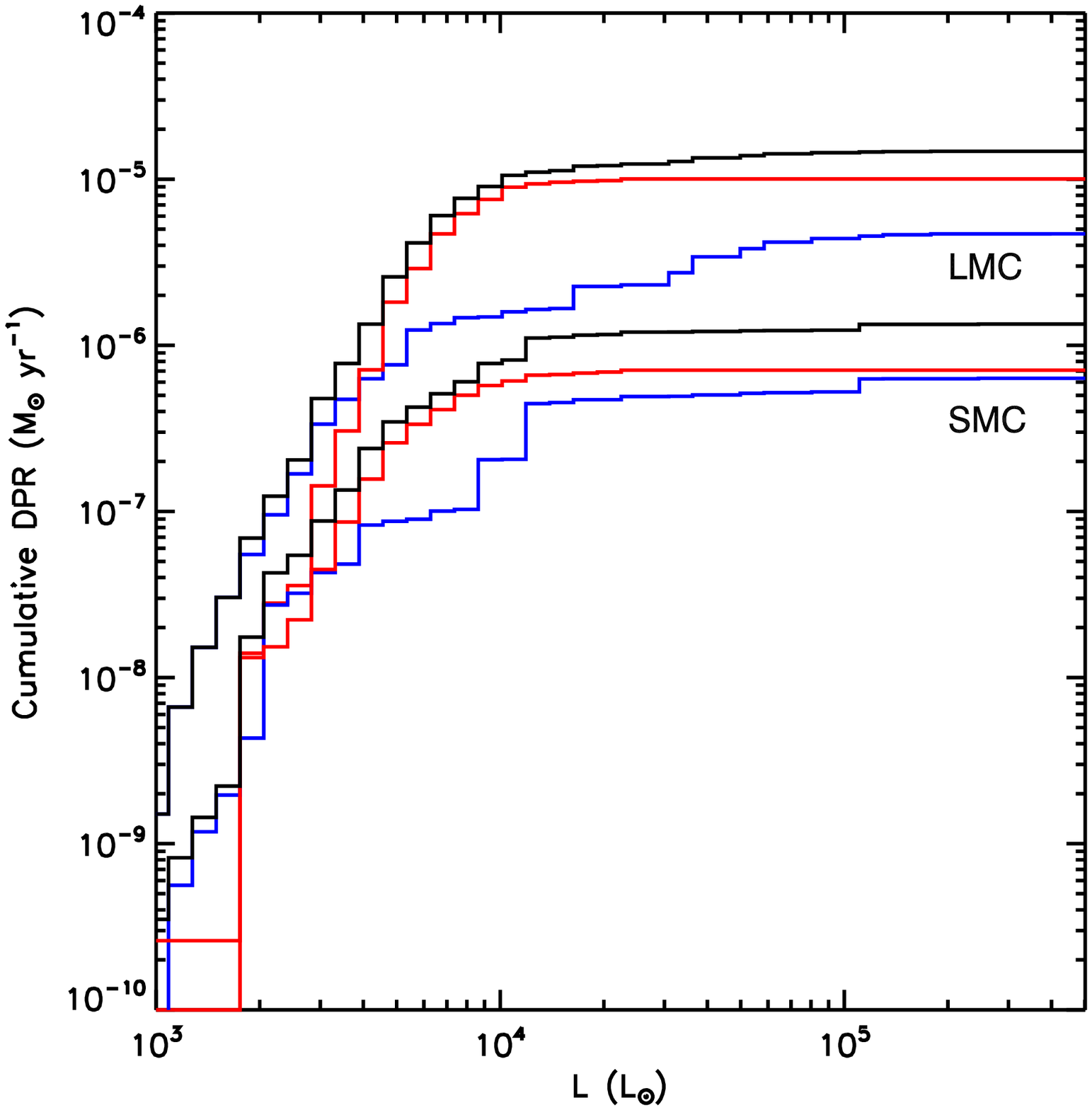}
 \caption{Cumulative dust-production rates (DPRs) of the LMC computed by \citetalias{R2012}, scaled according to Equation \ref{eqn:vexpscaling} (upper curves) and the SMC (this work, lower curves) as a function of DPR (left) and luminosity (right). For each galaxy, the plots show the trend for each chemical type (blue: O--rich, red: C--rich) as well as for the total (black). The SMC cumulative DPR includes FIR objects from Groups 1--4.
 \label{fig:cumdpr}}
\end{figure*}

To compute the dust budget for the SMC, we include all FIR fits in Groups 1--4. We ignore any non-FIR sources with relative uncertainty in the DPR $>$ 1. Fig. \ref{fig:cumdpr} compares the cumulative dust-production rate of SMC sources for each GRAMS class, computed in this work, with the \citetalias{R2012} estimates for the LMC. As in Table \ref{tab:compareDPRs_B2012_R2012}, for a direct comparison we scale the LMC DPRs using Equation \ref{eqn:vexpscaling} and a gas:dust ratio of 500 (200) for O--rich (C--rich) stars. The cumulative DPR is plotted versus the DPR and the luminosity. For each chemical type, the shape of the curve is the same for both galaxies, except that the SMC numbers are lower in each case. For each galaxy, the O--rich sources start contributing at low DPRs, until the mass loss from carbon stars kicks in. Eventually, the carbon-star DPR overtakes the O--rich one. However, the point at which this happens is at a lower DPR for the SMC, due to its lower metallicity. The cumulative DPRs for both O--rich and C--rich sources achieve their maximum values at lower DPRs in the SMC case, reinforcing the conclusion that the SMC lacks high-DPR sources.\\

Table \ref{tab:compareDPRs_B2012_R2012} shows that our total DPRs for C--AGB and x--AGB stars agree with the values from \citetalias{B2012}. While the total for O--AGB and a--AGB stars is lower than that in \citetalias{B2012} by a factor of $\approx$1.5, this deficit is offset by a higher contribution from RSGs, so that our global non-FIR dust budget estimate is nearly the same as that derived from the mid-IR excess. This is not surprising, as the excess--DPR relation in \citetalias{B2012} was calibrated using GRAMS fits to a small number of SMC sources. The difference in DPRs is probably due to the differing source counts; our sample has a smaller (large) number of O--AGB and a--AGB (RSG) sources. Our mean DPR for FIR objects is about 2.5 times higher than that estimated based on the 8 \mic\ excess; it is clear from Fig. \ref{fig:validFIRfits} that this excess flux may not reflect the significant amount of 24 \mic\ dust emission, resulting in a lower DPR. Our total DPR for FIR objects is therefore higher than the \citetalias{B2012}, making the global dust input to the SMC from AGB and RSG stars (including FIR objects) about 1.4 times higher than the excess-based determination.\\

In their paper, \citet{Matsuuraetal2013} selected C-AGB and O-AGB candidates from the [8.0] {\it vs.} [3.6]--[8.0] CMD and the \ks\--[24] {\it vs.} \ks\--[8.0] colour-colour diagram. They then used a colour--mass-loss rate relation, calibrated using a sample with detailed modelling, to compute the mass-loss rates for the entire population. They found a global DPR of $3\times 10^{-6}$ \msunperyr\ for O--rich AGB and RSG stars, and a total of $4\times 10^{-6}$ \msunperyr\ for carbon stars. Using a gas:dust ratio of 200 for both chemical types, they computed a gas mass-loss rate of $6\times 10^{-4}$ and $8\times 10^{-4}$ \msunperyr\ for O--AGB/RSG and carbon stars respectively.\\
We expect our numbers to be different for three main reasons. First, the sets of optical constants used to compute radiative transfer fits to the calibration stars is different from the sets used in GRAMS. This can introduce a systematic discrepancy of up to (4--6)$\times$, as previously mentioned. Second, the calibration of the \citet{Matsuuraetal2013} colour--mass-loss rate relation is based on a small number of stars, and outliers can skew the relationship. Finally, their classification scheme could assign a chemical type different from the GRAMS classification. We accounted for this last factor by applying their colour classifications to our sample and selecting the corresponding (O--rich or C--rich) best-fit (unscaled) GRAMS DPR for each source. We then find totals of $3.4\times 10^{-6}$ and $9.8\times 10^{-7}$ \msunperyr\ for O--AGB/RSG and carbon stars respectively. The global DPR for carbon stars is smaller by a factor of four, which is consistent with the choice of optical constants. Our estimate of the O--rich contribution is higher than that of \citet{Matsuuraetal2013}, as they classified the FIR objects as possible post-AGBs/PNe or YSOs (see their fig. 5).

\subsection{Is there a dust budget `crisis' in the SMC?}
\label{subsec:dustbudgetcrisis}

\begin{table}
\scriptsize
\centering
\begin{minipage}{180mm}
\caption{Comparison of total dust-production rates by chemistry.\label{tab:compareDPRs_all}}
\begin{tabular}{@{}lll@{}}
\hline
\hline
 & Total DPR, O--rich\footnote{O--rich class includes O-AGB, a-AGB and RSG stars.}& Total DPR, C--rich\\
\cline{2-3}\\
& \multicolumn{2}{c}{(\msunperyr)}\\
\hline
LMC & & \\
\hspace{0.1cm} \citet{R2012}& & \\
\hspace{1.5cm} (DPRs scaled\footnote{DPRs scaled according to Equation \ref{eqn:vexpscaling}})& $4.7\times 10^{-6}$& $1.0\times 10^{-5}$\\
SMC & & \\
\hspace{0.1cm} \citet{B2012}&$1.1\times 10^{-7}$&$(7.5-8.4)\times 10^{-7}$\\
\hspace{0.1cm} \citet{Matsuuraetal2013}&$3.0\times 10^{-6}$ & $4.0\times 10^{-6}$\\
\hspace{0.1cm} This paper (DPRs scaled)& $6.3\times 10^{-7}$& $7.1\times 10^{-7}$\\
\end{tabular}
\end{minipage}
\end{table}

Table \ref{tab:compareDPRs_all} summarises the DPR estimates from \citetalias{R2012}, \citetalias{B2012}, \citet{Matsuuraetal2013}, and the present work. Including the FIR objects, we find a global DPR of  \dustbudgetOrich\ and \dustbudgetCrich\ \msunperyr\ for stars with O--rich and C--rich dust chemistries. These numbers are comparable to those of \citetalias{B2012}, and are therefore also consistent with the present AGB dust-production rates predicted by \citet{Schneideretal2014} based on theoretical models, using the old ATON yields (dashed line in their fig. 4). In fact, the somewhat higher value for O--rich dust injection derived in this paper shows better agreement with the \citet{Schneideretal2014} results for silicate dust.\\

The time-scale at which stellar sources replenish ISM dust can be estimated by comparing the current ISM dust mass to the DPR from evolved stars. As a way of incorporating the respective uncertainties, we determine the maximum (minimum) replenishment time-scale using the maximum (minimum) ISM dust mass and minimum (maximum) DPR estimates. We first estimate the time-scale assuming that AGB stars are the sole contributors to the ISM dust. The global AGB DPR derived in this paper is $1.3\times 10^{-6}$ \msunperyr, and the discussion in Section \ref{subsec:dpruncertainties} estimates that it can be up to 13$\times$ higher. Using these lower and upper bounds on the DPR and the ISM dust mass estimate from \citet{Gordonetal2014} (\ismdustmass\ M$_\odot$), we compute a replenishment time-scale in the range \treplenish\ Gyr. If the ISM dust mass originated primarily from AGB stars, then the AGB dust injection rate would replenish the total ISM dust mass in a timeframe shorter than the age of the SMC \citep[$\sim$12 Gyr; e.g.,][]{HarrisZaritsky2004}. Since the upper bound on the time-scale is longer than the age of the Universe, we must consider the contribution from other dust sources.\\

Besides AGB and RSG stars, supernovae (SNe) are a major stellar source of dust. In addition, it is possible that grain growth in the cold ISM is necessary to make up the difference. While the inclusion of SNe dust reduces the ISM replenishment time-scale to a more reasonable range, the mass of dust created by SNe spans a large range, as does the mass destroyed by the reverse shock \citep[{\it e.g.},][]{Sugermanetal2006,Matsuuraetal2011,Gallagheretal2012,Andrewsetal2015,OwenBarlow2015}. \citet{Temimetal2015} set an upper limit on the DPR from SMC core-collapse SNe at \CCSNeDPR\ \msunperyr, assuming no dust destruction. \citetalias{B2012} use  the observed SN rate and dust mass estimates for various SMC SNe to estimate a lower limit of \SNeDPRlowerlimit\ \msunperyr, which is comparable to the global AGB/RSG DPR estimate in this paper. With these lower and upper bounds on the SNe dust-injection rate, the AGB+SN replenishment time-scale is in the range \treplenishboostwithSNe\ Gyr. However, the supernova rate in the SMC is higher over the past 12 Myr than at previous times, so the lower limit to the time-scale is perhaps too optimistic. The estimate above can be further refined by incorporating a treatment of dust destruction. However, the extent of dust destruction by SNe is highly uncertain \citep[e.g.,][]{Silviaetal2012,Temimetal2015,Lakicevicetal2015,Lauetal2015}.\\

\citetalias{B2012} estimate a dust lifetime in the SMC of $(0.38-0.86)$ Gyr, which is consistent with the lower limit for the replenishment time-scale estimated above including SN dust production. However, the dust grain lifetime may be much shorter in the SMC. For instance, \citet{Temimetal2015} find a lifetime of  \TemimetallifetimeO\ Myr (\TemimetallifetimeC\ Myr) for O--rich (C--rich) dust grains in the SMC. The AGB/RSG contribution over this time-scale would then be \AGBRSGdustmasscontribution\ M$_\odot$ of dust. This range incorporates the $(3-13)\times$ inflated estimate for the global DPRs due to differences in outflow speed and choice of optical constants. The total dust mass contribution is \dustmasscontributionwithSNe\ M$_\odot$ if the contribution from SNe is included. The most optimistic estimate for the dust injected still falls short of the estimated ISM dust mass.\\

Estimates of SNe dust destruction on a global scale do not take into account the differences in the spatial distribution of AGB/RSG stars and SNe. Fig. 1 in \citet{Cionietal2006} and Figs. 12--14 in \citetalias{B2011} show the distribution of AGB/RSG stars. AGB stars are smoothly distributed out to a radius of at least 2 deg from the centre of the bar \citepalias[Fig. 34 in][]{B2011}, while SNe are concentrated in the star-forming regions in the bar \citep[Fig. 2 in][]{Temimetal2015}. It follows from these figures that most low- and intermediate-mass stars migrate away from the regions of high dust destruction {\em before} they reach their dust-producing phase, suggesting that the lifetime of AGB-produced dust is longer than the average dust destruction rate in the SMC.\\

The upper bound to the replenishment time-scale, as well as the total injected dust mass estimate above, imply a non-stellar origin for dust in the SMC. This conclusion is consistent with that of \citetalias{B2012} but not with that of \citet{Schneideretal2014}. While our dust budget estimate is consistent with theirs within uncertainties, the main difference between this paper and \citet{Schneideretal2014} is that they estimate the {\em maximum} contribution to the existing mass, without considering a finite dust lifetime due to destruction in the ISM. The lifetimes derived by \citet{Temimetal2015} are significantly shorter than those in the Milky Way, and also to those estimated for the SMC in \citetalias{B2012}. This difference in dust timescales leads us to the same conclusion as \citet{Temimetal2015}.\\

The current ISM dust distribution is more centrally concentrated than the current AGB distribution \citep[see, {\it e.g.}, the maps in][]{B2011,Gordonetal2014}. This suggests that, if the dust is solely of AGB origin, it is either migrated away from its formation site or it formed from the earlier, more massive AGB stars which would have also been centrally located. In that case, however, it would be nearer to SNe \& would have a shorter lifespan. This argument points to ISM grain growth as a dust source, unless there is a significant amount of dust in as yet undiscovered AGB/RSG sources. Is it possible that there are sources with very short bursts of intense mass loss? For instance, \citet{deVriesetal2014} note that such an abrupt mass loss phase is required to explain the shorter-than-expected superwind observed in their OH/IR star sample.\\

\begin{figure}
 \includegraphics[width=84mm]{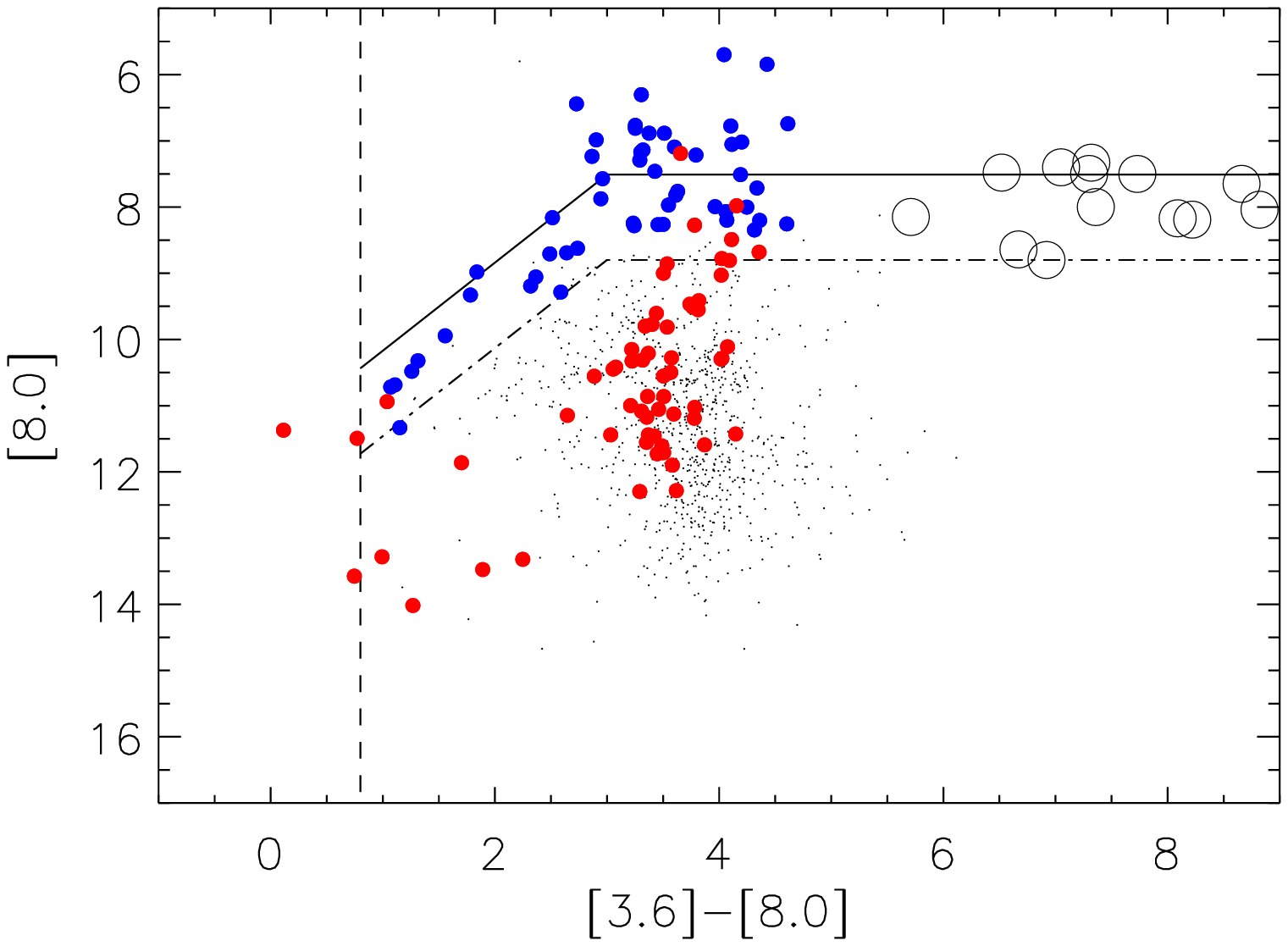}
 \includegraphics[width=84mm]{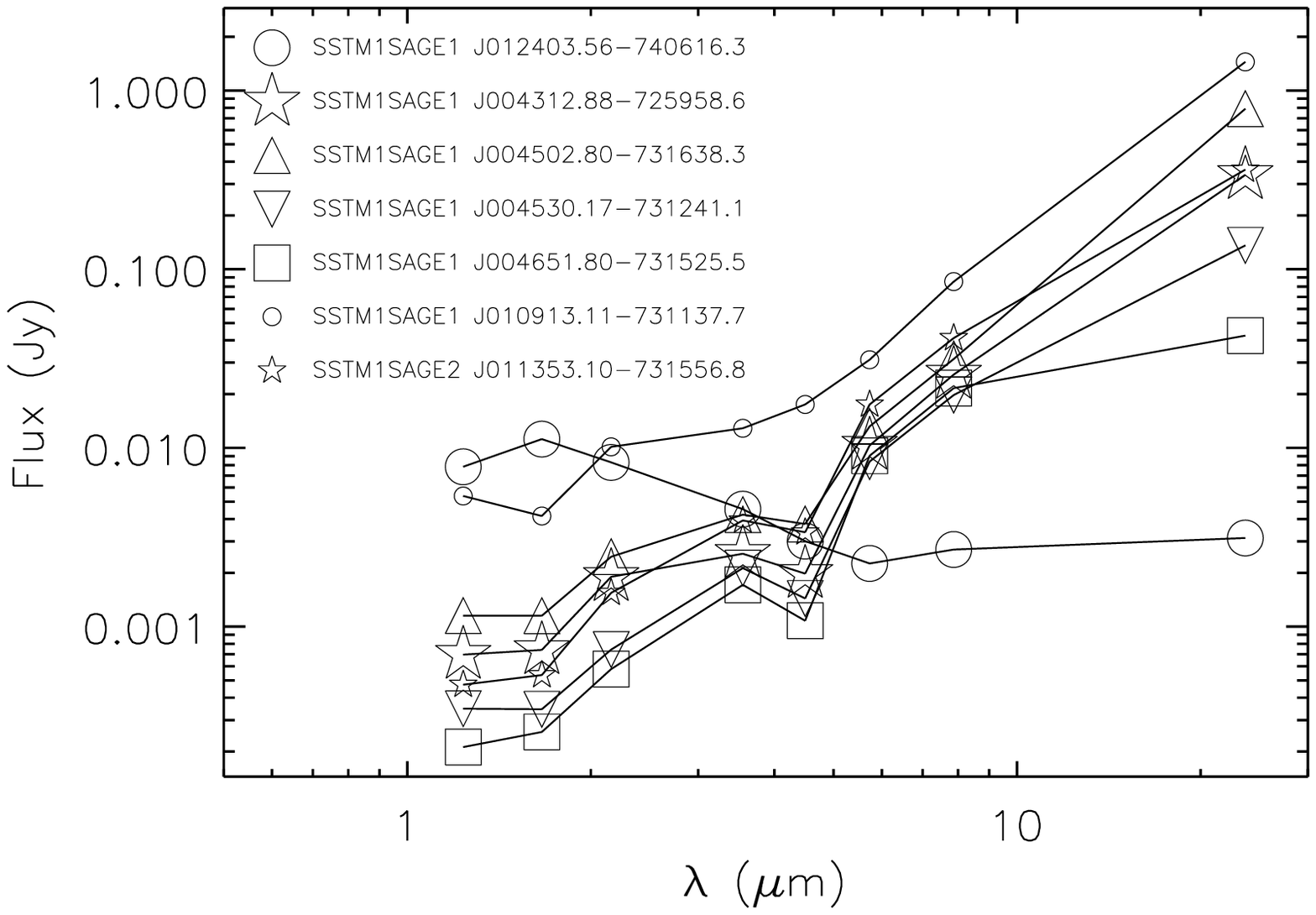}
 \caption{{\em Top:} IRAC CMD showing the locations of all 984 \citet{Sewiloetal2013} YSO candidates, including the `primary photometry' sources (black dots and blue circles) as well as those extracted using aperture photometry (red circles). The blue circles indicate Sewi{\l}o et al. sources that are also in our evolved star list. The open circles represent the \citet{Gruendletal2008} extreme carbon stars placed at SMC distance. \citetalias{B2011} classified sources redder than the dashed line and brighter than the solid lines as extreme AGB stars. These criteria have to be relaxed (dot-dashed line) in order to preserve the `extreme' classification of the Gruendl et al. stars. These relaxed criteria also flag all of the Sewi{\l}o et al. `primary photometry' sources already in our list (blue circles) as x--AGB stars. In addition, the extreme classification can be extended to seven of the aperture photometry sources (red circles). {\em Bottom:} SEDs of these seven aperture photometry sources, identified by their SMC-SAGE IRAC designations. \label{fig:Sewiloetal2013ysos}}
\end{figure}

The extreme carbon stars in the LMC were discovered by \citet{Gruendletal2008} when they reprocessed the SAGE images using aperture photometry to extract point sources. This discovery prompts us to ask: {\em are there candidate extreme carbon stars hidden in the SAGE-SMC data?} There is already a dataset available to help answer this question: \citet{Sewiloetal2013} identified $\sim$1000 high-quality YSO candidates, 62 of which did not previously exist in the SAGE-SMC IRAC point-source catalog. These sources were extracted by performing aperture photometry on the SAGE-SMC images, in a manner similar to Gruendl et al.'s LMC work. The \citet{Sewiloetal2013} stars are classified as either `probable' or `high probability' YSOs based on model fits. Fig. \ref{fig:Sewiloetal2013ysos} shows a [8.0] {\it vs.} [3.6]--[8.0] CMD with all 984 of the \citet{Sewiloetal2013} YSO candidates. The \NSewiloYSOsmatched\ sources matched to our list are overlaid in blue, and the entire list of sources from aperture photometry is shown in red. The dashed and solid lines depict the \citetalias{B2011} colour cuts used to identify extreme AGB candidates. When the \citet{Gruendletal2008} sources are placed on this diagram (after scaling to SMC distance), only a few are above the B2011 cut. Moreover, many of the evolved-star candidates (many classified on basis of their near-IR photometry) are below this cut and in the YSO-dominated region. If this cut is relaxed so as to include all of the extreme carbon star candidates (dot-dashed line) it classifies \NSewiloYSOsapphotxAGB\ of the aperture photometry sources as extreme AGB stars. \citet{Sewiloetal2013} classify four of these sources as `high probability' YSOs and the rest as `probable'. However, SED inspection shows that they are probably YSOs or background galaxies (Fig. \ref{fig:Sewiloetal2013ysos}, bottom panel). Despite relaxing our selection criteria, we do not find any promising candidates for very dusty AGB stars in the Sewi{\l}o et al. sample. It is also interesting to note that there are no SMC YSO candidates with colours as red as the \citet{Gruendletal2008} extreme carbon stars. Based on this brief investigation, we conclude that we have probably accounted for all the dust from evolved stars in the SAGE-SMC data.

\section{Summary}
\label{sec:conclusion}
Using a very careful selection procedure, we produce the most complete dataset of mass-losing evolved stars in the SMC to date. We fit dust radiative transfer models to these sources to resolve their chemical types and estimate their luminosities and dust-production rates (DPRs). We find that our chemical classification agrees well with those based on near- and mid-IR colours \citep{B2011}, and that the fraction of high-confidence O--rich and C--rich classifications is nearly identical to the determinations of \citet{R2012} for the LMC.\\

We find a total dust injection rate of \globaldustbudget\ \msunperyr\ from all SMC AGB and RSG stars. The extreme AGB stars, comprising \xnumberpercent\% of the sample by number, are the most significant (non-FIR) contributors to the global dust budget ($\approx$ \xdustpercent\%). The RSG contribution is similar to that in the LMC, about 5--10\%. Our results for the total dust budget from non-FIR sources is similar to that of \citetalias{B2012}. However, we find that the almost all of the FIR objects are better fit with O--rich models, and that the global dust budget increases by a factor of $\approx$1.5 upon including the FIR contribution. While some of these sources are spectroscopically confirmed, it is crucial to determine the AGB/RSG status as well as chemical type of the remaining sources in order to constrain the dust budget. Regardless of their chemical nature, these objects contribute significantly to the dust input, and it is therefore essential to obtain a complete inventory of such extremely dusty sources, which will become possible with the advent of the next generation instruments such as the James Webb Space Telescope.\\

The total input to the SMC from stellar sources based on mid-IR observations cannot be reconciled with the ISM dust mass. This points to a non-stellar origin for the SMC dust, unless there are as yet undiscovered evolved stars with very high dust-production rates. Owing to their red colours, such sources would be confused with YSOs. We find no evidence in the {\it Spitzer} data of such sources; however, they may be identified by future surveys.\\

The largest sources of uncertainty in determining the AGB/RSG dust budget are the unknown expansion speeds and the choice of dust optical constants, and can raise the dust budget estimates by {\em up to an order of magnitude}. Future observations with the Atacama Large (Sub)Millimetre Array will determine the expansion speeds of Magellanic Cloud stars and, together with improved laboratory experiments and models for dust opacities, help resolve these uncertainties.\\

\section*{Acknowledgements}
We are grateful to the anonymous referee for their thorough and helpful comments. We thank Lynn Carlson for her insight on separating YSOs from AGB candidates in colour-magnitude and colour-colour diagrams. We acknowledge funding from the NAG5-12595 grant and the SAGE-LMC {\it Spitzer} grant 1275598. MLB is supported by the NASA Postdoctoral Program at the Goddard Space Flight Center, administered by ORAU through a contract with NASA. FK acknowledges financial support by the Ministry of Science and Technology in Taiwan, under grant codes MOST103-2112-M-001-033- and MOST104-2628-M-001-004-MY3. MM acknowledges the NASA ADAP grant NNX13AE36G. BAS acknowledges the NASA ADAP grant NNX13AD54G. The authors would also like to thank Bernie Shiao at STScI for his invaluable assistance with the SAGE database.
This work is based on observations made with the {\it Spitzer} Space Telescope, which is operated by the Jet Propulsion Laboratory, California Institute of Technology, under NASA contract 1407. The publication makes use of data products from the Wide-field Infrared Survey Explorer, which is a joint project of the University of California, Los Angeles, and the Jet Propulsion Laboratory/California Institute of Technology, funded by the National Aeronautics and Space Administration. This research has also made use of the SAGE CASJobs database, which is made possible by the Sloan Digital Sky Survey Collaboration; SAOImage DS9, developed by the Smithsonian Astrophysical Observatory; the Vizier catalog access tool, CDS, Strasbourg, France; the SIMBAD database, operated at CDS, Strasbourg, France; the cross-match tool available through the US Virtual Astronomical Observatory, which is sponsored by the NSF and NASA; and NASA's Astrophysics Data System Bibliographic Services.

\bibliographystyle{mnras}

\clearpage

\appendix

\section{Fits for FIR objects in Groups 5--7}
\begin{figure*}
\includegraphics[width=50mm]{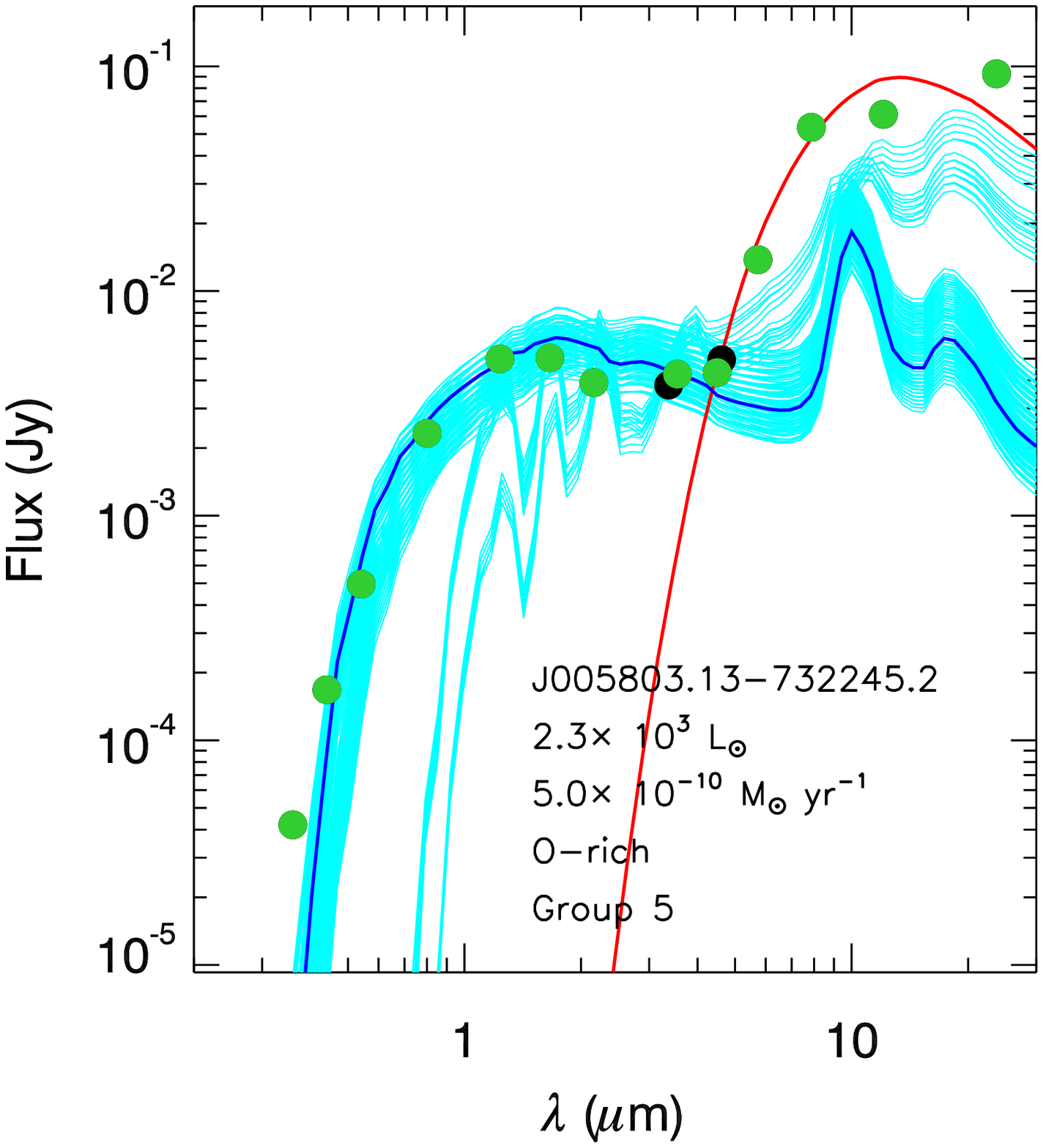}\includegraphics[width=50mm]{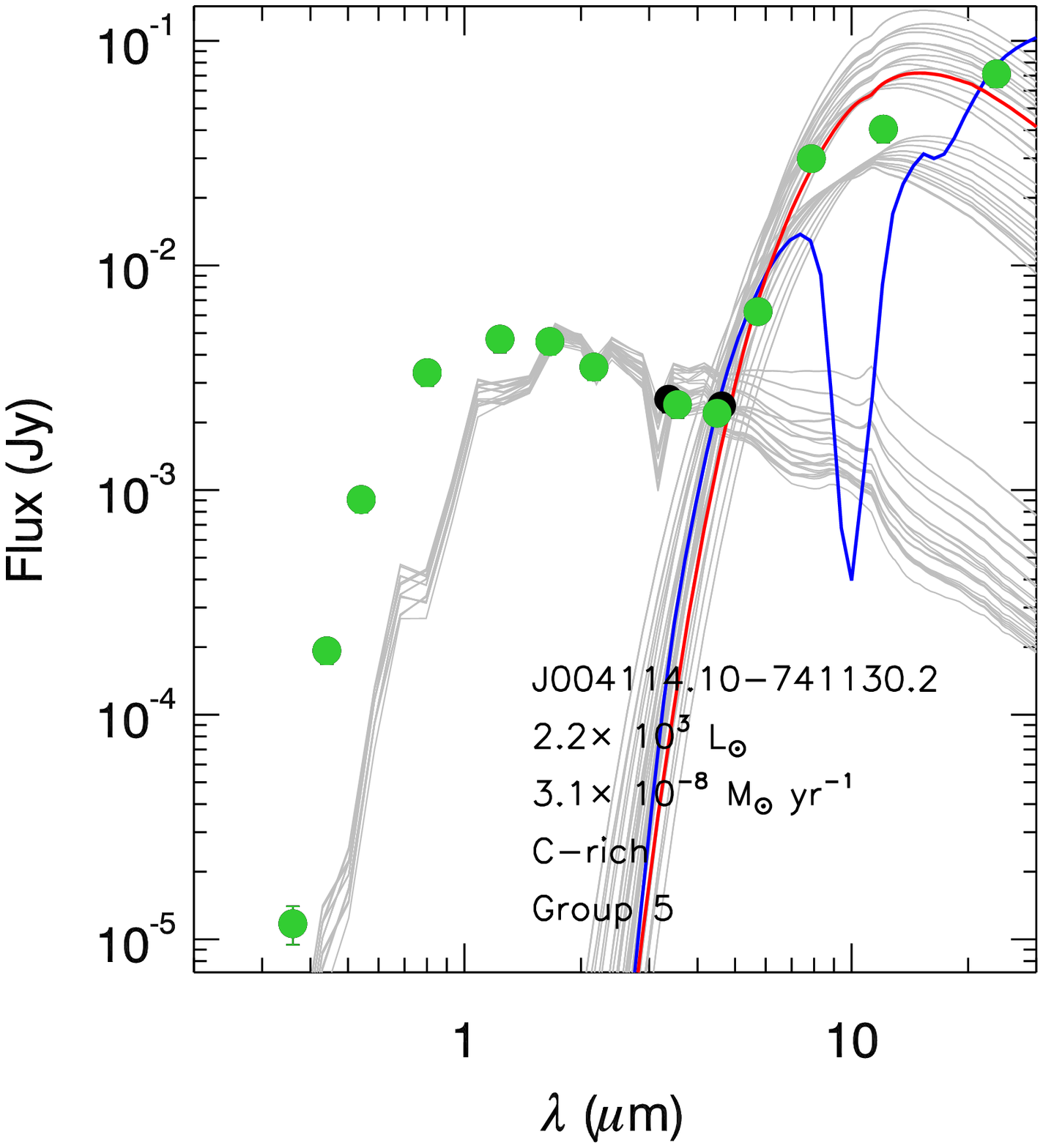}\includegraphics[width=50mm]{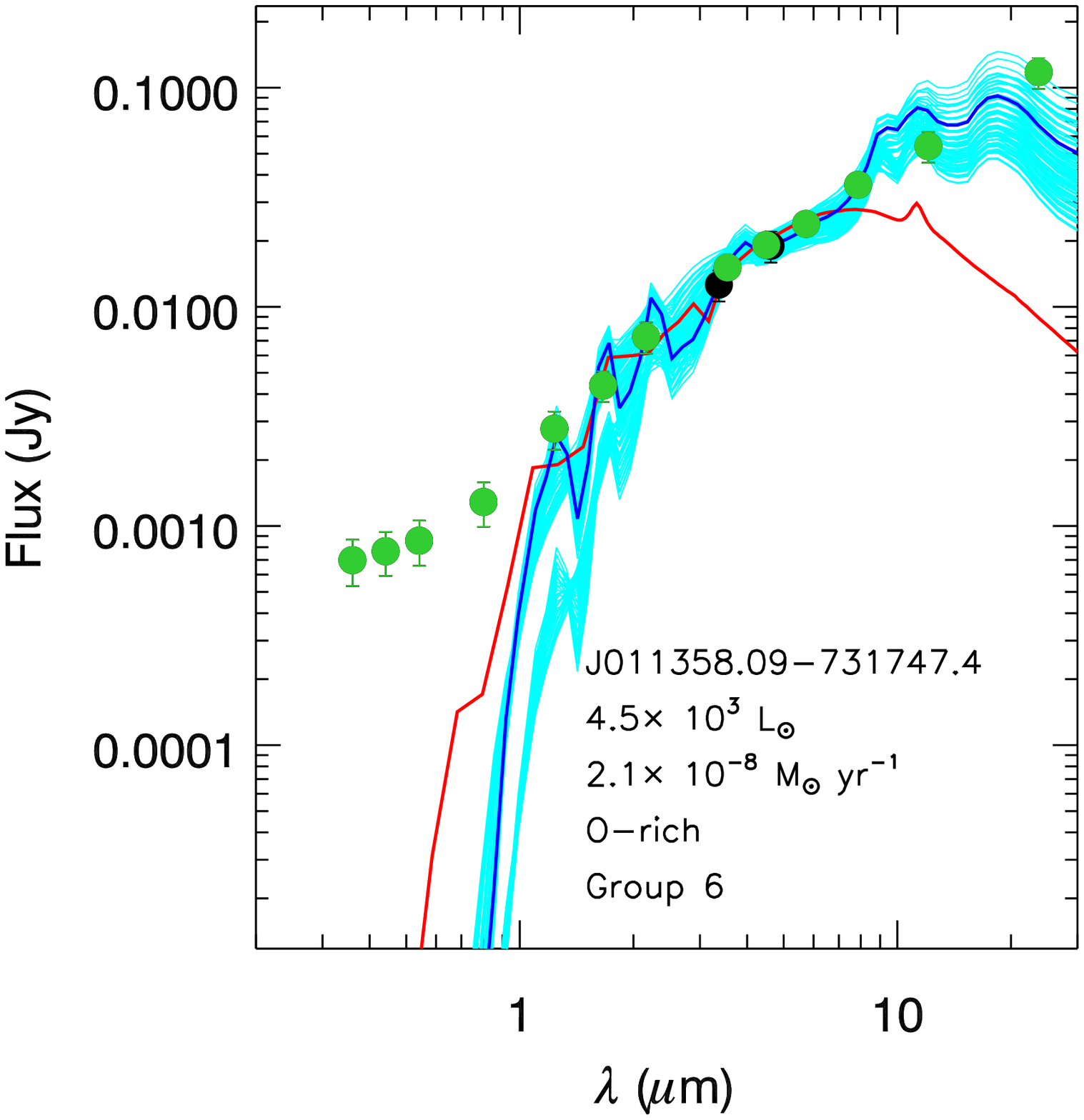}\\
\includegraphics[width=50mm]{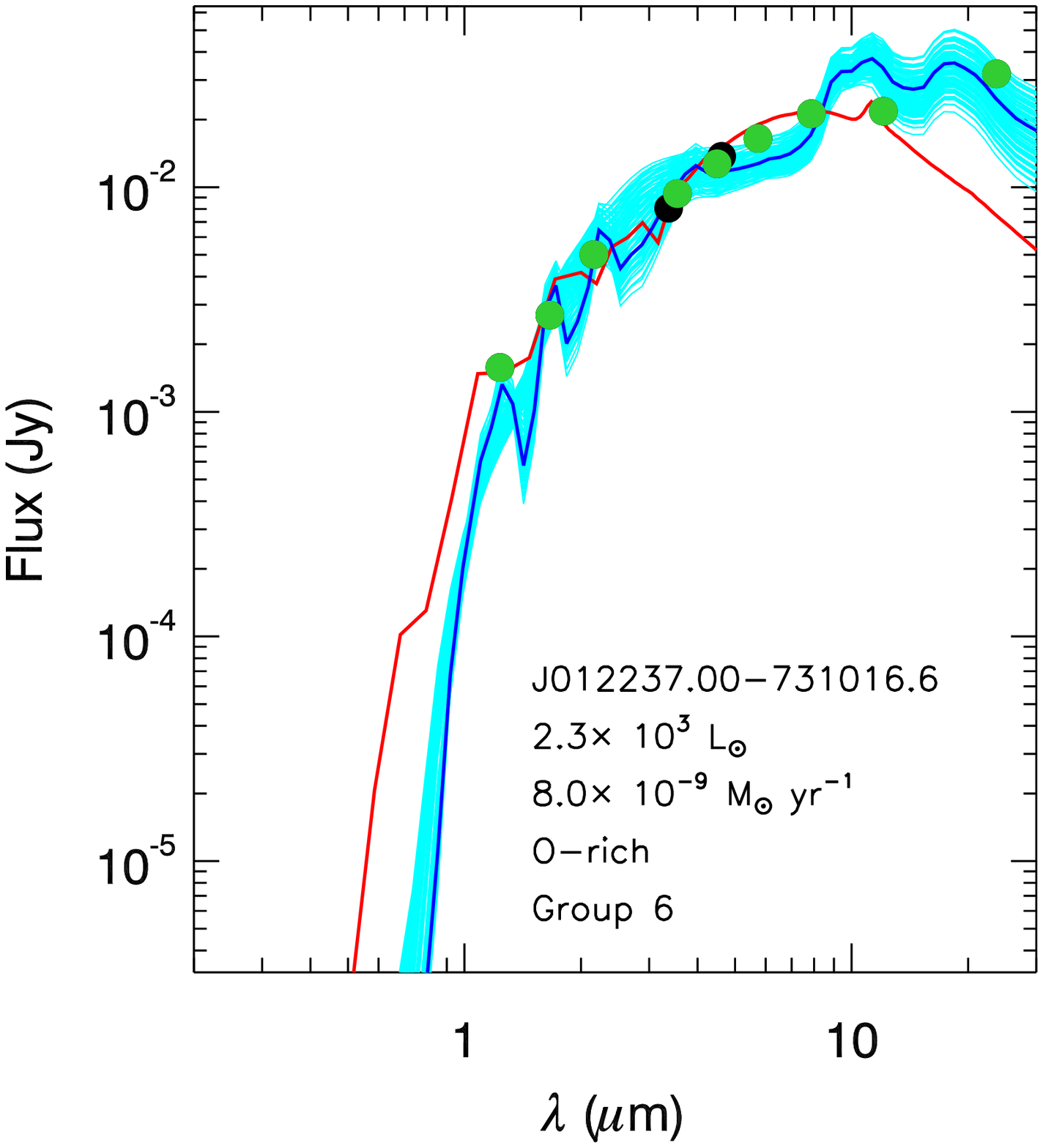}\includegraphics[width=50mm]{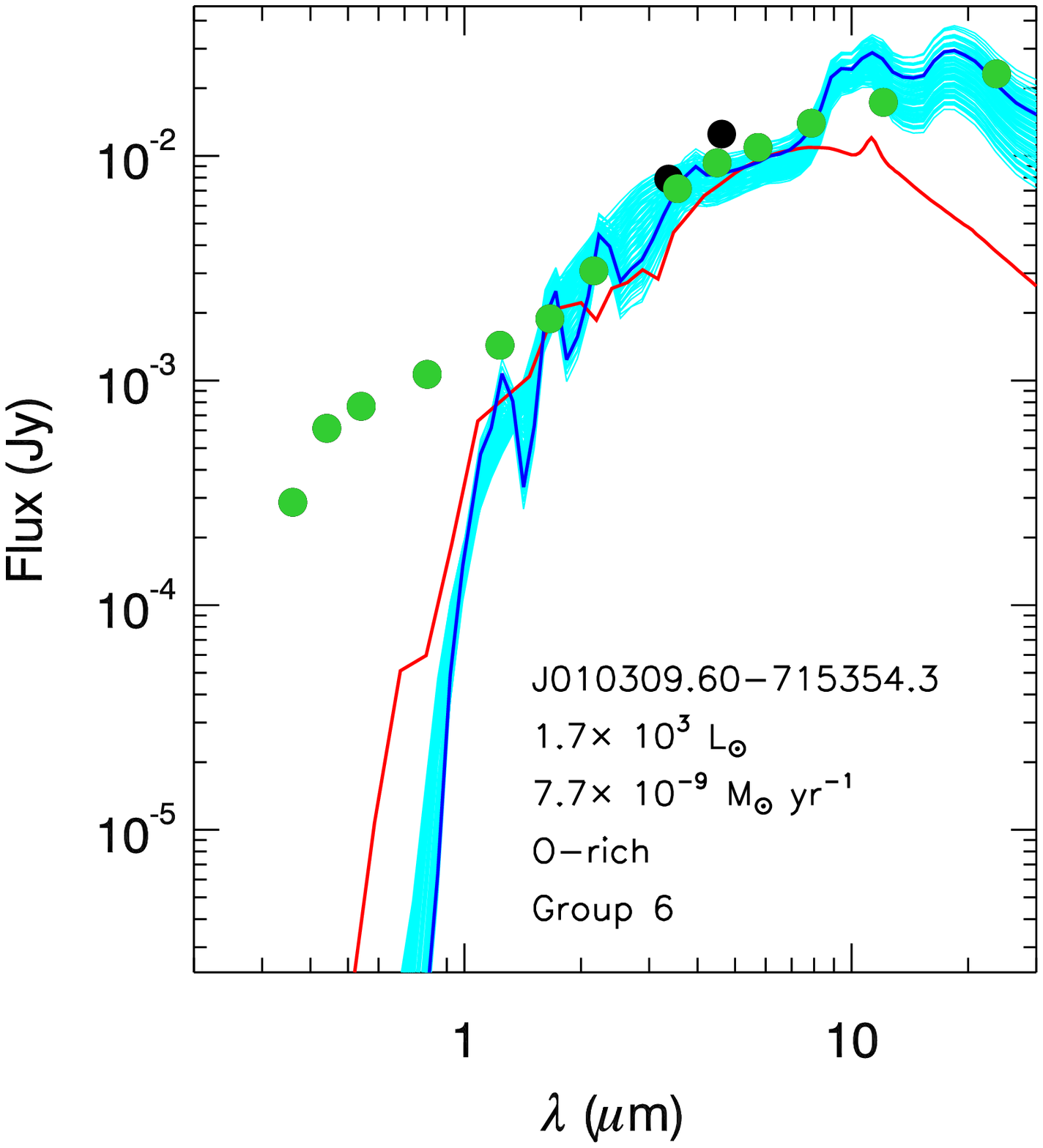}\includegraphics[width=50mm]{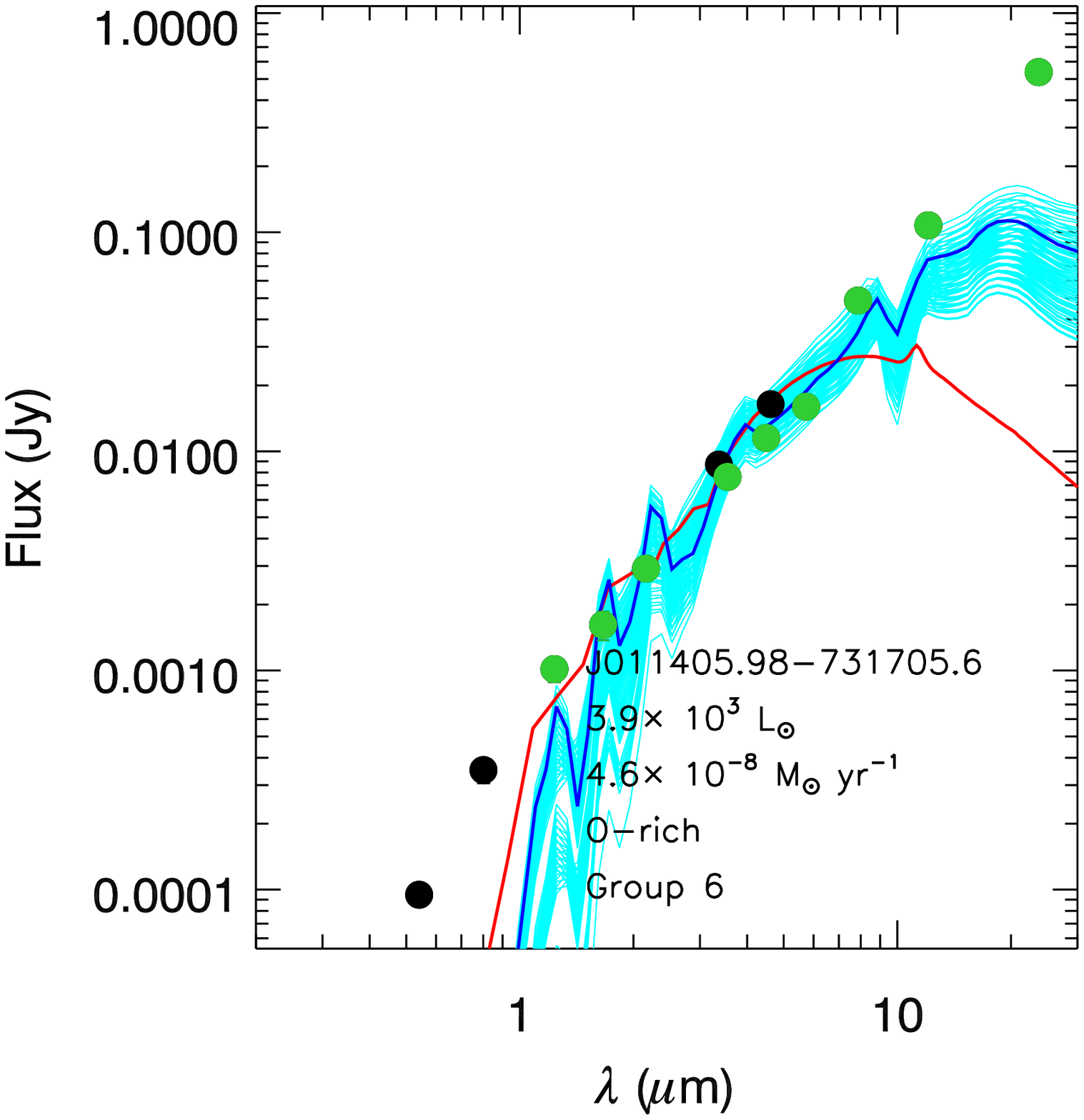}\\
\includegraphics[width=50mm]{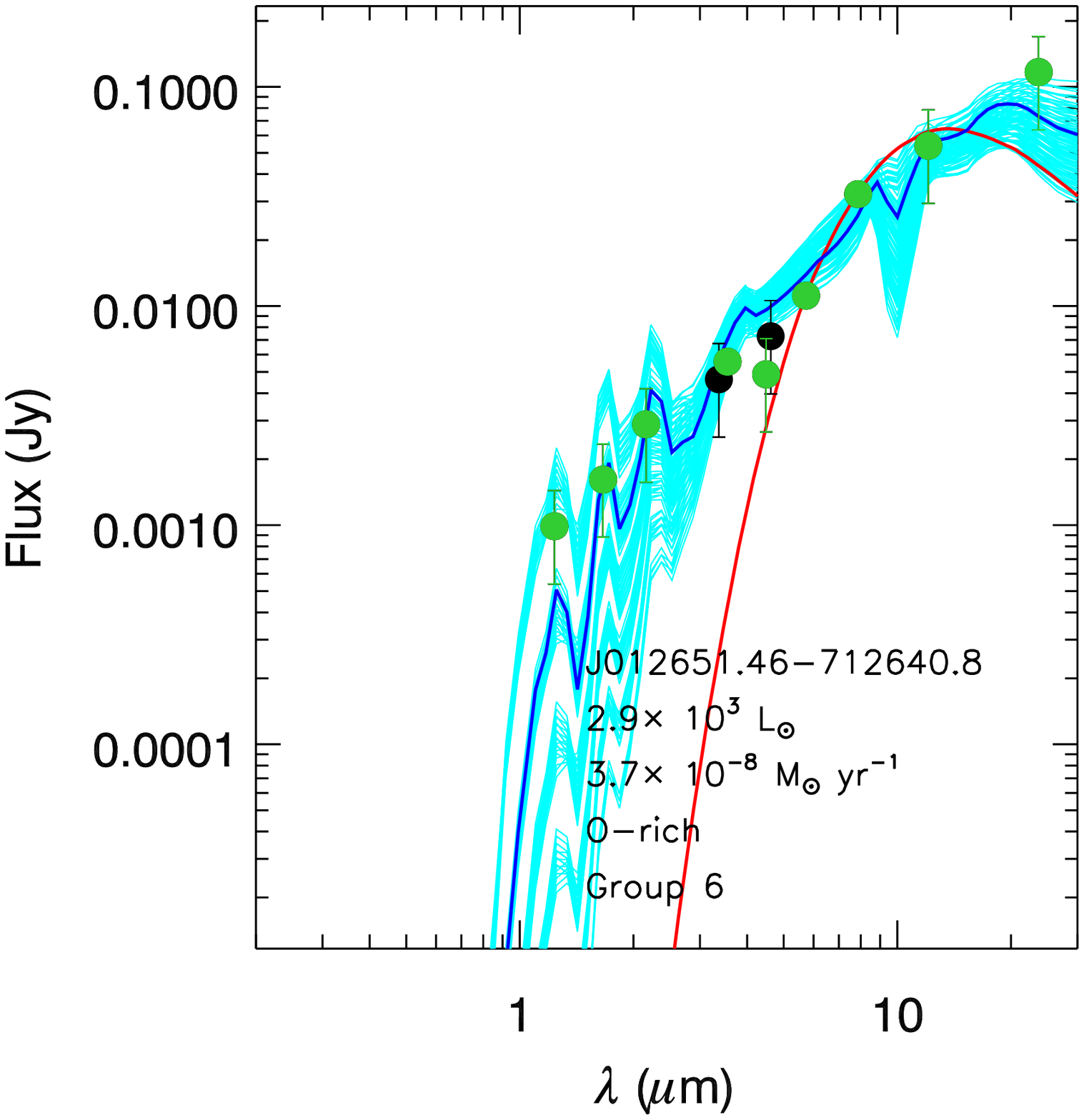}\includegraphics[width=50mm]{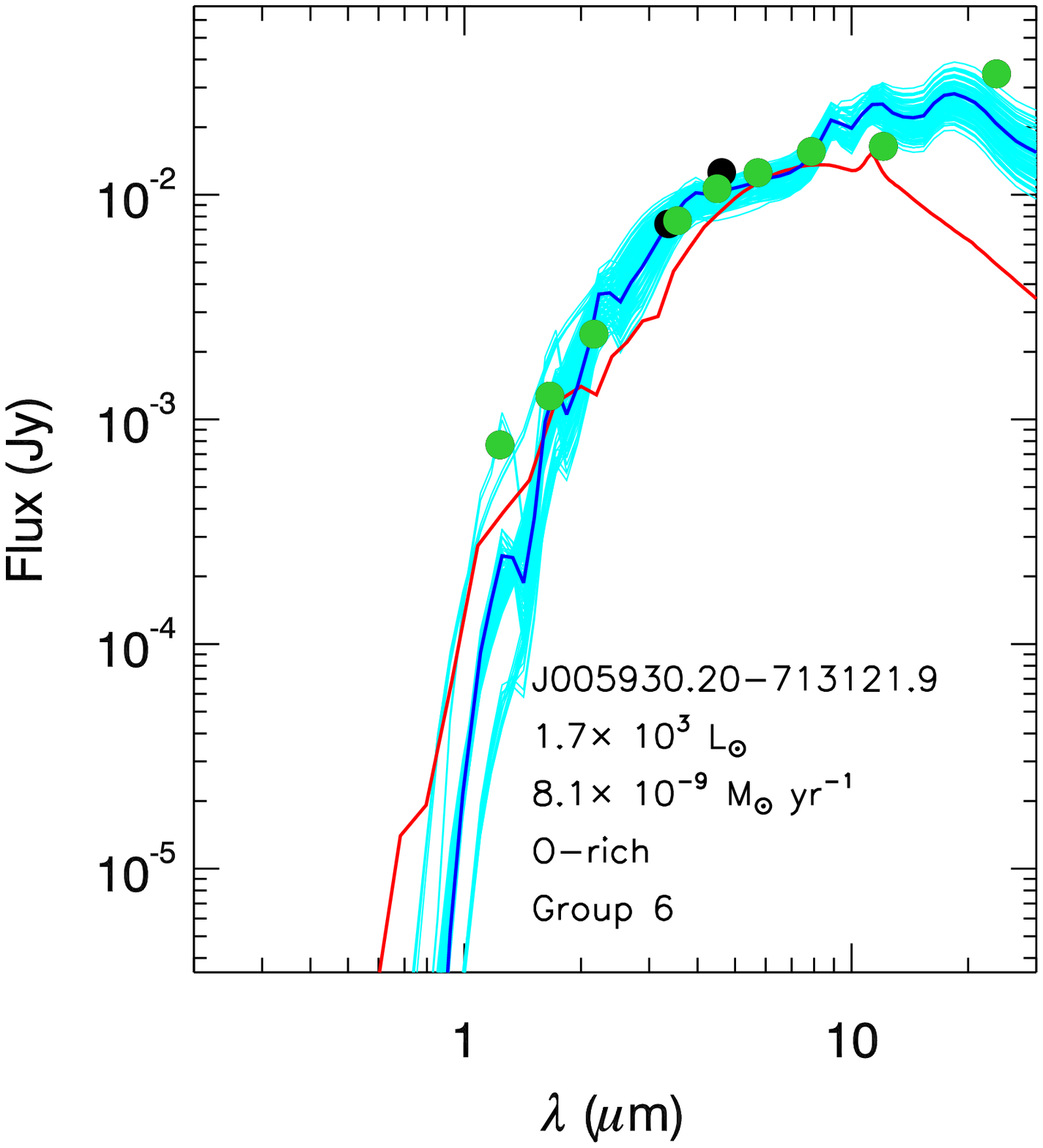}\includegraphics[width=50mm]{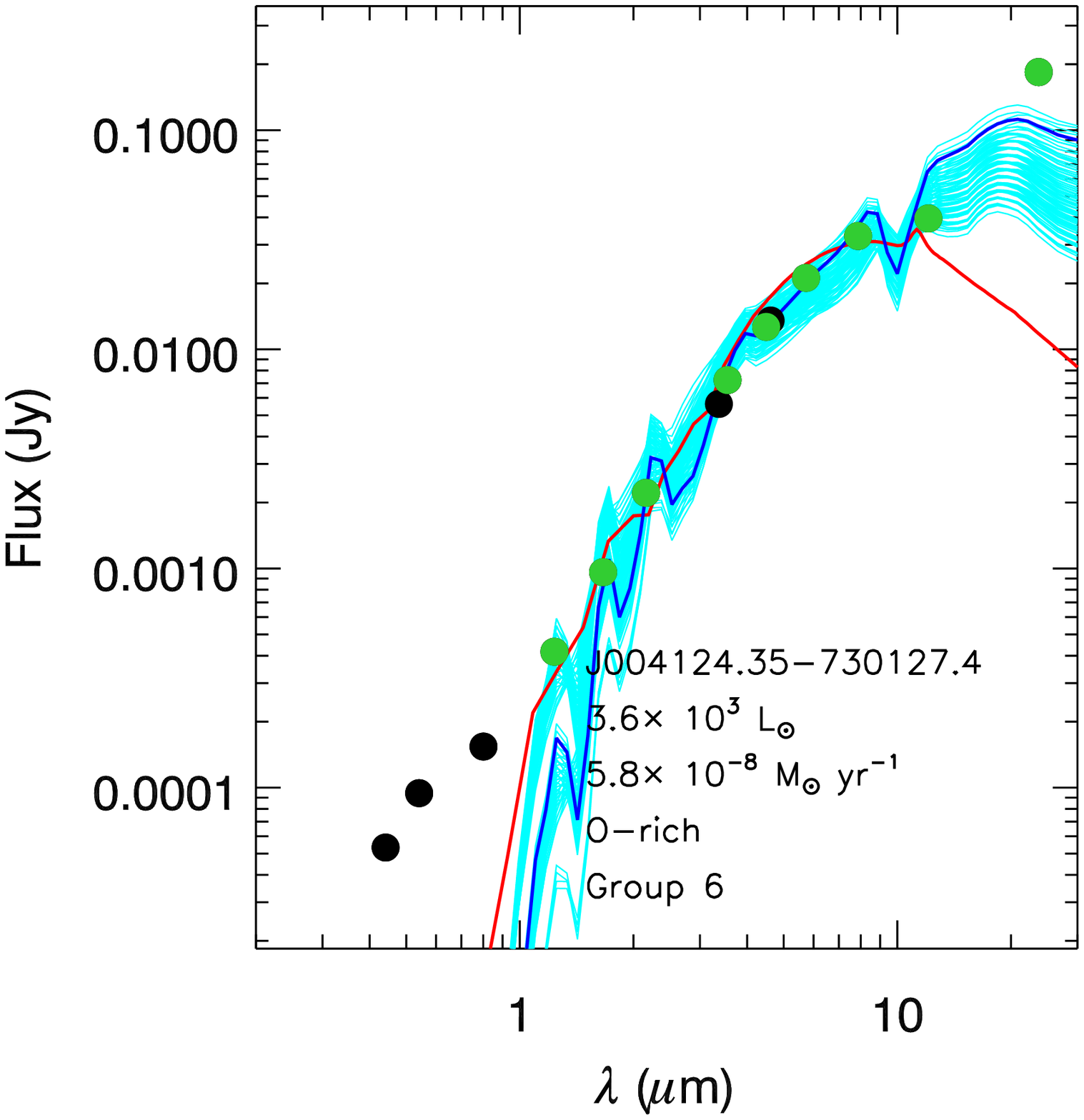}\\
\includegraphics[width=50mm]{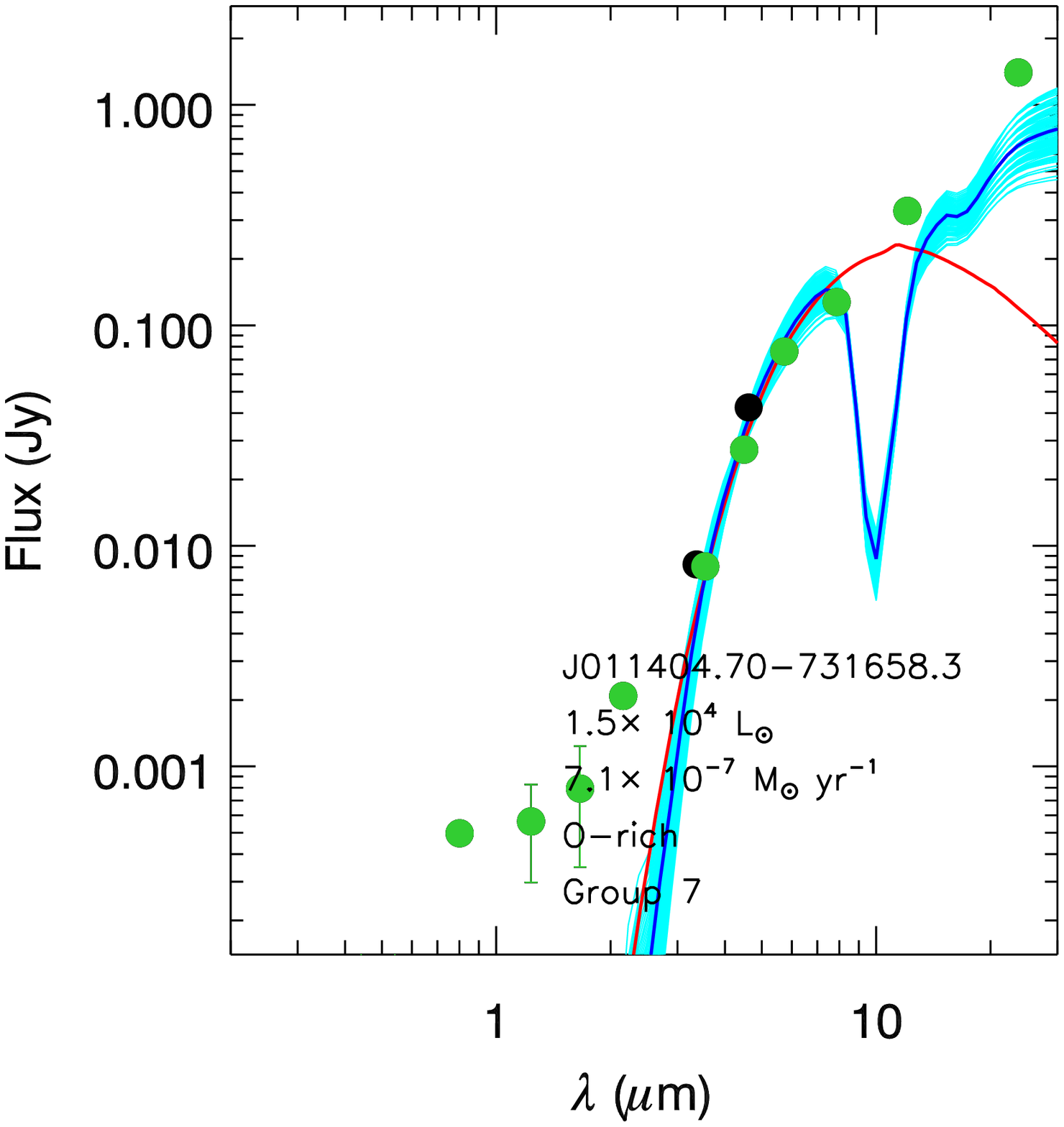}\includegraphics[width=50mm]{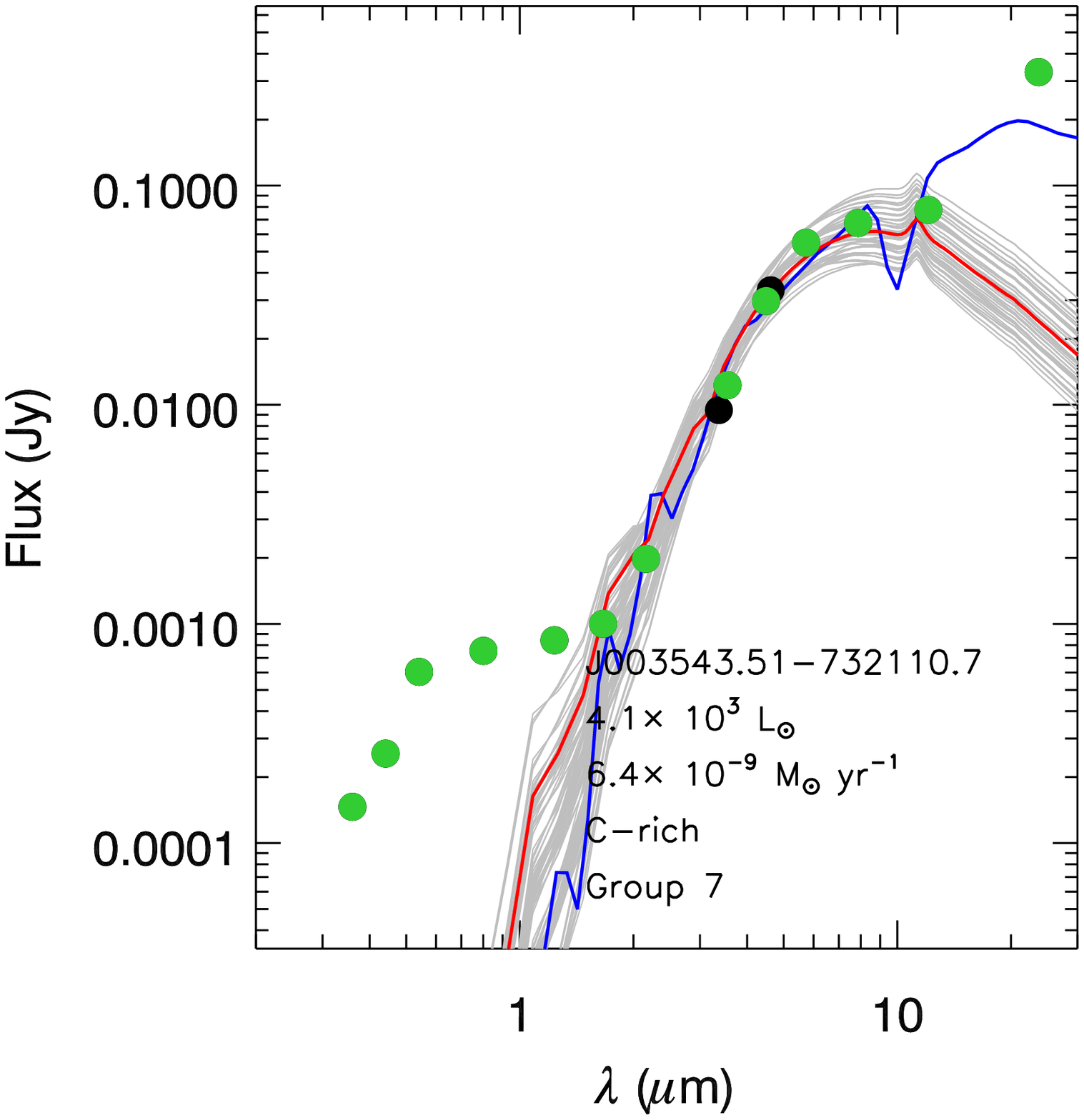}\includegraphics[width=50mm]{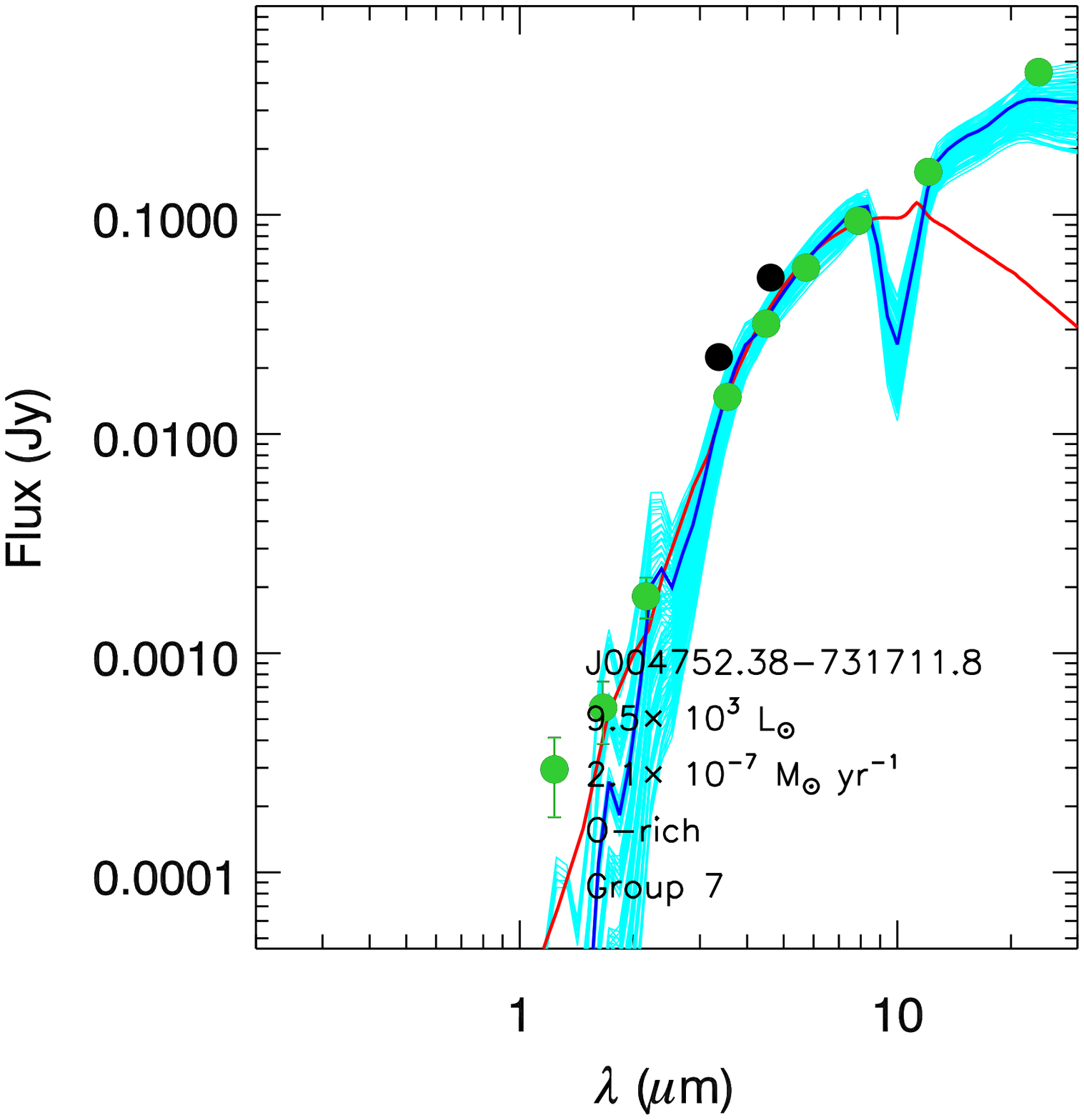}
\caption{Fits for objects in FIR Groups 5--7 (best-fit O--rich: blue, best-fit C--rich: red). The 100 acceptable models of the best-fit chemical type are also shown (cyan: O--rich, grey C--rich). These objects either have steep SEDs, or show a second peak in the mid-IR. The fit quality in all cases is poor.\label{fig:invalidFIRfits}}
\end{figure*}

\renewcommand{\thefigure}{\arabic{figure} (Cont.)}
\addtocounter{figure}{-1}

\begin{figure*}
\includegraphics[width=50mm]{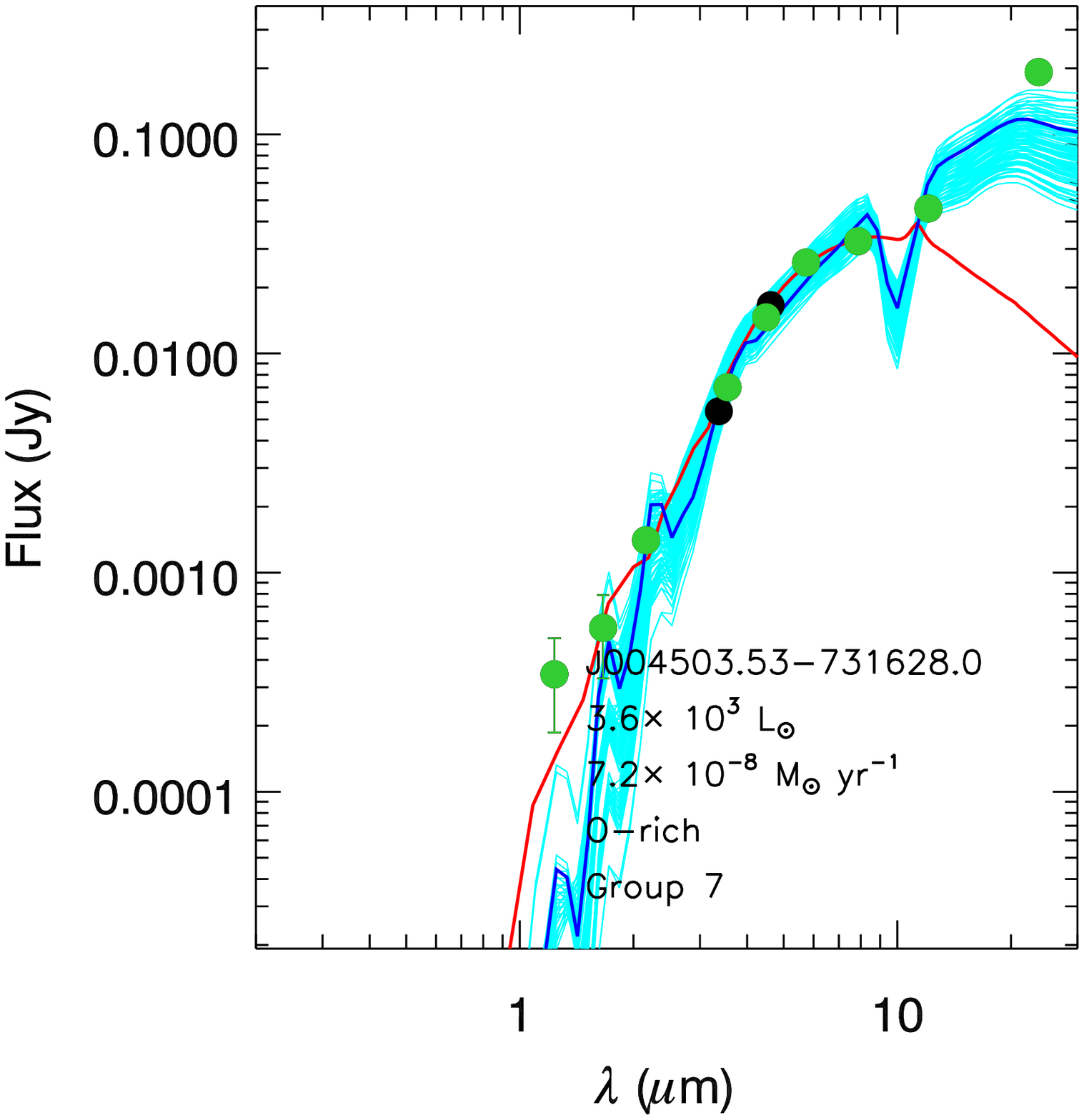}\includegraphics[width=50mm]{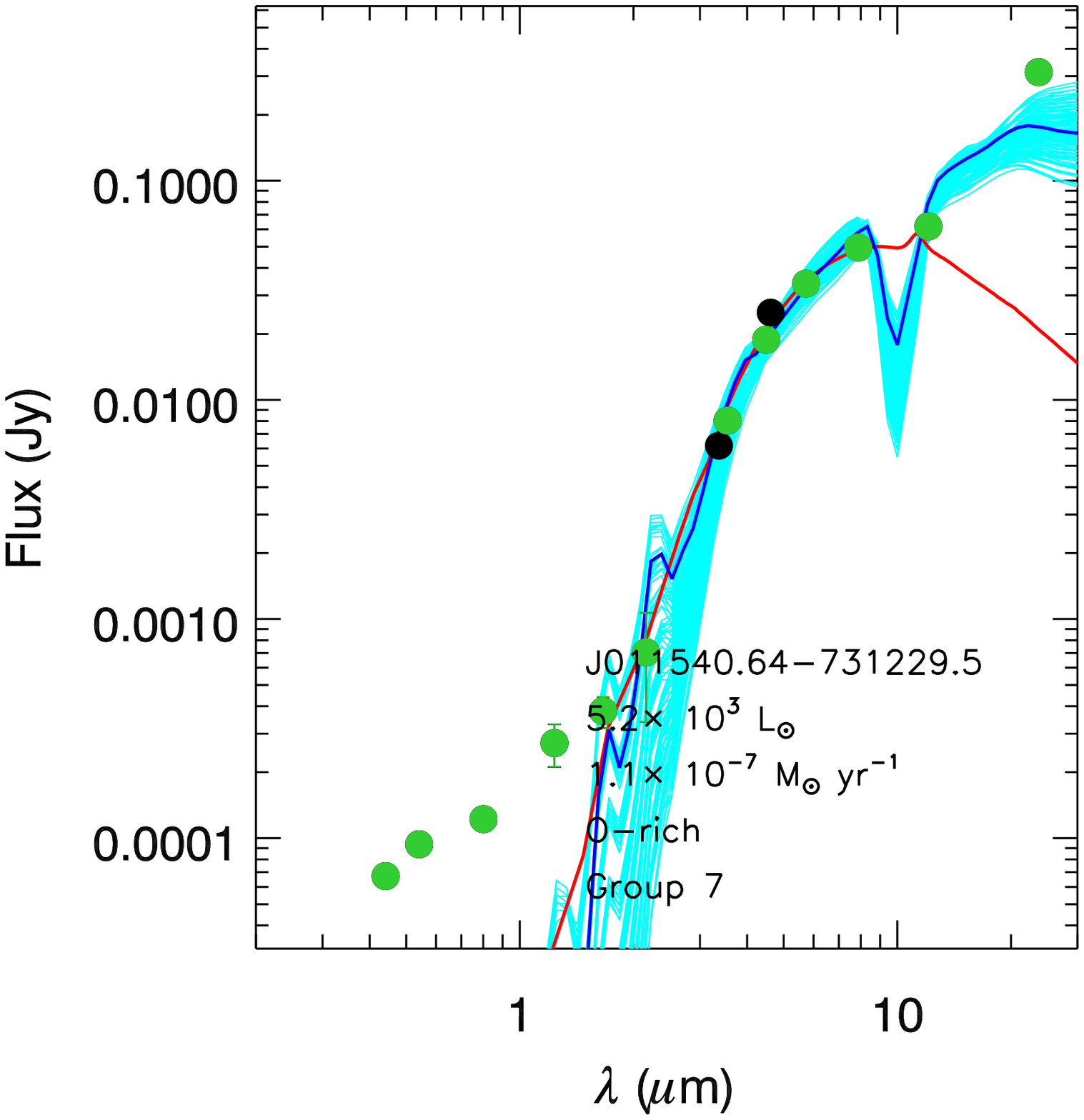}\includegraphics[width=50mm]{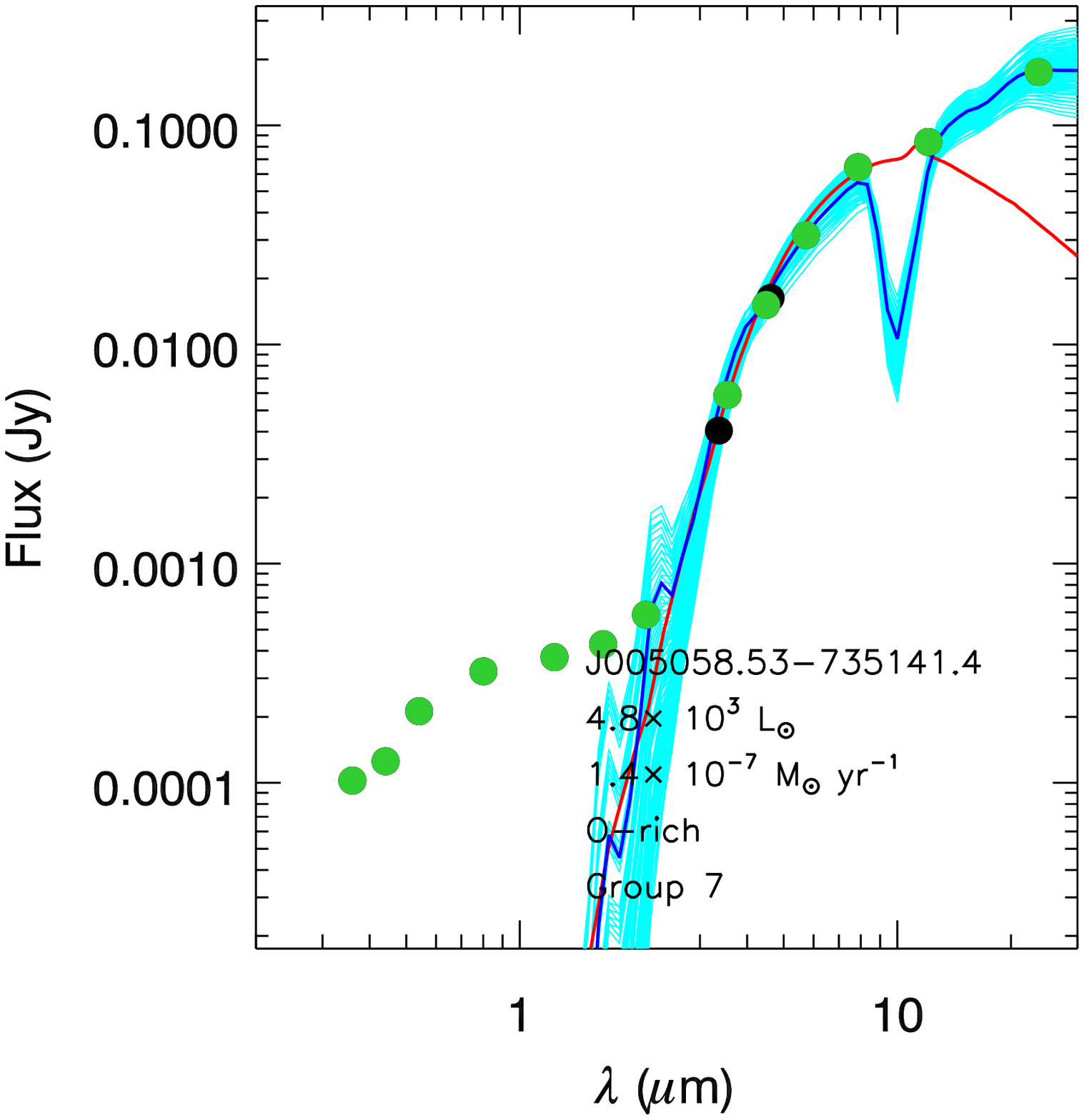}\\
\includegraphics[width=50mm]{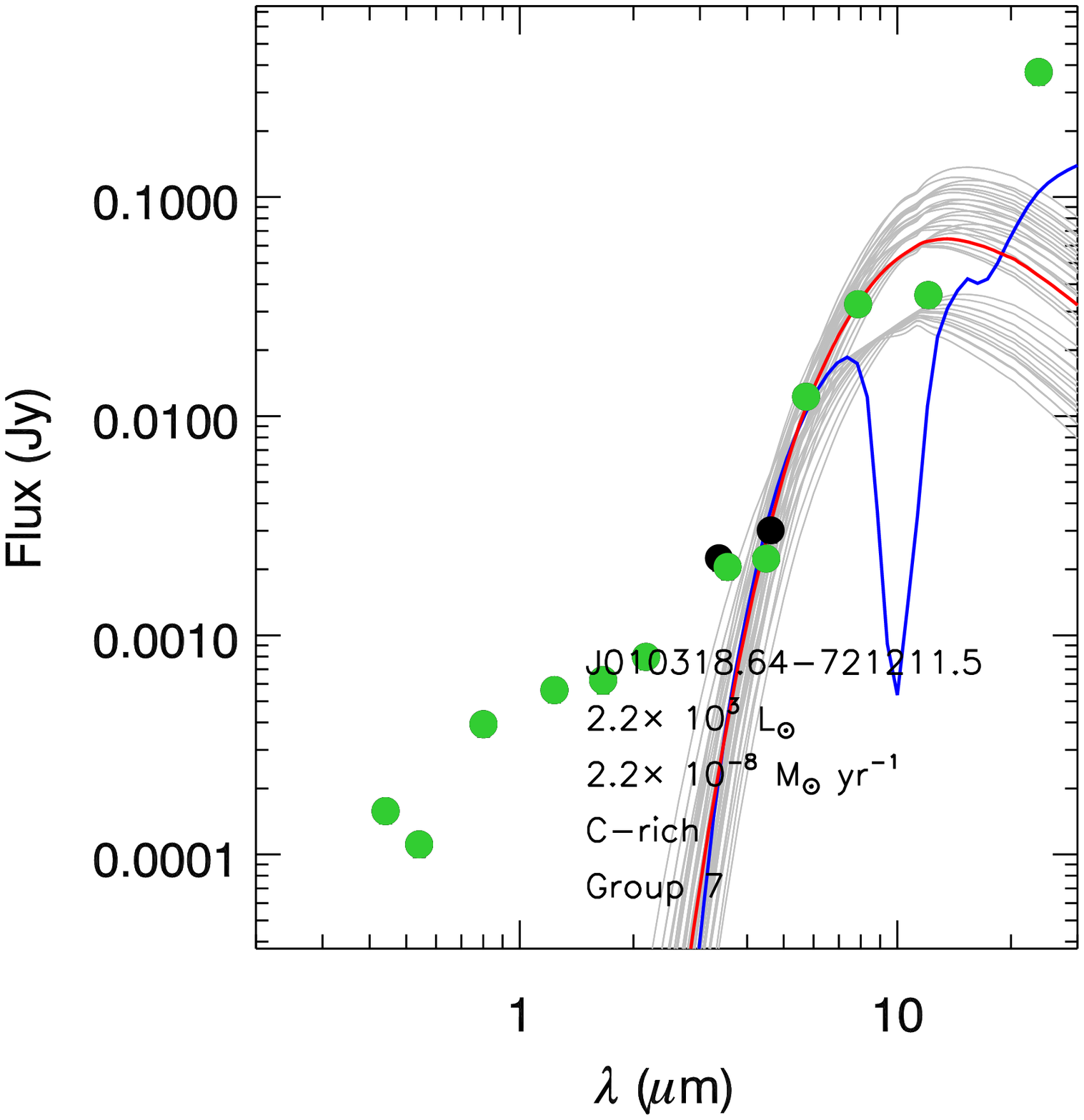}
\end{figure*}
\renewcommand{\thefigure}{\arabic{figure}}

Fig. \ref{fig:invalidFIRfits} shows the fits to FIR objects in Groups 5--7. The total DPR of this sample is about $4.6\times 10^{-6}$ \msunperyr, comparable to the total from the valid fits (Section \ref{subsec:FIRfitsdiscussion}). However, the fit quality is poor in general. Moreover, based on their SED shapes, they are unlikely to be AGB/RSG stars.

\label{lastpage}
\end{document}

%% file: smcsedfit_definitions.tex
\newcommand{\msunperyr}{M$_\odot$~yr$^{-1}$}
\newcommand{\globaldustbudget}{$(1.3\pm 0.1)\times 10^{-6}$}
\newcommand{\xdustpercent}{66}
\newcommand{\xnumberpercent}{7}
\newcommand{\mic}{$\mu$m}

\newcommand{\cumulativeDPRallLMC}{$2.1\times 10^{-5}$}
\newcommand{\ismdustmass}{$(8.3 \pm 2.1)\times 10^4$}
\newcommand{\excessbaseddustbudget}{$(8.6-9.5)\times 10^{-7}$}
\newcommand{\colorbaseddustbudget}{$7\times 10^{-6}$}
\newcommand{\ks}{{\it K$_{\rm s}$}}
\newcommand{\alambda}{0.154, 0.139, 0.120, 0.071, 0.029, 0.017, and  0.010}
\newcommand{\NOGLEmatches}{5\,068}
\newcommand{\NIRSFmatches}{7\,617}
\newcommand{\NSthreeMCmatches}{4\,179}
\newcommand{\NAKARImatches}{463}
\newcommand{\NWISEmatches}{8\,765}
\newcommand{\Nmisclassified}{88}
\newcommand{\NSMCSpecnotES}{23}
\newcommand{\NSMCSpecES}{81}
\newcommand{\NrawO}{2486}
\newcommand{\NvetO}{2485}
\newcommand{\NrawC}{1714}
\newcommand{\NvetC}{1714}
\newcommand{\NrawX}{353}
\newcommand{\NvetX}{341}
\newcommand{\NrawaO}{1198}
\newcommand{\NvetaO}{1198}
\newcommand{\NrawRSG}{3935}
\newcommand{\NvetRSG}{3850}
\newcommand{\NrawFIR}{95}
\newcommand{\Nrawev}{9781}
\newcommand{\Nvetev}{9621}
\newcommand{\Nvetignoreoptical}{3\,044}
\newcommand{\Nvetignoreopticalhighchisq}{264}
\newcommand{\NvetnonFIRflaggedinvalid}{7}
\newcommand{\Nvetchangechemtype}{seven}
\newcommand{\chiratiolowo}{0.20}
\newcommand{\chiratiohio}{0.35}
\newcommand{\chiratiolowc}{0.39}
\newcommand{\chiratiohic}{0.64}
\newcommand{\Nhicono}{1913}
\newcommand{\Nhiconc}{1558}
\newcommand{\percenthicono}{67}
\newcommand{\percenthiconc}{75}
\newcommand{\NOrichaboveLAGB}{84}
\newcommand{\NRSGaboveLAGB}{80}
\newcommand{\bolcorra}{$2.1\pm 0.1$}
\newcommand{\bolcorrb}{$1.0\pm 0.1$}
\newcommand{\bolcorrc}{$-0.25\pm 0.04$}
\newcommand{\NhiDPR}{236}
\newcommand{\NhiDPRO}{31}
\newcommand{\totalDPRhiDPRO}{$2.6\times 10^{-7}$}
\newcommand{\totalDPRhiDPRC}{$7.2\times 10^{-7}$}
\newcommand{\meanDPRhiDPRO}{$8.3\times 10^{-9}$}
\newcommand{\meanDPRhiDPRC}{$3.5\times 10^{-9}$}
\newcommand{\NhiDPRxAGB}{223}
\newcommand{\NvetFIR}{33}
\newcommand{\chisq}{$\chi^2$}
\newcommand{\GroupIIIandIVtotalDPRifallO}{$3.9\times 10^{-7}$}
\newcommand{\GroupIIIandIVtotalDPRifallC}{$4.8\times 10^{-8}$}
\newcommand{\GroupItoIVtotalDPR}{$(0.45\pm 0.01)\times 10^{-6}$}
\newcommand{\GroupVtoVIItotalDPR}{$(1.5\pm 0.03)\times 10^{-6}$}

\newcommand{\dustbudgetOrich}{$6.3\times 10^{-7}$}
\newcommand{\dustbudgetCrich}{$7.1\times 10^{-7}$}
\newcommand{\treplenish}{$(3.6-78)$}
\newcommand{\CCSNeDPR}{$5.1\times 10^{-4}$}
\newcommand{\SNeDPRlowerlimit}{$0.01\times 10^{-4}$}
\newcommand{\treplenishboostwithSNe}{$(0.1-44)$}
\newcommand{\TemimetallifetimeO}{$(54\pm 32)$}
\newcommand{\TemimetallifetimeC}{$(72\pm 43)$}
\newcommand{\AGBRSGdustmasscontribution}{$(125-1900)$}
\newcommand{\dustmasscontributionwithSNe}{$(300-64\,000)$}
\newcommand{\NSewiloYSOsmatched}{53}
\newcommand{\NSewiloYSOsapphotxAGB}{7}
\defcitealias{B2011}{B2011}
\defcitealias{B2012}{B2012}
\defcitealias{R2012}{R2012}